\documentclass[aps, prb, letterpaper, 10pt, twocolumn]{revtex4-2} 
\usepackage[T1]{fontenc}
\usepackage[utf8]{inputenc}
\usepackage{amsmath}
\usepackage{amsfonts}
\usepackage{amssymb}
\usepackage{times}
\usepackage{bbm, dsfont}
\usepackage[breaklinks,colorlinks,bookmarks=false,citecolor=blue,linkcolor=blue,urlcolor=blue]{hyperref}
\usepackage{orcidlink}

\newcommand{\nn}{\nonumber}
\newcommand{\cv}[1]{\left(#1\right)}
\newcommand{\cvb}[1]{\left[#1\right]}
\newcommand{\cvc}[1]{\left\{#1\right\}}
\newcommand{\cvv}[1]{\left\vert #1\right\vert}
\newcommand{\ket}[1]{\vert#1\rangle}
\newcommand{\bra}[1]{\langle#1\vert}
\newcommand{\ketbra}[2]{\ket{#1}\!\bra{#2}}
\newcommand{\braket}[1]{\langle#1\rangle}
\newcommand{\iden}{\mathbbm{1}}

\newcommand{\tra}{\text{Tr}}
\newcommand{\vb}[0]{V$_\text{B}^-$ }

\begin{document}

\title{Synchronous manipulation of nuclear spins via boron vacancy centers in hexagonal boron nitride}

\author{Fattah Sakuldee\,\orcidlink{0000-0001-8756-7904}}
\email{fattah.sakuldee@sjtu.edu.cn}
\affiliation{Wilczek Quantum Center, School of Physics and Astronomy, Shanghai Jiao Tong University, 200240 Shanghai, China}

\author{Mehdi Abdi\,\orcidlink{0000-0002-9681-5751}}
\email{mehabdi@gmail.com}
\affiliation{Wilczek Quantum Center, School of Physics and Astronomy, Shanghai Jiao Tong University, 200240 Shanghai, China}

\begin{abstract}
We develop a method for entangling operations on nuclear spins surrounding a negatively charged boron vacancy (VB-center) point defect in hexagonal boron nitride (hBN). To this end, we propose to employ the electron spin of a VB-center as a control qubit. We show that in the presence of a background magnetic field and by applying control pulses, one can collectively manipulate the state of the nuclei with $\hat{U}_z$ and $\hat{U}_x$ rotations. These rotations can serve for implementing the synchronous three-qubit $X$, $Z$, and the Hadamard gates. Through our numerical analyses considering realistic system parameters and the decoherence effects, we demonstrate that these gates can be executed with high fidelities. Furthermore, as an example for the application of our toolbox, we utilize these collective gates to prepare the highly entangled GHZ states among the three nuclear spins with a fidelity of $0.99$. By including the electron decoherence effects, we find that the relative deviations of the gate fidelities from the noisy terms are negligibly small, proving the noise-resilience of our protocols. Our work can serve as the foundation for exploiting the nuclear spins in hBN in future quantum technological applications.
\end{abstract}

\date{\today}
		
\maketitle

\section{Introduction}\label{sec:introduction}
Nuclear spins are reliable solid-state candidates for storing quantum information, because of their long coherence time~\cite{Morton2008, McCamey2010, Muhonen2014}.
Despite the difficulties in their accessibility for read-write and possibly other manipulations, several techniques have so far been proposed and even successfully implemented~\cite{Witzel2007, Schuetz2014}.
Devising versatile methods for not only transferring a quantum state to or from nuclear spins, but further accessibility for applying various gates can significantly push forward the idea of fault-tolerant quantum information processing in such platforms~\cite{Rong2015, Abobeih2022}.
Such methods mostly rely on invoking a quantum mediator, namely, an electron spin interacting with the nuclei~\cite{Denning2019}.
The color centers in solids are one of the most reliable objects that can provide such access, thanks to the exhaustive control over the electron spin quantum mediator~\cite{Parthasarathy2023}. In particular, diamond, several breakthroughs have been achieved leading to profound developments in controlling nitrogen and carbon nuclear spins due to the high controllability of the NV-centers~\cite{Zhao2011,Kolkowitz2012,Taminiau2012,Smeltzer2009,Busaite2020, ChenQ2017,Haase2018,Perlin2019,Abobeih2019,Goldman2020,Tratzmiller2021,Soshenko2021,JiangJ2022,MunueraJavaloy2023,Zeng2024,Xu2024,Mizuno2024,Bartling2025}.
In particular, for the application in quantum computation, it has been recently shown that universal one-qubit and two-qubit gates can be realized with NV-centers~\cite{Bartling2025,Jaeger2024}, and the implementation an error correction onto such gate operations has been devised~\cite{Dulog2024}. Expansion of such fine controls to similar physical systems can significantly help to advance the field. Among the others, controlling nuclear spins in hexagonal structures of boron nitride via the embedded color centers seems a promising option~\cite{Lee2025}.

The experimental observation of color centers in the layered hexagonal boron nitride (hBN) structures~\cite{Tran2015, Gottscholl2021, Stern2022, Mu2022} and the subsequent theoretical investigations~\cite{Abdi2018, Sajid2020} have identified highly controllable electron spins and raised the quest for employing them for various quantum purposes.
The negatively charged boron vacancy centers (\vb), which in this paper we simply refer to as VB-center, are shown to be accessible for initialization and control via optical and microwave pulses~\cite {Gottscholl2020}.
The utilization of these centers in quantum applications has attracted great interest from the quantum community in the past few years.
It is a promising physical platform for several purposes raging from a host for quantum emitters~\cite{Cakan2025, Wolfowicz2021, Chejanovsky2016, Proscia2018, Exarhos2019} to quantum manipulation of other degrees of freedom and quantum sensing~\cite{Tabesh2021, Liu2019, Abdi2021, Kianinia2020, Shaik2021, Gao2021, Mendelson2021, Guo2022, Vaidya2023, Das2024}.
Even though the basics of a qubit in the manipulation (initialization, control, and readout) of the electron spin of a VB-center in hBN has been established (see e.g. Refs.~\cite{Gottscholl2021,Gottscholl2020,Wong2015}), a reliable and scalable quantum information processing strategy still requires further investigations and is an ongoing agenda.
There remain several open possibilities to incorporate other degrees of freedom into the processing, which would lead to a more fruitful deployment of the potential of the mentioned system.
Among the others, the nuclear spins in hBN offer a rather unique feature in systems hosting color centers.
That is, all the stable isotopes of both nitrogen and boron have non-zero nuclear spins forming a lattice.
This natural proximity of nuclei to the color center results in a large interaction with the electron spin, and yet in a regular fashion.
Namely, given the lattice structure, the nuclear spins form an orderly spin structure, rather than being randomly located.
These two features---strong coupling and the order---can be exploited to efficiently manipulate the quantum state of the nuclear spins.
It is the purpose of this work to partially address this issue by putting forth a quantum toolbox composed of control gates for the collective manipulation of the three nearest nuclear spins surrounding a VB-center in hBN.

The strong hyperfine interaction between nuclear spins and the electron spin compared to competing interactions~\cite{Ivady2020, Gao2021}, alongside the possibility to hyperpolarize the nuclear spins~\cite{Gao2022, Tabesh2023} as well as the great controllability of the electron spin in VB-center~\cite{Gottscholl2021}, all offer a promising opportunity for devising an indirect control over the nuclear spins via the VB-centers in hBN. 
The same principle can also be applied to the writing and reading processes of nuclear spin registers. Namely, since the central spin can be manipulated and read out efficiently \cite{Gottscholl2021,Murzakhanov2022}, one can employ two-qubit gates to transfer the state between the nuclear spins and the electron spin, allowing the writing and reading processes of nuclear spins via the central electron spin as the mediator. 
In this work, we propose two crucial practical schemes in such perspective in the language of quantum computation, namely a synchronous nuclear spin qubit gates implementation and a preparation of three-qubit entangled state---the Greenberger-Horne-Zeilinger (GHZ) states.

We first demonstrate that the three $^{14}$N nitrogen nuclear spins---in an isotopic purified sample---with the strongest coupling strength to the central spin, can be considered as three copies of an identical spin, which can be further reduced to three identical copies of a qubit.
By employing the symmetric nature of the hyperfine interaction and the dynamical control techniques, we conceive the entangling gates, in which the reduced action on each nuclear spin is a rotation with the same phase across all the spins.
In other words, such three identical qubits can be manipulated synchronously while assuming identical states.
Such rotations are then exploited to conceive the Pauli gates $X$ and $Z$ as well as the Hadamard gate.
These provide the basic tools for quantum information processing from a quantum computation perspective. 
In addition to the synchronous control, we also consider the application of these entangling gates for the preparation of GHZ states, which are highly entangled states with various applications, e.g., for quantum sensing.
Such states can be attained by the concatenation of the spin initialization, the entangling gates, and a central spin measurement.
The protocol leads to reduced operators on the nuclear spins subspace mapping the nuclear initial state to two GHZ states with different phases, where one of them can be transformed into the other via proper post-processing.
The implication of the GHZ states opens another way to utilize the nuclear spins in hBN for quantum sensing. 

The paper is organized as follows.
In Sec.~\ref{sec:formulation} we discuss the model of three nuclear spins coupled to a central electron spin and the mechanism for the realization of entangling gates used in our protocols.
We use a two-level subspace of the electron spin as a control qubit subspace to manipulate the nuclear spins of the three adjacent nitrogen atoms.
Later, we introduce an effective qubit subspace for the spin-1 system in Sec.~\ref{sec:spins_algebra}, where we demonstrate that the entangling gates can be considered as Pauli gates on such an effective subspace of each nuclear spin. The gate fidelities for realistic parameters of the system are numerically studied in Sec.~\ref{sec:Gates_Fidelity}. 
In Sec.~\ref{sec:GHZ_prepatation}, we utilize the entangling gates for the preparation of GHZ states, in which a high fidelity of the process can also be obtained.
We then consider the effect of electron dephasing in Sec.~\ref{sec:dephasing}, and finally, a conclusion is given in Sec.~\ref{sec:conclusion}.

\section{Formulation}\label{sec:formulation}
In this section, we discuss the relevant model of an electron spin and three ${}^{14}$N nuclear spins cluster in a boron vacancy center in hexagonal boron nitride.
We show that by focusing on only a two-level subspace of the central spin, one can construct collective entangling gates, where the effective phase on all nuclear spins is approximately uniform.
Such collective gates are allowed by neglecting the fast-rotating terms, and by introducing a control term to the Hamiltonian, where the remaining coupling coefficients after the manipulation are uniform across all nuclear spins.

\subsection{Boron vacancy in hexagonal boron nitride} \label{sec:hBN}
\begin{figure}[t]
	\centering
	\includegraphics[width=\columnwidth]{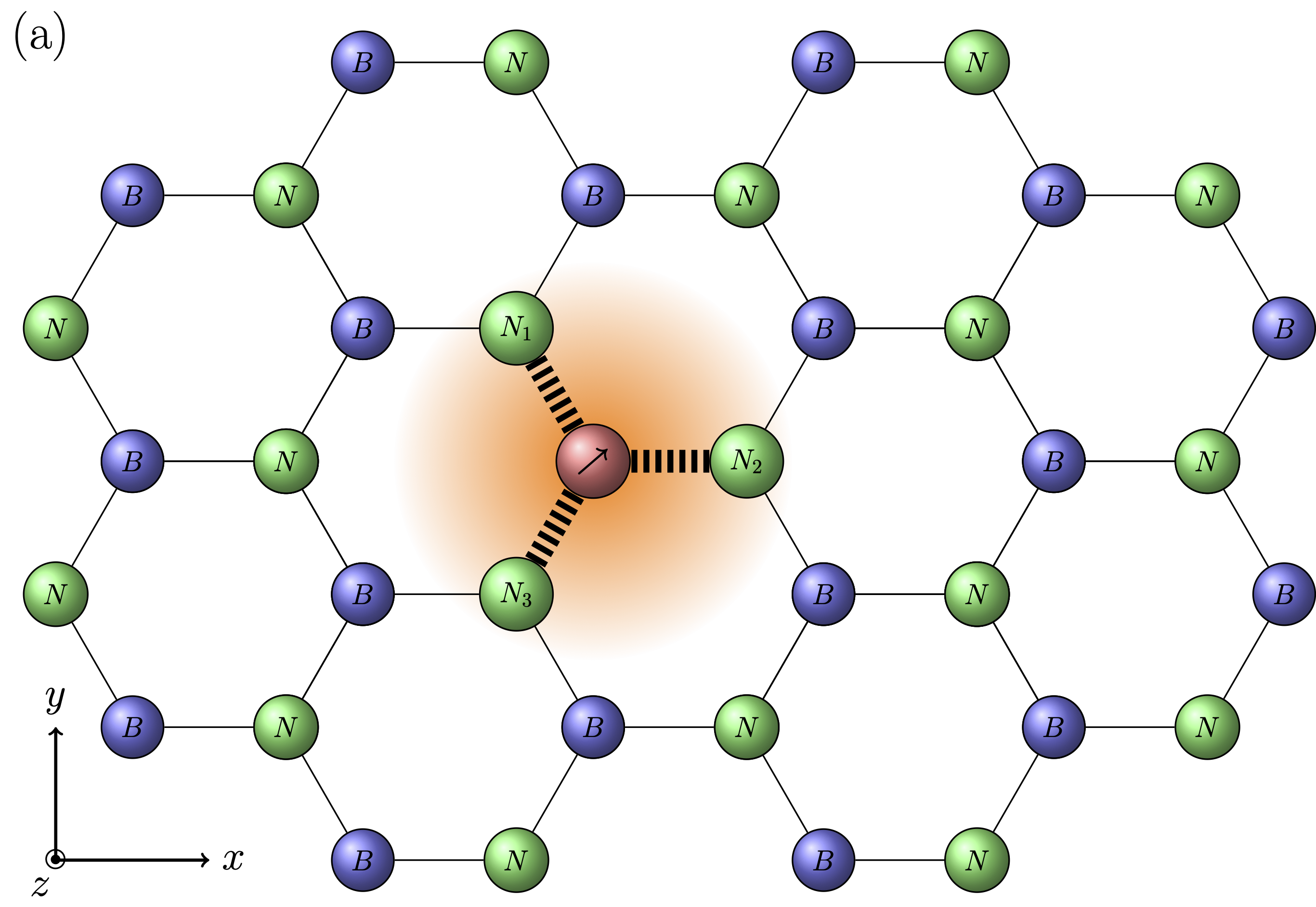}\\
    \vspace{0.5cm}
    \includegraphics[width=0.47\columnwidth]{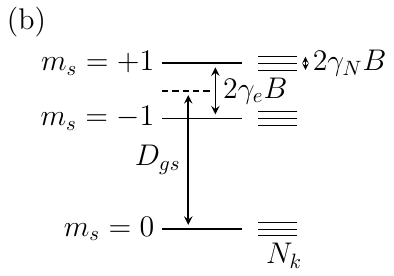}\hfill
    \includegraphics[width=0.47\columnwidth]{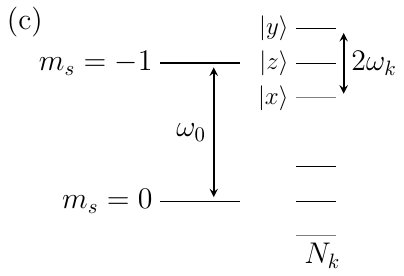}
	\caption{The geometry of VB-center in hBN (a): the electron spin of the boron vacancy (red ball) strongly couples to the three nearest nitrogen nuclear spins (green balls $N_1$ to $N_3$). Such strong interactions are enabled by Fermi contact (illustrated by the orange area around the VB-center). Other interactions of the electron spin with other sites are mainly from dipolar interactions, which are significantly weaker. A magnetic field is applied perpendicular to the lattice plane, i.e., the $z$-direction. (b) Energy level structure of the electron and either of the nuclei, which is symmetric under permutation of $\{N_k\}$ for $k=1,2,3$. (c) The effective energy level structure in the reduced Hilbert space picture is shown, where the Larmor frequencies of spin $N_k$ are shifted to $\omega_k = \gamma_NB + A_k^{zz}/2$ for the magnetic field in the $z$-direction.
    }\label{fig:hBN}
\end{figure}

The nuclear spin of the three nearest neighbors of a negatively charged boron vacancy color center in hBN (VB center) strongly couples with the electron spin via Fermi contact (and small contribution of dipolar) interaction \cite{Ivady2020,Gottscholl2021}.
The coupling is anisotropic and thus allows one to systematically manipulate the nuclear spins when a proper external magnetic field is applied.
The Hamiltonian describing the whole system is $\hat{H}_{sys} = \hat{H}_e + \hat{H}_n + \hat{H}_i$ with the following components ($\hbar = 1$):
	\begin{equation}
		\hat{H}_e = D_{gs}\hat{S}_z^2 + \gamma_e \mathbf{B}\cdot\mathbf{S} \label{eq:H_e},
	\end{equation}
where $D_{gs}$ is the zero-field splitting of VB-center groundstate triplet and $\gamma_e$ is the electron gyromagnetic ratio.
Here, $\mathbf{B}$ is the vector of background magnetic field, while $\mathbf{S} {=} (\hat{S}_x,\hat{S}_y,\hat{S}_z)^\mathrm{T}$ is the vector of spin-$1$ operators.
The nuclear spins are also affected by the background magnetic field and the electric quadrupole interaction.
They also couple to each other through magnetic dipole-dipole and electric quadrupole interactions, which are negligibly small compared to the other frequencies.
Hence, the Hamiltonian describing the dynamics of the nuclear spins reads
	\begin{equation}
		\hat{H}_n = - \sum_k\gamma_k \mathbf{B}\cdot\mathbf{I}_k \label{eq:H_n},
	\end{equation}
where $\gamma_k$ and $\mathbf{I}_k$ are the nuclear gyromagnetic ratio and the spin operators vector, respectively. The electron spin of the VB-center couples to the nuclei via two different mechanisms.
For the nearest-neighbor nuclear spins (three nitrogen atoms), the interaction is dominated by the Fermi contact, which is independent of the magnetic field orientation. 
This contribution is determined by the overlap of the wave function of the electron spin and the nuclear spins, namely, the coupling between the electron spin and the nuclear spin $k$ is proportional to $\vert\Psi(\mathbf{r}_k)\vert^2 \mathbf{S}\cdot\mathbf{I}_k$ where $\Psi(\mathbf{r}_k)$ is the electron wavefunction at the location of the $k$th nuclear spin~\cite{Tabesh2023,Doherty2013}. Given the localized nature of the defect orbitals, see e.g.~\cite{Abdi2018}, only contributions from the three nearest nuclear spins are relevant, whereas the interactions with the other sites are negligibly small, see Fig.~\ref{fig:hBN}(a) for an illustration. 
The electron also couples to them as well as the farther spins in a dipole-dipole fashion, but the coupling strength is at least one order of magnitude smaller~\cite{Ivady2020}.
Therefore, we consider the four-spin system composed of the VB-center and the three nearest-neighbor nitrogen nuclei as a computational unit.
The interaction Hamiltonian reads
	\begin{equation}
		\hat{H}_i =  \sum_k\mathbf{S} \cdot\mathbb{A}_k\cdot\mathbf{I}_k \label{eq:H_i},
	\end{equation}
where $\mathbb{A}_k$ is the hyperfine tensor of the $k$th spin.
Here, we have limited the interaction to the hyperfine interaction between the \vb center and the three nearest neighbor nuclei and have omitted the other interactions. This is justified because of the much smaller coupling rates for the other interactions, as mentioned above. The experimental data as well as the computational works suggest an anisotropy in the coupling of the VB-center to the surrounding nuclei~\cite{Ivady2020, Liu2022}. We will shortly show how this property can be exploited for the manipulation of the nuclear spins through the electron.

In fact, the hyperfine interaction matrix $\mathbb{A}_k$ inherits the geometric $D_{3h}$ symmetry of the point defect and assumes the following form~\cite{Ivady2020}:
	\begin{equation}
		\mathbb{A}_k = \left(\begin{array}{ccc}
			A_k^{xx} & A_k^{xy} & 0\\
			A_k^{xy} & A_k^{yy} & 0\\
			0 & 0 & A_k^{zz}
		\end{array}\right)\label{eq:A_k}.
	\end{equation}
This can be represented as $\mathbb{A}_k = \mathbb{A}^\perp_k\oplus\cvb{A_k^{zz}}$.
Therefore, the hyperfine interaction can be seen as a composition of a parallel $\hat{S}_z \sum_kA_k^{zz}\hat{I}_k^z$ and an in-plane  $\sum_k\mathbf{S}_\perp\cdot\mathbb{A}^\perp_k\cdot\mathbf{I}_k^\perp$ interaction.
Here, we have introduced the in-plane spin operators as $\mathbf{S}_\perp {=} (\hat{S}_x,\hat{S}_y)^\mathrm{T}$ and $\mathbf{I}_k^\perp = (\hat{I}_k^x,\hat{I}_k^y)^\mathrm{T}$.
If the magnetic field vector is chosen such that the free evolution part of the Hamiltonian, i.e. $\hat{H}_{e}+\hat{H}_n$, contains only in-plane or parallel spin components, the dynamics governed by the total Hamiltonian $\hat{H}_{sys}$ can then be considered as two separate independent branches of the evolution of parallel and in-plane components.
Hence, a possibility of controlling these components independently is provided.
This could be the case, for example, when one applies the magnetic field in the parallel direction or the direction lying in the lattice plane.

In this work, we consider the magnetic field $\mathbf{B}$ pointing in the direction of the VB-center axis of symmetry, i.e. $\mathbf{B} {=} B\hat{z}$.
Also, we assume that one can probe and manipulate the \vb electron-subspace spanned by $\cvc{\ket{0},\ket{-1}}$ the set of two lowest eigenstates of $\hat{S}_z$, which can be achieved by employing proper microwave drive frequencies, see e.g. Refs.~\cite{Gottscholl2021, Gao2022}.
By introducing the Pauli matrix $\hat{\sigma}_z {=} \ketbra{0}{0} - \ketbra{-1}{-1}$ and similarly for $\hat\sigma_x$ and $\hat\sigma_y$, the relevant Hamiltonian, up to a global phase, in the two-level subspace reads
    \begin{equation}
		\hat{H} =  - \dfrac{\omega_0}{2}\hat{\sigma}_z - \sum_k (\omega_k - \dfrac{A_k^{zz}}{2}\hat{\sigma}_z)\hat{I}_k^z + \dfrac{1}{\sqrt{2}}\sum_k\boldsymbol{\sigma}_{\!\perp}\!\cdot\! \mathbb{A}^\perp_k \cdot \mathbf{I}_k^\perp,
		\label{eq:H_main}
	\end{equation}
where $\omega_0 = D_{gs} - \gamma_eB$ and $\omega_k = \gamma_n B + A_k^{zz}/2$ are effective Larmor frequencies of the electron and the $k$th nuclear spins, respectively [Figs.~\ref{fig:hBN}(b) and \ref{fig:hBN}(c)].
Here, $\boldsymbol{\sigma}_\perp {=} (\hat{\sigma}_x,\hat{\sigma}_y)^\mathrm{T}$ is a vector of the effective spin operators for the in-plane components.

\subsection{Dynamical Control} \label{sec:DD_control}
Now we introduce control drives to the system that rotate the electron spin in the aforementioned subspace.
The drive Hamiltonian thus reads $\hat{H}_{\rm C}\cv{t} = h_x\cv{t}\hat{\sigma}_x + h_y\cv{t}\hat{\sigma}_y$ with the functions $h_x\cv{t}$ and $h_y\cv{t}$ that need to be properly chosen.
The total controlled Hamiltonian can then be described as
	\begin{equation}
		\hat{H}_{\rm tot}\cv{t} = \hat{H} + \hat{H}_{\rm C}\cv{t} \label{eq:H_0+C}.
	\end{equation}
In our study, we assume that only one of the controls in a rotating frame with respect to $\hat{H}_e$ is applied at a time and that both functions $h_x$ and $h_y$ are square pulses in that frame.
These two assumptions simplify the problem and guarantee the filtered Hamiltonian form in the following parts of this article (see Appendices~\ref{appen:H_rotate-derivation} and \ref{appen:filter-insertion} for details). 

With these, one can rewrite the Hamiltonian Eq.~\eqref{eq:H_0+C} in the rotating frame with respect to the free Hamiltonian
$\hat{H}_0 {=} {-}\frac{\omega_0}{2}\hat{\sigma}_z - \sum_k\omega_k\hat{I}_k^z + \hat{H}_{\rm C}\cv{t}$
in the following form
	\begin{align}
		\widetilde{H}\cv{t} &\approx \hat{H}_{\rm eff}\cv{t} + \hat{H}_{\rm RW, odd}\cv{t} + \hat{H}_{\rm CRW}\cv{t} \label{eq:H_rotate}.
	\end{align}
Here, the effective Hamiltonian is the approximation for $\widetilde{H}\cv{t}$ and is given by the following
	\begin{align}
		\hat{H}_{\rm eff}\cv{t} &= a_zF_z\cv{t}\hat{\sigma}_z\sum_k\hat{I}_k^z + a_\perp F_x\cv{t} \cos\Delta t\hat{\sigma}_x\sum_k\hat{I}_k^x\nn\\ 
			&\phantom{=.}+ a_\perp F_y\cv{t} \cos\Delta t\hat{\sigma}_y\sum_k\hat{I}_k^y.
			\label{eq:H_eff}
	\end{align}
This is found by replacing the individual coupling coefficients in the first three terms with their averages:
	\begin{align*}
		\Delta &= \sum_k\cv{\omega_0 - \omega_k}/3,
		\\
		a_z &= \dfrac{1}{6}\sum_kA_k^{zz},~~a_\perp = \dfrac{1}{6\sqrt{2}}\sum_k\cv{A_k^{xx} + A_k^{yy}}.
	\end{align*}
In Eq.~\eqref{eq:H_eff} the functions $F_\alpha\cv{t}$ with $\alpha =x,y,z$ are filter functions induced by the control Hamiltonian $\hat{H}_{\rm C}\cv{t}$.
Meanwhile, the rotating term with odd modulations is
    \begin{equation}
        \hat{H}_{\rm RW, odd}\cv{t} = \sum_kA_k^\perp\sin\Delta_k t\cv{F_y(t)\hat{\sigma}_y\hat{I}_x
        -F_x(t) \hat{\sigma}_x\hat{I}_k^y} \label{eq:H_RW-odd},
    \end{equation}
and $\hat{H}_{\rm CRW}\cv{t}$ includes all counter-rotating terms.
Both these latter contributions in the dynamics become vanishing by a proper choice of the filter functions, see below. 
Note that the homogeneity assumed here in introducing $\Delta,$ $a_z$, and $a_\perp$ results from the symmetric nature of the hyperfine interaction~\cite{Gao2022,Ivady2020}.
The derivation for the rotated Hamiltonian Eq.~\eqref{eq:H_rotate} and its full expression is given in Appendix~\ref{appen:H_rotate-derivation} while the filter function formulation can be found in Appendix~\ref{appen:filter-insertion}.

To obtain collective gates with desired rotations on the nuclear spins, one could choose the filter functions such that only one of the interactions in $\hat{H}_{\rm eff}$ survives and the other two interactions as well as the extra terms $\hat{H}_{\rm RW, odd}\cv{t} + \hat{H}_{\rm CRW}\cv{t}$ vanish.
For instance, to single out the term $\hat{\sigma}_x\sum_k\hat{I}_k^x$ one should select the filter functions such that $F_z\cv{t}$, $F_y\cv{t}$, and $\hat{H}_{\rm RW, odd}\cv{t} + \hat{H}_{\rm CRW}\cv{t}$ average out to zero, that will be exemplified in the following.

In this paper, we discuss the construction of the following gates without loss of generality
	\begin{subequations}
	\begin{align}
		\hat{U}_z\cv{\varphi} &= \exp\{-i\varphi\hat{\sigma}_z\sum_k\hat{I}_k^z\} \label{eq:U_z}\\
		\hat{U}_x\cv{\varphi} &= \exp\{-i\varphi\hat{\sigma}_x\sum_k\hat{I}_k^x\} \label{eq:U_x}.
	\end{align}
	\end{subequations}	
Rotations about the $z$-axis require no control pulses and can be approximately obtained by choosing the magnetic field to be strong.
In other words, when $\cvv{\Delta}{\gg} |A_k^{\alpha\beta}|/\sqrt{2}$ for $\alpha,\beta = x,y,$ all interactions except for the $\hat\sigma_z\hat{I}_k^z$ terms in Eq.~\eqref{eq:H_rotate} become counter-rotating and thus negligible.
In this case, arbitrary rotation angles are attainable by setting the evolution time to $t=\varphi/a_z$ such that 
	\begin{equation*}
		\exp\{-i\int_0^t \widetilde{H}\cv{t} dt\} \approx \exp\{-ia_zt\hat{\sigma}_z\sum_k\hat{I}_k^z \} = \hat{U}_z\cv{\varphi}.
	\end{equation*}
In particular, the oscillation terms in $\hat{H}_{\rm eff}\cv{t}$ and $\hat{H}_{\rm RW,odd}\cv{t}$ with the frequency $\Delta$ can be considered as fast rotating terms, which are averaged out in the dynamics.
Since $\hat{H}_{\rm CRW}\cv{t}$ contains oscillatory terms with a frequency $\Sigma {=} \sum_k\cv{\omega_0+\omega_k}/3$ which are larger than $\Delta$, the contribution from $\hat{H}_{\rm CRW}\cv{t}$ can also be neglected, see Appendix~\ref{appen:H_rotate-derivation}. 
 
The construction of $\hat{U}_x$ is more complicated and it demands an active control ($\hat{H}_{\rm C}(t) \neq 0$). In this work, we will employ a conventional Carr-Purcell-Meiboom-Gill (CPMG) sequence discussed in the following.

\subsection{Preparation of $\hat{U}_x$} \label{eq:Ux_prep}
The CPMG sequence is implemented by the insertion of $\pi-$rotation about $\sigma_y.$ For a sequence of period $T=2\pi/\Omega,$  such an impulse is put at the time $T/4$ and $3T/4,$ resulting in a set of filter functions:
	\begin{equation}
		F_x\cv{t} = \sum_{k=0}^\infty\dfrac{4\sin\cv{k\pi/2}}{k\pi}\cos\cv{k\Omega t} \label{eq:F_x},
	\end{equation}
$F_y\cv{t}=1$ and $F_z\cv{t} = F_x\cv{t}.$ Here, the pulse sequence is expressed in the rotating frame with respect to $\hat{H}_e.$ 
Now we choose the period such that the electron Larmor frequency is an odd multiple of the fundamental frequency $\Delta = p\Omega.$ Suppose that the sequence completes at the $N$th period. One then has
	\begin{align}
		&\exp\big(-i\int_0^{NT} \widetilde{H}\cv{t} dt\big)\nn\\ 
			&
            \exp{\bigg\{}{-i}\cv{\dfrac{4\sin\cv{p\pi/2}a_\perp N}{\Delta}}\hat{\sigma}_x\sum_k\hat{I}_k^x + C(NT){\bigg\}},
   \label{eq:Ux_construct}
	\end{align}
where $C\cv{NT} = -i\int_0^{NT} \hat{H}_{\rm CRW}\cv{t} dt$ is the residue of counter-rotating terms.
The terms with the filter functions $F_z\cv{t}$ and $F_y\cv{t}$ are averaged out to zero, while the integral of $\hat{H}_{\rm RW, odd}\cv{t}$ vanishes from the orthogonality of trigonometric functions. 

The contribution $C\cv{NT}$ is relatively small compared to the main contribution since the effective coupling coefficients are small compared to $a_\perp$ and the average of the oscillation terms with counter-rotating frequencies $\Sigma_k = \omega_0 + \omega_k$ are off-resonant. We avoid long-term behavior when such an effect can accumulate and alter the observed results. The exact form of the remaining term $C\cv{NT}$ is complicated, but it can be made small by choosing an appropriate $p$ and other parameters, e.g., $B$ and $N.$ In other words, with appropriate choices of parameters, one can safely remove the contribution $C\cv{NT}$ in some relevant situations, which will be exemplified by numerical calculation in Sec.~\ref{sec:Fidelity}. 

With this scheme, the target rotation $\hat{U}_x\cv{\varphi}$ is attained with a finite resolution.
Namely, for a given rotation angle $\varphi$ in the preparation with $N$ periods of CPMG pulses, one has $\cvv{\varphi} = N\delta_\Delta$ with $\delta_\Delta = 4a_\perp/\cvv{\Delta}$.
This suggests that $\Delta$ exhibits a resolution parameter and $N$ becomes the counting factor for a given phase $\varphi.$
To achieve a reasonable resolution, one can choose the magnetic field to be large.
However, having a large $\Delta$ may cause another practical issue, namely, the frequency of the control pulses $\Omega = \Delta/p$ will need to become large.
This widely depends on the limitation of the experimental setups, meaning that the ability to vary $\varphi$ is also limited by the capability of the control sequences implementation.

\section{Spins Algebras and Collective Gates}\label{sec:spins_algebra}
Before demonstrating the gate implementations, let us visit the algebraic structure of nuclear spins. We begin by discussing local manipulations of a two-level subspace embedded in each nuclear spin manifold and show that such manipulations can be prepared collectively from our devised gates. Later, we demonstrate that one can have a collective Pauli group on three nuclear two-level systems, and lastly, we use the preparation of GHZ states in the nuclei to exemplify the potential of our framework.

\subsection{Two Level Sub-Systems}\label{sec:TWS}
First, for clarity, let us define several notations as follows.
On the electron spin, let $\ket{\pm} {=} \cv{\ket{0} \pm \ket{-1}}/\sqrt{2}.$
Let $\ket{m_I}_k$ denote an eigenstates of $\hat{I}_k^z$ associated with the eigenvalues $m_I=0,\pm 1$ for the spin $k.$
Now, for the nuclear spins, we introduce $\ket{z}_k {=} \ket{0}_k,$ $\ket{x}_k {=} \cv{\ket{1}_k - \ket{-1}_k}/\sqrt{2}$ and $\ket{y}_k {=} \cv{\ket{1}_k + \ket{-1}_k}/\sqrt{2}$.
With these nomenclatures, one observes that
	\begin{subequations}
	\begin{align}
		\hat{I}_k^x &= \hat{X}_k^{yz} := \ket{y}_k\bra{z} + \ket{z}_k\bra{y}\label{eq:Ix-k}\\
		\hat{I}_k^y &= -\hat{Y}_k^{zx} := i\ket{z}_k\bra{x} - i \ket{x}_k\bra{z}\label{eq:Iy-k}\\
		\hat{I}_k^z &= \hat{X}_k^{xy} := \ket{x}_k\bra{y} + \ket{y}_k\bra{x}\label{eq:Iz-k}.
	\end{align}
	\end{subequations}	
In other words, each spin operator acts as an $X$ gate or $Y$ gate on two orthogonal states apart from the state associated with the operator label, e.g., the operator $\hat{I}_k^x$ swaps between $\ket{y}_k$ and $\ket{z}_k$ but maps $\ket{x}_k$ to null. 

By employing Euler's formula for spin-$1$ operators, this observation leads to
	\begin{align}
		\exp\cv{-i\varphi\hat{I}_k^x} &= \hat{P}_k^{x} + \cos\varphi\hat{P}_k^{yz} - i\sin\varphi\hat{X}_k^{yz} \label{eq:expIx-k}\\
		\exp\cv{-i\varphi\hat{I}_k^y} &= \hat{P}_k^{y} + \cos\varphi\hat{P}_k^{zx} + i\sin\varphi\hat{Y}_k^{zx} \label{eq:expIy-k}\\
		\exp\cv{-i\varphi\hat{I}_k^z} &= \hat{P}_k^{z} + \cos\varphi\hat{P}_k^{xy} - i\sin\varphi\hat{X}_k^{xy} \label{eq:expIz-k},
	\end{align}
where $\hat{P}_k^{\alpha} = \ket{\alpha}_k\bra{\alpha}$ are the projective operators and $\hat{P}_k^{\alpha\beta} = \hat{P}_k^{\alpha} + \hat{P}_k^{\beta}$ for $\alpha,\beta = x,y,z.$ 
From these expressions one can interpret that an exponential operator generated by $I_k^\alpha$ maintains the state $\ket{\alpha}_k,$ whereas it shuffles the other two states, with the transition amplitudes depending on the phase $\varphi$ as seen in the coefficients of $\hat{P}_k^{\alpha\beta}$ and $\hat{X}_k^{\alpha\beta}.$ Note that one can rewrite Eqs.~\eqref{eq:U_z} and \eqref{eq:U_x} as
	\begin{subequations}
	\begin{align}
		\hat{U}_z\!\cv{\varphi} &= \ketbra{0}{0}\!\otimes\!\prod_{k=1}^3e^{-i\varphi\hat{I}_k^z} + \ketbra{-1}{-1}\!\otimes\!\prod_{k=1}^3e^{i\varphi\hat{I}_k^z} \label{eq:U_z-dephasing-form},\\
		\hat{U}_x\!\cv{\varphi} &= \ketbra{+}{+}\!\otimes\!\prod_{k=1}^3e^{-i\varphi\hat{I}_k^x} + \ketbra{-}{-}\!\otimes\!\prod_{k=1}^3e^{i\varphi\hat{I}_k^x} \label{eq:U_x-dephasing-form}.
	\end{align}
	\end{subequations}
In this sense, it is possible to employ the entangling gates $\hat{U}_z\cv{\varphi}$ and $\hat{U}_x\cv{\varphi}$ to manipulate only two of the three states on each spin, with an aid of the electron state $\ket{0}$ or $\ket{+}$ as an auxiliary qubit. 
The effect on three spins is the same, thanks to the structure of the conditioned map on the nuclear spins part. That is, for a given electron state $\ket{0}$ or $\ket{+},$ all nuclear spins will be rotated with the same phase $\varphi$ synchronously. 

With the above formulation, one can employ any pair of states in $\cvc{\ket{x}_k,\ket{y}_k,\ket{z}_k}$ and use them as basis states for a qubit, allowing us to construct a three qubits subspace from the system of three nuclear spins.
Moreover, for any nuclear spin $k,$ since $\hat{U}_\alpha\cv{\varphi}$ for $\alpha =x,y,z$ concerns only one fixed axis $\alpha,$ one can manipulate each type of qubit subspace separately, synchronously on three nuclear spins at once, while leaving the components in the perpendicular subspace invariant.
For instance, one can choose a Hilbert subspace spanned by $\cvc{\ket{y}_k,\ket{z}_k}_k$ (which throughout this paper we will call either $yz$-subspace or $\mathcal{H}_{yz}$ for short) on which either $\hat{U}_z\cv{\varphi}$ or $\hat{U}_x\cv{\varphi}$ can be employed to manipulate the state within the subspace. This can be done by setting the initial state in $\mathcal{H}_{yz},$ or for the case of an initial mixed state, the density matrix on such Hilbert subspace $\mathcal{H}_{yz}.$
A similar framework is applicable for the other two $\mathcal{H}_{zx}$ and $\mathcal{H}_{zx}$ as well.
In the following, and throughout this article, we use $\mathcal{H}_{yz}$ as an example and demonstrate how one can construct three-qubit synchronous gates.

\subsection{Synchronous Pauli Group}\label{sec:Pauli}
Now we use the observations above for the preparation of qubit $Z$, $X$, and Hadamard gates on the subspace $\mathcal{H}_{yz}$.
For instance, one can consider the nuclear spins subspace spanned by $\cvc{\ket{y}_k,\ket{z}_k}_k$ and treat it as a collective two-level Hilbert space.
In our analyses we adopt the compact notation $\hat{A}_{\alpha\beta} = \hat{A}_1^{\alpha\beta}\hat{A}_2^{\alpha\beta}\hat{A}_3^{\alpha\beta}$ for a collective of the operator $\hat{A}.$ 
We call this type of operator, in which the action on each nuclear spin is identical to that of each other, a synchronous gate.
For the operator $\hat{X}_{yz},$ it can be easily obtained from the restriction of the $\hat{U}_x\cv{\pi/2}$ on the mentioned subspace $\mathcal{H}_{yz}.$
Specifically, we have
	\begin{align}
		\hat{U}_x\cv{\pi/2}&\cv{\iden\otimes\hat{P}_{yz}}\nn\\
			&= \ketbra{+}{+}\otimes\prod_{k=1}^3\cvb{\cv{\hat{P}_k^{x} - i\hat{X}_k^{yz}}\hat{P}_k^{yz}}\nn\\
			&\phantom{=.} + \ketbra{-}{-}\otimes\prod_{k=1}^3\cvb{\cv{\hat{P}_k^{x} + i\hat{X}_k^{yz}}\hat{P}_k^{yz}}\nn\\
		 	&= i\cv{\hat{\sigma}_x\otimes\hat{X}_1^{yz}\hat{X}_2^{yz}\hat{X}_3^{yz}} = i\hat{\sigma}_x\otimes\hat{X}_{yz} \label{eq:X_yz}.
	\end{align}
The projection on the right in the first expression is employed to represent the preparation of the nuclear spin initial state in the $yz$-subspace. 
With another local gate $-i\hat{\sigma}_x$ on the electron part, the restriction of the action $(-i\hat{\sigma}_x\otimes\iden)\hat{U}_x\cv{\pi/2}$ on the $yz$ nuclear collective subspace, i.e.,
	\[(-i\hat{\sigma}_x\otimes\iden)\hat{U}_x\cv{\pi/2}\cv{\iden\otimes\hat{P}_{yz}} = \iden\otimes\hat{X}_{yz}\] 
can be considered as a collective $X$ gate as expected. 

For the $Z$ gate on the same subspace, from the relation
	\begin{align}
		\hat{U}_z\cv{\pi}&\cv{\iden\otimes\hat{P}_{yz}}\nn\\ 
			&= \ketbra{0}{0}\otimes\prod_{k=1}^3\cvb{\cv{\hat{P}_k^{z} - \hat{P}_k^{xy}}\hat{P}_k^{yz}}\nn\\
			&\phantom{=.} + \ketbra{-1}{-1}\otimes\prod_{k=1}^3\cvb{\cv{\hat{P}_k^{z} - \hat{P}_k^{xy}}\hat{P}_k^{yz}}\nn\\
		 	& = -\iden\otimes\hat{Z}_{yz} \label{eq:Z_yz},
	\end{align}
or the restriction of $-\hat{U}_z\cv{\pi}$ on the $yz$ nuclear collective subspace can be considered as a collective $Z$ gate as claimed. With these $\hat{X}_{yz}$ and $\hat{Z}_{yz},$ one automatically has a product Pauli group on three qubits, with a relation 
	\[\hat{Z}_{yz}\hat{X}_{yz}=-\hat{X}_{yz}\hat{Z}_{yz},\] 
setting gates set for the collective three-qubit computation. 

In addition to $X$ and $Z$ gates, by varying the phase $\varphi,$ one can also use the same technique to prepare a Hadamard gate 
	\begin{equation}
		\hat{H}_{yz} = \hat{H}_1^{yz}\hat{H}_2^{yz}\hat{H}_3^{yz} = 2^{-3/2}\prod_{k=1}^3\cv{\hat{P}_{yz} + i\hat{X}_{yz}} \label{eq:Hadamard_yz}.
	\end{equation}
This can be prepared from Eq.~\eqref{eq:expIx-k}, whose reduced action on $yz$ collective subspace $(\ketbra{-}{-}\otimes\hat{P}_{yz})\hat{U}_x\cv{\pi/4}$ resembles the gate $\hat{H}_{yz}.$ In particular, one can write
		\begin{align}
		\hat{U}_x\cv{\pi/4}&(\ketbra{-}{-}\otimes\hat{P}_{yz})\nn\\
			&= \ketbra{-}{-}\otimes\prod_{k=1}^3\cvc{\cvb{\hat{P}_k^{x} + \frac{\hat{P}_k^{yz} + i\hat{X}_k^{yz}}{\sqrt{2}}}\hat{P}_k^{yz}}\nn\\
		 	&= \ketbra{-}{-}\otimes\hat{H}_{yz} \label{eq:H_yz}.
	\end{align}
Suppose that one can rotate the central spin initial state to $\ket{-},$ the action Eq.~\eqref{eq:Hadamard_yz} on the nuclear spins simply follows Eq.~\eqref{eq:H_yz} as claimed. A similar technique is also implementable for the preparation of a rotation gate
	\begin{equation}
		\hat{R}_{yz}\cv{\varphi} = \prod_{k=1}^3\cv{\cos\varphi\hat{P}_{yz} + i\sin\varphi\hat{X}_{yz}} \label{eq:M-theta_yz},
	\end{equation}
for various phase $\varphi.$ With Pauli groups generated by $\hat{X}_{yz}$ and $\hat{Z}_{yz}$ and the collection of rotation gates $\hat{R}_{yz}\cv{\varphi}$ above, in principle, one is allowed to manipulate the three nuclear spin qubits synchronously in an arbitrary fashion. This technique will be promising for fault-tolerant information encoding.

In summary, through $\hat{U}_\alpha\cv{\varphi}$ and the initialization of all nuclear spin states to an identical state in a particular subspace $\mathcal{H}_{\alpha\beta},$ one can employ the three nuclear spins as three copies of identical qubit where one can manipulate these three qubits synchronously. 
Since one can employ one subspace $\mathcal{H}_{\alpha\beta}$ at a time and any two subspaces overlap, e.g. $\mathcal{H}_{\alpha\beta}\cap\mathcal{H}_{\beta\gamma} \neq \emptyset,$ one can utilize the whole system Hilbert space of the system by a concatenation of the gates on different subspaces, e.g. $\hat{X}_{\alpha\beta}\hat{X}_{\beta\gamma}.$ This allows us to simplify the manipulation of the three qutrits into qubit manipulation problems. Another interesting topic is to extend the consideration from the construction of Pauli gates to other gates, which, combined with the Pauli gates and rotation gates we introduced, constitute a set of universal qubit gates.
Further development in this direction is out of the scope of the current work and invokes a dedicated investigation.

\section{Gate Fidelities}\label{sec:Gates_Fidelity}
Next, we numerically calculate the fidelity of our constructed gates using the real parameters observed in the experiments to signify that our protocols are promising in real-world applications. 
To determine the quality of the derived gates, we investigate the exact evolution governed by $\widetilde{H}(t)$ given in Eq.~\eqref{eq:H_rotate}. To gauge quality of the gates we consider a relative average gate fidelity defined by $F_R(\hat{E},\hat{V}) {=} F(\hat{V}^\dagger\hat{E})/F(\hat{V}^\dagger\hat{V})$, where $F(\hat{E}) {=} \int\! d\psi \big\vert\braket{\psi\vert\hat{E}\vert\psi}\big\vert^2$ is the average gate fidelity of the channel $\mathcal{E} = \hat{E}\cdot\hat{E}^\dagger$~\cite{Nielsen_Chuang_2010, Gilchrist2005}.
The integration is performed over the uniform (Haar) measure $d\psi$ on the full Hilbert space of electron and nuclear spins.
Here, $\hat{E}$ and $\hat{V}$ are operators on such Hilbert space, and $\hat{V}$ is a projected unitary operator on $\iden{\otimes}\mathcal{H}_{yz}\equiv \mathcal{H}_{yz},$ namely $\hat{V}=(\iden{\otimes}\hat{P}_{yz})\hat{V}(\iden{\otimes}\hat{P}_{yz})$.
We use the relative average gate fidelity rather than the average gate fidelity since we are working on the collective qubit subspace where each nuclear spin is reduced to a two-level system as it was formulated in Sec.~\ref{sec:TWS}.
Note that the target operator $\hat{V}$ is not a full rank but a unitary operator on the concerned subspace, and thus, the renormalization of the average is necessary.

For the preparation of the $X$ gate, in parallel to the relation Eq.~\eqref{eq:X_yz} we set 
	\begin{align*}
		\hat{E} &= \hat{U}_x\cv{\pi/2}\cv{\iden\otimes\hat{P}_{yz}}\nn\\
		\hat{V} &= i\hat{\sigma}_x\otimes\hat{X}_{yz}.
	\end{align*}
As mentioned in the previous section, the preparation of $\hat{U}_x$ involves various parameters, which one can choose to achieve a better quality of the gate. Throughout this article, the numerical results were performed using the QuTip package \cite{QuTip1,QuTip2} on Python. In particular, we employ the built-in functions to compute average gate fidelities as well as the state fidelities in the following sections.

\begin{figure}[tb]
	\centering
	\includegraphics[width=\columnwidth]{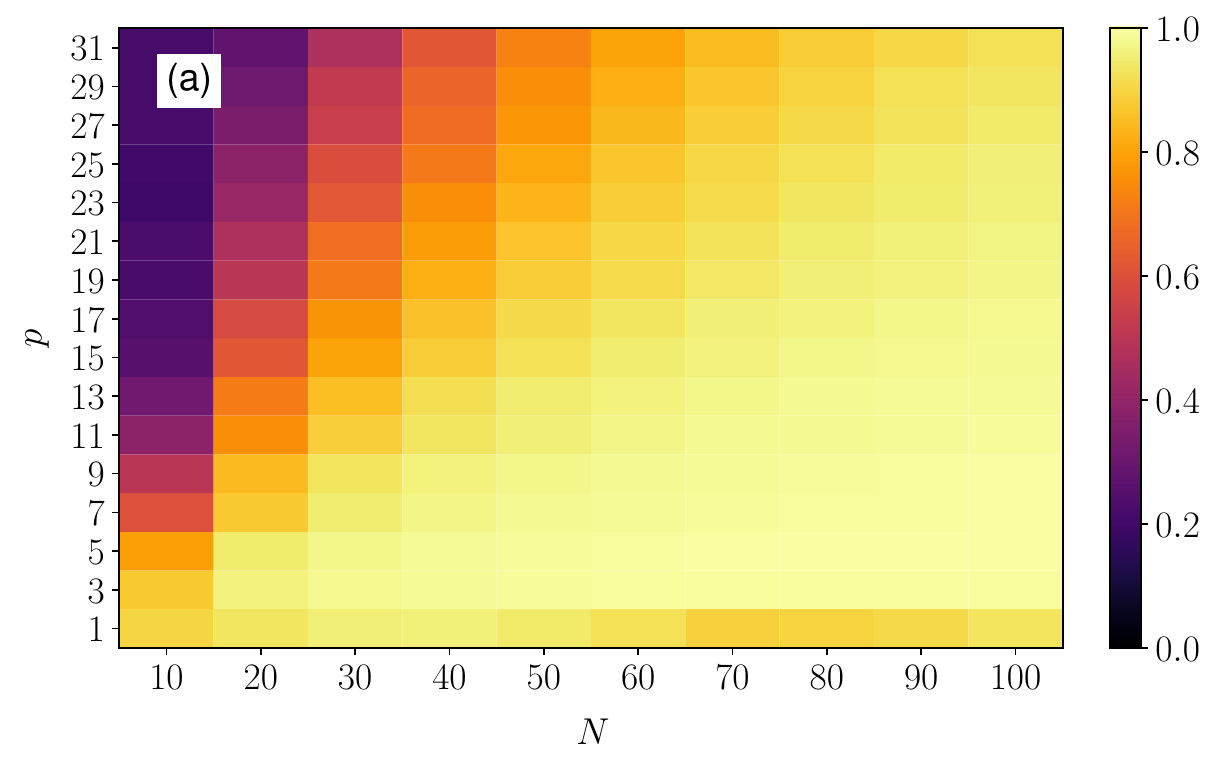}\hfill
	\includegraphics[width=\columnwidth]{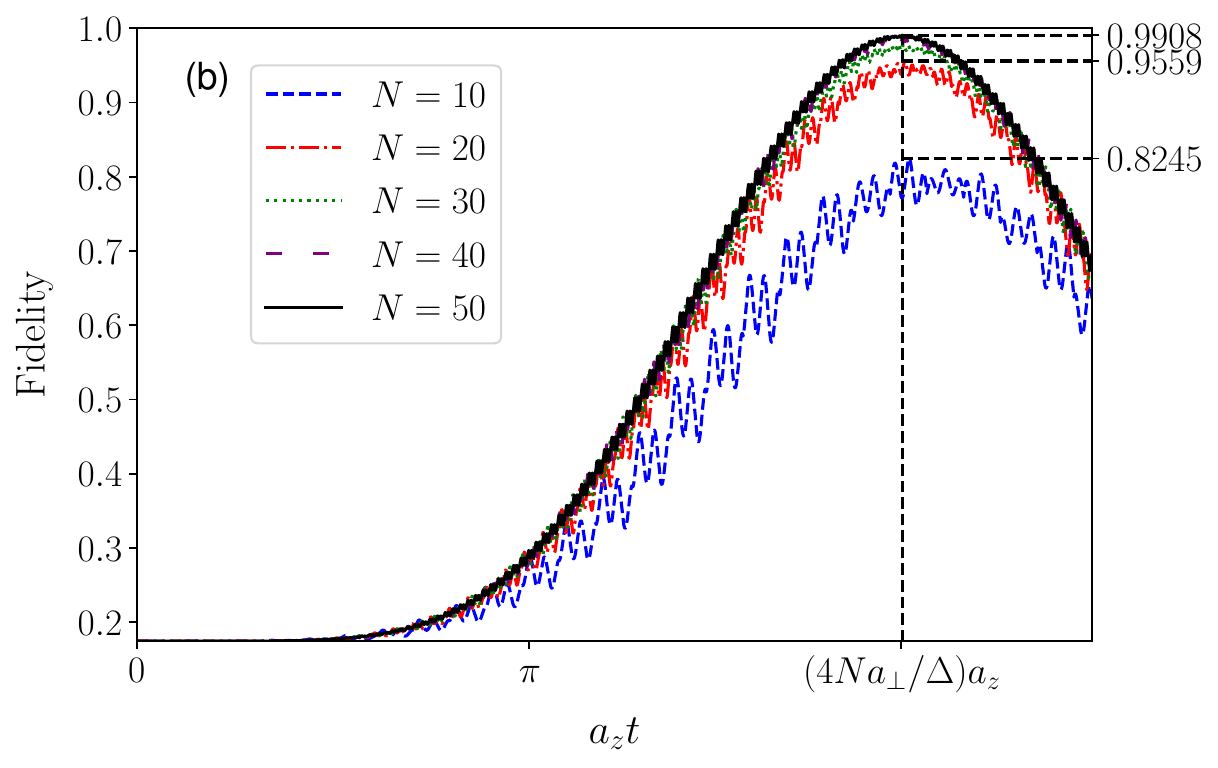}
	\caption{Relative average gate fidelity for the $X$ gate derived from the exact dynamic $\widetilde{H}(t)$ in Eq.~\eqref{eq:H_rotate}:
	(a) The maxima for various $N$ and $p$ values indicate the region where one can achieve fidelity close to unity (the light shade area). 
    The colors in (a) represent the values of fidelity ranging from $0$ (dark purple) to $1$ (light yellow).
	(b) Time evolution built-up of the $X$-gate for the first few instances for the case $p=5$. The corresponding magnetic field can be computed as a function of $N$ by Eq.~\eqref{eq:B_op}. One can achieve a fidelity as high as $0.9908$ for $N=50$ at the evolution time close to $t=4Na_\perp/\Delta = {40.43~\!\rm{ns}}$ (the vertical dashed line) predicted by the expected operator $\hat{U}_x\cv{\pi/2}.$ The horizontal dashed lines indicate the maxima for $N = 10,20$ and $50$ respectively.
    }\label{fig:Pauli-X-yz-gate}
\end{figure}
In Fig.~\ref{fig:Pauli-X-yz-gate}(a), the maxima of relative average gate fidelity for the $X$ gate are displayed for various $N$ and $p,$ where the magnetic field strengths are set from the relation between the revolution step $N$ and the target phase $\varphi=\pi/2.$
Technically, from the relation $\cvv{\varphi}= 4Na_\perp/\cvv{\Delta}$, we choose the magnetic field above the crossing point ($\Delta>0$):
	\begin{equation}
		B_{op}\cv{\varphi,N} = \dfrac{1}{\gamma_e-\gamma_n}\cv{\dfrac{4Na_\perp}{\varphi} + D_{gs} + a_z}.\label{eq:B_op}
	\end{equation}
One can observe in Fig.~\ref{fig:Pauli-X-yz-gate}(a) a region where the fidelity is very high, exceeding $0.98.$ 
We note that for finer resolution $\delta_\Delta,$ one needs to operate at higher magnetic field strength. 
For example, for higher number of revolution $N=100,$ one can have $B_{op}\cv{\pi/2,100} = 567$~mT and $B_{op}\cv{\pi/2,200} = 1.01$~T. In Fig.~\ref{fig:Pauli-X-yz-gate}(b), we demonstrate the evolutions of the fidelity for various $N,$ where one can observe the maxima around $t=4a_\perp N/\Delta,$ which is {$40.43~\!\rm{ns},$} predicted by the entangling gate $\hat{U}_x\cv{\pi/2}.$ At $N=50,$ $\Omega=\Delta/5,$ and with $B= 345$ mT, one can obtain $0.9908$ fidelity for the preparation of $\hat{X}_{yz}.$ The deviations of the optimal times---the times at which the maxima occur--- are relatively small and can be neglected for high enough $N.$
For instance, we observe that the deviation for $N=50$ is less than the resolution of the calculations. 
Therefore, we will use the parameters $p=5$ and $N=50$ for the examples of the $\hat{X}_{yz}$ preparation throughout the rest of the paper.

\begin{figure}[b]
	\centering
	\includegraphics[width=\columnwidth]{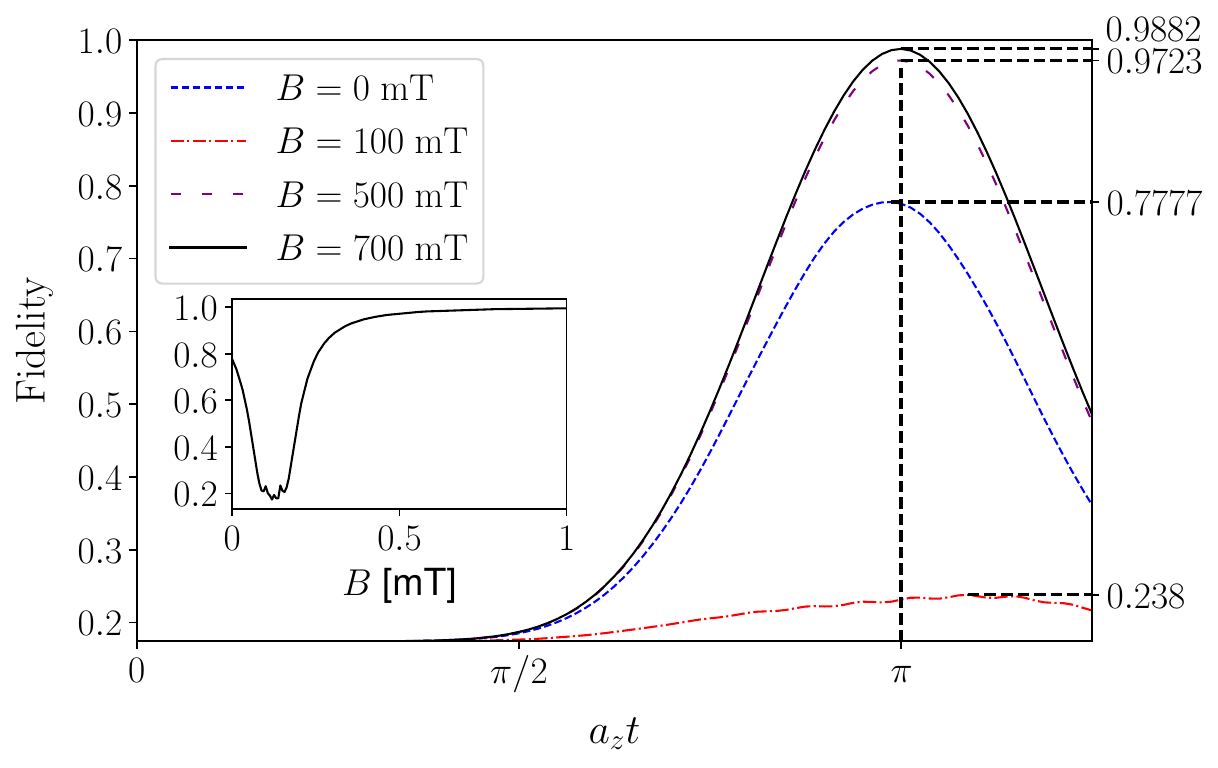}
	\caption{The relative average gate fidelity for the $Z$ gate derived from the exact dynamic $\widetilde{H}\cv{t}$ for various magnetic field. The inset shows the fidelity at the end of the evolution $t=\pi a_z^{-1}= {20.76~\!\rm{ns}}.$ for magnetic field strength between $0$ T and $1$ T. The horizontal dashed lines indicate the maxima, while the vertical dashed line represents the target optimal time $t=\pi a_z^{-1}$ predicted by $U_z\cv{\pi}.$ Here, no deviation from the target optimal time for large magnetic fields ($B = 500$ mT and $700$ mT) is observable. 
    }\label{fig:Pauli-Z-yz-gate}
\end{figure}

Similarly, for the preparation of $Z$ gate, in connection to Eq.~\eqref{eq:Z_yz} we set 
	\begin{align*}
		\hat{E} &= \hat{U}_z\cv{\pi}\cv{\iden\otimes\hat{P}_{yz}}\nn\\
		\hat{V} &= -\iden\otimes\hat{Z}_{yz}.
	\end{align*}
In Fig.~\ref{fig:Pauli-Z-yz-gate}, the relative average gate fidelity for the $Z$ gate given by this protocol is shown for various magnetic fields.
One can observe that at zero field, the relative average gate fidelity is small due to the contribution becoming non-negligible from the rotating and anti-rotating terms.
{Even for the zero field case where the value of Larmor frequency $\cvv{\Delta}\sim 3447 ~2\pi\times{\rm MHz}$ is much larger than the hyperfine coupling rates $\big\vert A_k^{\alpha\beta}\big\vert \sim 20 - 80 ~2\pi\times{\rm MHz}$ for $\alpha,\beta = x,y,$ the accumulation of errors from the counter-rotating terms and the odd rotating terms in Eq.~\eqref{eq:H_rotate} in the dynamics of in-plane interaction is still visible and results in a rather low relative average gate fidelity.}
At $B=100$ mT, $\Delta$ becomes comparable with the in-plane interaction, and the observed fidelity is drastically reduced.
Nevertheless, as predicted by our theory, the fidelity increases when the magnetic field is larger (see the inset in Fig.~\ref{fig:Pauli-Z-yz-gate}) and can reach about $0.9723$ for $B=500$ mT and $0.9882$ for $B=700$ mT. In addition, we also observe no deviation in optimal times from the target value $t=\pi a_z^{-1} = {20.76~\!\rm{ns}}$ for high magnetic fields $B = 500$ mT and $700$ mT.

\begin{figure}[t]
	\centering
	\includegraphics[width=\columnwidth]{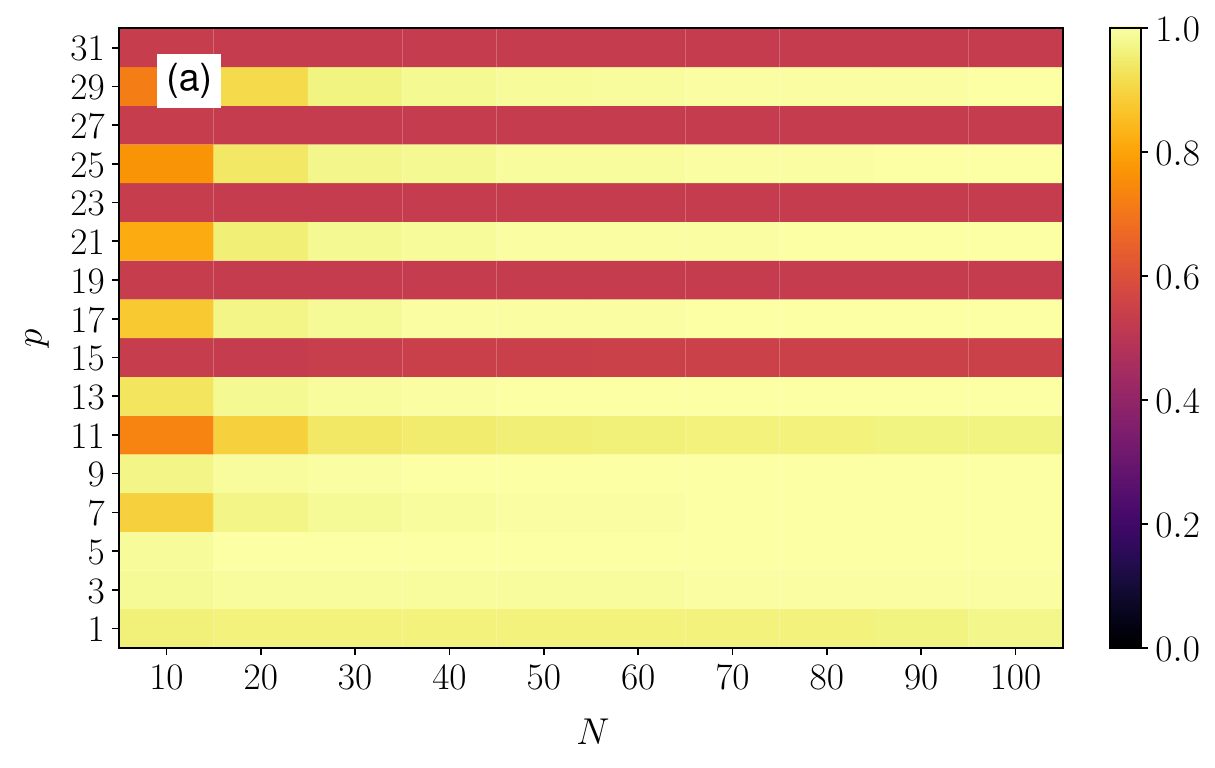}\hfill
	\includegraphics[width=\columnwidth]{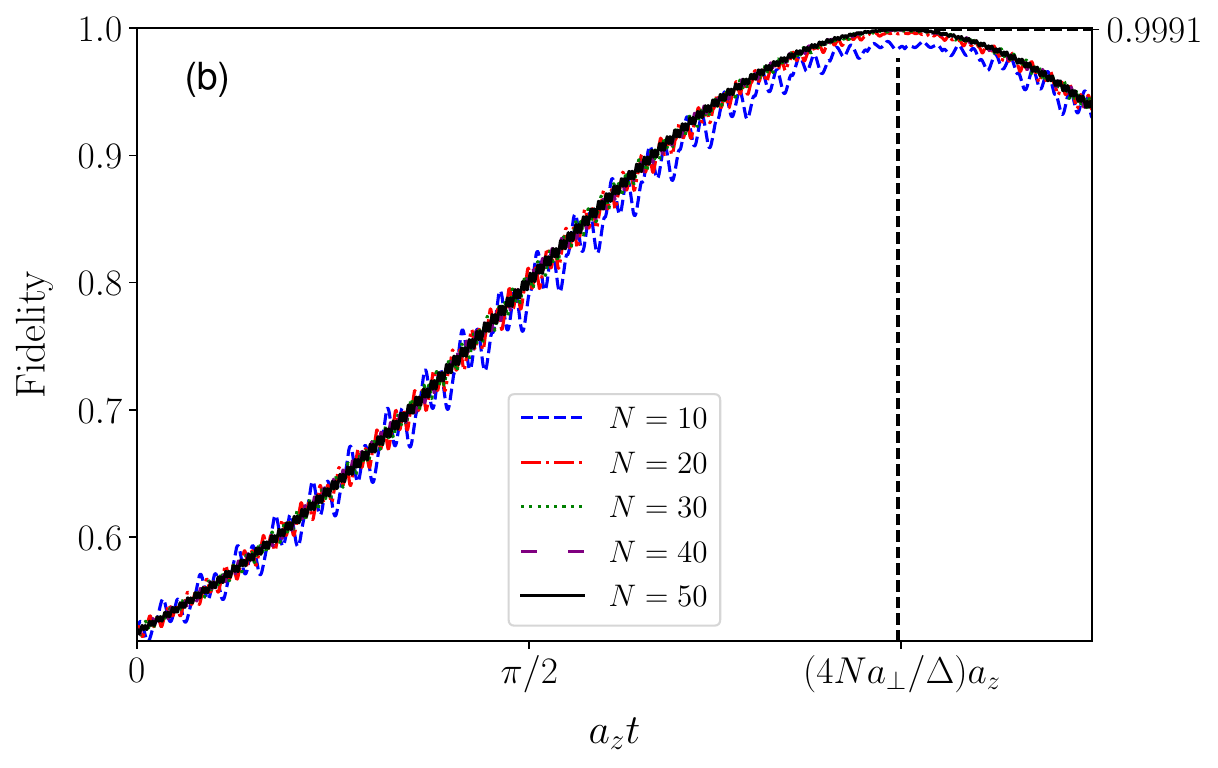}
	\caption{The maximum relative average gate fidelity for the Hadamard gate derived from the exact dynamic $\widetilde{H}\cv{t}$ according to Eq.~\eqref{eq:H_yz} in (a), for various $N$ and $p,$ indicates the region where one can achieve the fidelity around $0.99.$ The first few instances for the case $p=5$ are shown in (b), where one can achieve fidelity $0.9991$ for $N=50$ at the evolution time $t=4a_\perp N/\Delta = {20.21~\!\rm{ns}}.$ The descriptions for the presentations are similar to Fig.~\ref{fig:Pauli-X-yz-gate}.
    }\label{fig:Hadamard-yz-gate}
\end{figure}
Finally, for the Hadamard gate Eq.~\eqref{eq:H_yz} we set 
	\begin{align*}
		\hat{E} &= \hat{U}_x\cv{\pi/4}(\ketbra{-}{-}\otimes\hat{P}_{yz})\nn\\
		\hat{V} &= \ketbra{-}{-}\otimes\hat{H}_{yz}.
	\end{align*}
We note that in this case, in addition to the $yz$-subspace for nuclear spins, the initial state is also concentrated in the electron spin subspace with respect to the state $\ket{-}.$ 
The maximum relative average gate fidelity for this gate, given by this protocol, is shown in
Fig.~\ref{fig:Hadamard-yz-gate}(a) for various harmonic order $p$ and number of revolutions $N$ and hence the operational magnetic field, where in Fig.~\ref{fig:Hadamard-yz-gate}(b) the fidelity is shown as a function of time, for the parameters of $\hat{U}_x$ are set as $p=5$ for various $N$, where the maxima can be found about $t=4a_\perp N/\Delta = {20.21~\!\rm{ns}}.$ 
First, in Fig.~\ref{fig:Hadamard-yz-gate}(a), we can observe an oscillation in $p$ of the maximum fidelity, which shows its clear visibility for high $p.$ This comes from the factor $\sin\cv{p\pi/2}$ in Eq.~\eqref{eq:Ux_construct} which alters the sign of the phase $\varphi$ depending on $p,$ leading to the maximum value only for the case when $p-1$ divisible by $4.$ This oscillation appears as a global phase in the preparation of the $X$ gate and hence cannot be observed. 
Apart from that, even though the gate is prepared by the operator $\hat{U}_x$ as for the case of $X$ gate, one can observe a higher relative average gate fidelity in this case since the evolution period of $\hat{U}_x$ in this implementation is shorter, leading to lower accumulation of the extra terms removed from the approximation. Another factor is the effect of the initialization of the electron spin, which is a projective operator and can remove coherence propagation in the dynamics, leading to a lower deformation of the observed fidelity. 
Unlike the preparation of $X$ and $Z$ gates, these mentioned effects possibly amount to a slightly larger deviation of the optimal time. In particular, we find that such deviations are bounded by $0.01223~\!a_z^{-1},$ which is, however, negligibly small.

\section{GHZ States Preparations}\label{sec:GHZ_prepatation}
Apart from the collective gates implementation, it is possible to utilize the entangling gates Eqs.~\eqref{eq:U_z}-\eqref{eq:U_x} for a different purpose, e.g., the preparation of nuclear-entangled states. In this section, we demonstrate this aspect by using it for the preparation of the  
Greenberger-Horne-Zeilinger state on a three-qubit subspace $\mathcal{H}_{\alpha\beta}.$

\subsection{Initialization}
First of all, we assume that one can perform initialization of the nuclear spins state in $\ket{m_n}^{\otimes 3}$ for $m_I=\pm 1.$ This can be done via the combination of polarization to ground states and conditional rotation from our derived gates. For the polarization to ground state, e.g., the state $\ket{-1}^{\otimes 3},$ one can employ, for instance, a hyperpolarization procedure introduced in Ref.~\cite{Tabesh2023}, where one polarizes the electron spin and uses flip-flop interaction. Another method proposed in Ref.~\cite{Gao2022} is the application of continuous optical pumping close to the level crossing where the state $\ket{0}\otimes\ket{1}^{\otimes 3}$ can be initialized.
Therefore, in our following discussion, we assume that the nuclear initial spin state is prepared in $\ket{m_I}^{\otimes 3}$ for $m_I=\pm 1$.

According to the notations adopted in Sec.~\ref{sec:TWS}, one has 
	\begin{equation}
		\ket{m_I}_k = \dfrac{m_I\ket{x}_k + \ket{y}_k}{\sqrt{2}} \label{eq:ket-mn_in-xy}.
	\end{equation}
As discussed earlier, the application of the unitary operators Eqs.~\eqref{eq:Ix-k} or \eqref{eq:Iy-k} will keep one component unchanged, while rotating the other component to another orthogonal state.
For instance, we observe that
	\begin{equation*}
		e^{\pm i\varphi\hat{I}_k^x}\ket{m_I}_k = \dfrac{m_I\ket{x}_k + \cos{\varphi}\ket{y}_k \pm i\sin{\varphi}\ket{z}_k}{\sqrt{2}},
	\end{equation*}
and in particular for $\varphi =\pi/2$ it becomes 
	\begin{equation}
		e^{\pm i(\pi/2)\hat{I}_k^x}\ket{m_I}_k = \dfrac{m_I\ket{x}_k \pm i\ket{z}_k}{\sqrt{2}} \label{eq:mn-rotate-to-xz},
	\end{equation}
for $m_I=\pm 1.$
This property can be exploited for the preparation of GHZ states.

\subsection{Formulations}\label{sec:GHZ_theory}
The Greenberger-Horne-Zeilinger state or the GHZ state \cite{Greenberger1990} of three qubits is written as
	\begin{equation}
		\ket{\text{GHZ}}_\nu = \dfrac{\ket{\bar{0}\bar{0}\bar{0}} + e^{i\nu}\ket{\bar{1}\bar{1}\bar{1}}}{\sqrt{2}} \label{eq:GHZ_basic},
	\end{equation}
where the qubit $k$ is described by some Hilbert space spanned by $\cvc{\ket{\bar{0}}_k,\ket{\bar{1}}_k}_k$ and $\nu$ is an arbitrary phase. Now we demonstrate as an example that one can use the entangling gate $\hat{U}_x\cv{\pi/2}$ to prepare a GHZ state on a subspace $zx$ of the three nuclear spins. This can be done through the conditional rotation Eq.~\eqref{eq:U_x-dephasing-form}, where the nuclear spins are rotated in different directions for the different electron spin control states. Together with a preparation and a measurement of electron spin in a superposition state, which results in the superposition of the rotated states on the nuclear part, it can be seen that the nuclear reduced state for each measurement outcome will become the GHZ state.

In particular, from Eq.~\eqref{eq:mn-rotate-to-xz}, let us define 
	\begin{align}
		\ket{\bar{0}}_k &= e^{-i(\pi/2)\hat{I}_k^x}\ket{m_I}_k = \dfrac{m_I\ket{x}_k - i\ket{z}_k}{\sqrt{2}}\label{eq:0_beta-gamma},\\
		\ket{\bar{1}}_k &= e^{i(\pi/2)\hat{I}_k^x}\ket{m_I}_k = \dfrac{m_I\ket{x}_k + i\ket{z}_k}{\sqrt{2}}\label{eq:1_beta-gamma},
	\end{align}
for each spin $k.$ 
Note that $\ket{\bar{0}}_k$ and $\ket{\bar{1}}_k$ are orthonormal states, and additionally $\{\ket{\bar{0}}_k,\ket{\bar{1}}_k\}_k$ form a basis of the $zx$-subspace. Let the state be initialized to $\ket{0}\otimes\ket{m_I}^{\otimes 3}$ as previously discussed. 
Since $\ket{0}=(\ket{+} + \ket{-})/\sqrt{2},$ by applying the rotation gate one gets 
	\begin{equation}
		\hat{U}_x\cv{\pi/2}\ket{0}\otimes\ket{m_I}^{\otimes 3} = \dfrac{\ket{+}\otimes\ket{\bar{0}\bar{0}\bar{0}} + \ket{-}\otimes\ket{\bar{1}\bar{1}\bar{1}}}{\sqrt{2}}\label{eq:U_pi-over-2-superpose}.
	\end{equation}
Next, one can measure the central spin state with the basis $\cvc{\ket{0}, \ket{-1}},$ where the measurement results will herald the nuclear spin state. Namely, if the central spin state after the measurement is found to be $\ket{0},$ the nuclear spin state will be left in the state $\ket{\text{GHZ}}_{0},$ whereas it will be $\ket{\text{GHZ}}_{\pi}$ if the electron spin is found at the state $\ket{-1}.$ 

Remarkably, if the mentioned measurement is realized by a projection onto the state $\ket{0}$ or $\ket{-1},$ the state of the nuclear spins is mapped to GHZ states in both successful and non-successful measurements, with different phases $\nu=0$ and $\nu=\pi,$ that can be heralded from the central spin measurement outcome. For the negative outcome branch, such phase in $\ket{\text{GHZ}}_{\pi}$ can be corrected by further applying a post-processing operation $\hat{U}_y\cv{\pi/2},$ resulting that
	\begin{equation}
	 	\hat{U}_y\cv{\pi/2}\ket{-}\otimes\ket{\text{GHZ}}_{\pi} = (-i\hat{\sigma}_y\ket{-}){\otimes}\ket{\text{GHZ}}_{0}. \label{eq:GHZpi-to-GHZ0}
	\end{equation} 
The relation above comes from the observation that $\ket{\overline{0}}_k\bra{\overline{0}} - \ket{\overline{1}}_k\bra{\overline{1}} = m_I\hat{Y}^{zx}_k,$ which can be prepared from $\hat{U}_y\cv{\pi/2}$ in the similar fashion as for Eq.~\eqref{eq:X_yz}.

Indeed, the entangling gate $\hat{U}_{x}\cv{\pi/2}$ with some post-processing can be used to prepare GHZ states on the  $zx$-subspace.  Such states can be further transferred to other collective subspaces by applying another step of rotation. For example, one can write 
	\begin{align*}
		\hat{U}_z\cv{\pi/2}&\cv{\ket{0}\otimes\ket{\text{GHZ}}_{\nu}}\\ 
			&= \ket{0}\otimes{\prod_k\cv{\hat{P}_k^z - i\hat{X}_{xy}}\ket{\text{GHZ}}_{\nu}}\\ 
			&= \ket{0}\otimes\ket{\widetilde{\text{GHZ}}}_{\nu}
	\end{align*}
where $\ket{\widetilde{\text{GHZ}}}_{\nu}$ takes the same form as $\ket{\text{GHZ}}_{\nu}$ with the replacement of $\ket{x}_k$ by $-i\ket{y}_k,$ i.e., the GHZ state on the  $yz$-subspace. 
We note that the introduction of this transformation will introduce a degradation of the concerned fidelity, hence, in case one expects operational qubits in $yz$-subspace, in practice the overall fidelity including the effect of the transformation $\hat{U}_z\cv{\pi/2}$ should be taken into account.

The procedure for GHZ state preparation can be generalized to the writing process on the nuclear spin collective qubit. For instance, using $\{\ket{\bar{0}\bar{0}\bar{0}}, \ket{\bar{1}\bar{1}\bar{1}}\}$ as a basis for the nuclear collective qubit, and replacing the electron spin state $\ket{0}$ on the left-hand side of Eq.~\eqref{eq:U_pi-over-2-superpose} by $\alpha\ket{+} + \beta\ket{-},$ the coefficients in the resulting nuclear spin collective states after the measurement of the electron spin on $\{\ket{0},\ket{-1}\}$ basis will be $\alpha\ket{\bar{0}\bar{0}\bar{0}} \pm \beta\ket{\bar{1}\bar{1}\bar{1}}$ where the sign depends on the measurement outcome. For the reading process, one could employ the same entangling gates to conditionally transfer the nuclear spin collective state to the electron spin state. For example, one can see that $\hat{U}_z(\pi/12)\hat{U}_x(\pi/2)\ket{0}{\otimes}\cv{\alpha\ket{\bar{0}\bar{0}\bar{0}} \pm \beta\ket{\bar{1}\bar{1}\bar{1}}}= (\tilde{\alpha}\ket{0} - im_I\tilde{\beta}\ket{1})\otimes\ket{\xi}$, where $\tilde{\alpha} = (\alpha + \beta)/\sqrt{2}$ and $\tilde{\beta} = (\alpha - \beta)/\sqrt{2}$, while $\ket{\xi}$ is an irrelevant nuclear spin collective state. This suggests that the synchronous gates allow us to use the three nuclear spins as a collective qubit whose manipulation can be done via the electron spin mediator.

\subsection{Fidelity of the GHZ States Preparation}\label{sec:Fidelity}
Here, we numerically demonstrate the fidelity in the preparation of GHZ states by our protocol. 
We set the initial state as $\ket{\psi_0} = \ket{0}\otimes\ket{1}^{\otimes 3},$ and, for simplicity, we assume a short and perfect Pauli gates $\hat{\sigma}_z$ and $\hat{\sigma}_x$ (and also their corresponding Hadamard gates) for the electron spin. 
For the mixed state, we use the state fidelity defined as~\cite{Jozsa1994} 
	\begin{equation}
		F\cv{\rho,\sigma} = \tra\cv{\sqrt{\sqrt{\sigma}\rho\sqrt{\sigma}}} \label{eq:def_fidel_mixed}
	\end{equation}
between the outcome states from our protocol and the target GHZ states. We compute the fidelity between $\ket{\text{GHZ}}_{0,\pi}$ and the nuclear spins state 
	\begin{subequations}
	\begin{align}
		\rho_+\cv{t} &= \tra_{cs}\cvb{(\ketbra{0}{0}\otimes\iden)\hat{U}_z(t)\ketbra{\psi_0}{\psi_0}\hat{U}^\dagger_z(t)}, \label{eq:rho-z_p-t}\\
        \rho_-\cv{t} &= \tra_{cs}\cvb{(\ketbra{-1}{-1}\otimes\iden)\hat{U}_z(t)\ketbra{\psi_0}{\psi_0}\hat{U}^\dagger_z(t)}, \label{eq:rho-z_m-t}
	\end{align}
    \end{subequations}
where $\tra_{cs}$ is a partial trace over the central spin qubit. We write 
	\[F_\nu^\pm\cv{t} = F\cv{\rho_\pm\cv{t},\rho^{\text{GHZ}}_\nu},\] 
where $\rho^{\text{GHZ}}_\nu = \ket{\text{GHZ}}_\nu\bra{\text{GHZ}},$ for the fidelity at time $t$ for considered pair of states.

\begin{figure}[tb] 
	\includegraphics[width=\columnwidth]{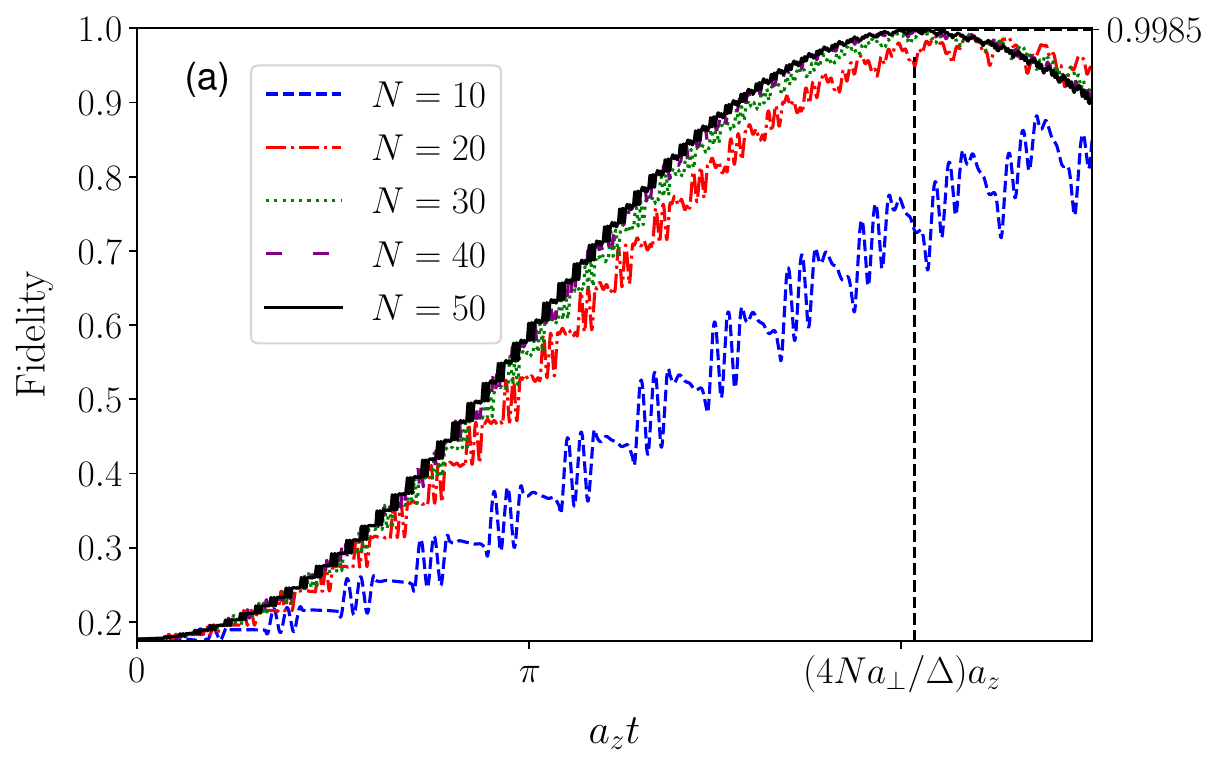}
	\includegraphics[width=\columnwidth]{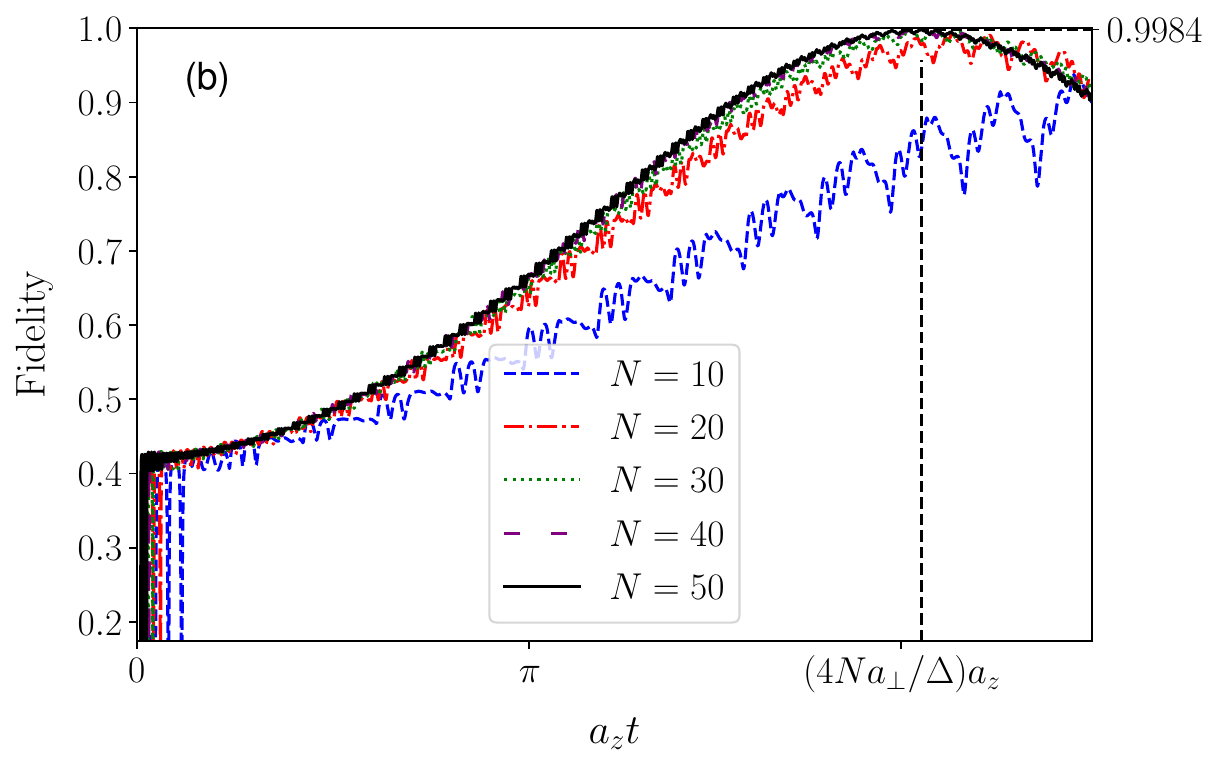}
	\caption{Fidelity of nuclear spins state after the operation Eqs.~\eqref{eq:rho-z_p-t} and \eqref{eq:rho-z_m-t} on $\ket{\psi_0} = \ket{0}\otimes\ket{1}^{\otimes 3}$ for a measurement result $\ket{0}$ with respect to the GHZ state $\ket{\text{GHZ}}_0$ in (a), with various numbers of revolutions $N$ and with $p=5.$ The similar results are shown for the negative measurement result $\ket{-1}$ in (b) with the referent state $\ket{\text{GHZ}}_\pi.$ Note that the operational magnetic field strengths are given by the expression Eq.~\eqref{eq:B_op}. For both cases, unlike the gate fidelities, one can observe slight deviations of the optimal times (vertical dashed lines) from the predicted value $t=4Na_\perp/\Delta = {40.43~\!\rm{ns}}.$ 
 }\label{fig:GHZ-prep}
\end{figure}
In Fig.~\ref{fig:GHZ-prep}, the fidelity $F_\nu^\pm\cv{t}$ is illustrated with the same parameters given for the preparation of the $X$ gate. One can see that our protocol can produce GHZ states with high fidelity larger than $0.99.$ 
As we discussed for the fidelity of the Hadamard gate, the fidelity, in this case, is larger than the protocol for the preparation of $\hat{X}_{yz},$ with both initialization and measurement operators in this case being projective, leading to a removal of interference effects. The fidelity can become distorted when $\Delta$ is not sufficiently large, for instance, when $B = 169$ mT ($N=10$) At a sufficiently large field, e.g. $B = 345$ mT ($N=50,$) our approximation of the entangling gate works well and can be employed for the preparation of the GHZ states as expected.
For $\nu=0,$ the fidelity is $0.9985$ at the optimal time and $0.9963$ at $t=4Na_\perp/\Delta,$ while for $\nu=\pi$ the fidelity is $0.9984$ at the optimal time and $0.9955$ at $t=4Na_\perp/\Delta.$ 
We observe larger deviations in optimal times from the referent value $t=4Na_\perp/\Delta$ compared to the preparation of $X$ gate, i.e. $0.1055~\!a_z^{-1}$ for $\nu=0$ and $0.1651~\!a_z^{-1}$ for $\nu=\pi.$ In practice, without additional information, there is no instructive method to determine the optimal time, and one can only expect the gate at the time $t=4Na_\perp/\Delta,$ i.e., at the time of completion of $N$ cycles of the control pulse sequence. Hence, in the realistic implementation, this deviation should always be taken into account. Nonetheless, thanks to the slow changes around the optimal point, the fidelities observed at $t=4Na_\perp/\Delta = {40.43~\!\rm{ns}}$ can be large, exceeding $0.99$ as reported above. The total preparation time of the GHZ states, including the re-initialization Eq.~\eqref{eq:mn-rotate-to-xz}, which is done by the $X$ gate, will then be about ${80.85~\!\rm{ns}}.$

\section{Electron Spin Dephasing}\label{sec:dephasing}

We have discussed the gates and GHZ states implications with a perfect condition, where each nuclear spin only interacts with the central spin, and the whole cluster is isolated from the external bath. In this section, we will demonstrate the contribution of an electron spin dephasing to the obtained fidelities. 
The dephasing can arise from the hyperfine interactions with the nuclear spin bath and the thermal phonons~\cite{Tabesh2025,Gottscholl2020,Ye2019}. This effect may also result from the deformation of the host material, where per se can also be employed to extend the coherence time \cite{Lee2022}. One can also expect that the impurity or imperfection of the sample might contribute to the coherence time, as suggested for NV-center in diamond \cite{Zhao2012,Park2022,Radishev2021,Chrostoski2022} and color centers in SiC \cite{Anderson2022}. However, studies in this aspect are still open for the hBN system. 

The dephasing can be modeled by a dissipative generator in the Lindblad equation for the whole system density matrix $\rho,$ i.e,
	\begin{equation}
		\dfrac{d\rho}{dt} = -i\cvb{\widetilde{H}(t),\rho} + \frac{\Gamma}{2}\cv{2\hat{L}\rho\hat{L}^\dagger -\hat{L}^\dagger\hat{L}\rho - \rho\hat{L}^\dagger\hat{L}} \label{eq:dephasing_term},
	\end{equation}
where $\hat{L} = \hat{\sigma}_z\otimes\iden$ defines a dephasing generator in the concerned two-level electron spin subspace, and $\Gamma>0$ is the pure dephasing rate of the electron spin.
The experimentally reported dephasing time depends on the temperature.
At room temperature, its typical value is around $\Gamma^{-1} \sim 2 ~\!\rm{\mu s}$ and can be larger for lower temperature, see e.g. Refs.~\cite{Gottscholl2021,Ramsay2023}.
Here, we set the $\Gamma^{-1} = 2~\!\rm{\mu s}$ for the value at room temperature and $4 ~\!\rm{\mu s}$ at a lower temperature in our illustration. 
Another factor that possibly degrades the fidelity of the gates or prepared GHZ states is the electric quadrupole interaction of the nuclear spins which modifies Eq.~\eqref{eq:H_n} to $\hat{H}_n = \sum_k -\gamma_k\mathbf{B}\cdot\mathbf{I}_k +Q_k(\hat{I}_k^z)^2$ with {$Q_k \approx 0.383~2\pi\times{\rm MHz}$} according to the experimental observations~\cite{Gottscholl2021}.
We have also included this interaction in our numerical considerations and found that the effect is negligibly small and does not appreciably affect the gate and state fidelities.
In particular, the relative deviations (defined below) contributed by the quadruple interaction are less than $0.25\%$ for the gate fidelities, and they are bounded by $0.015\%$ for the state fidelities.

\begin{figure}[t]
	\centering
	\includegraphics[width=\columnwidth]{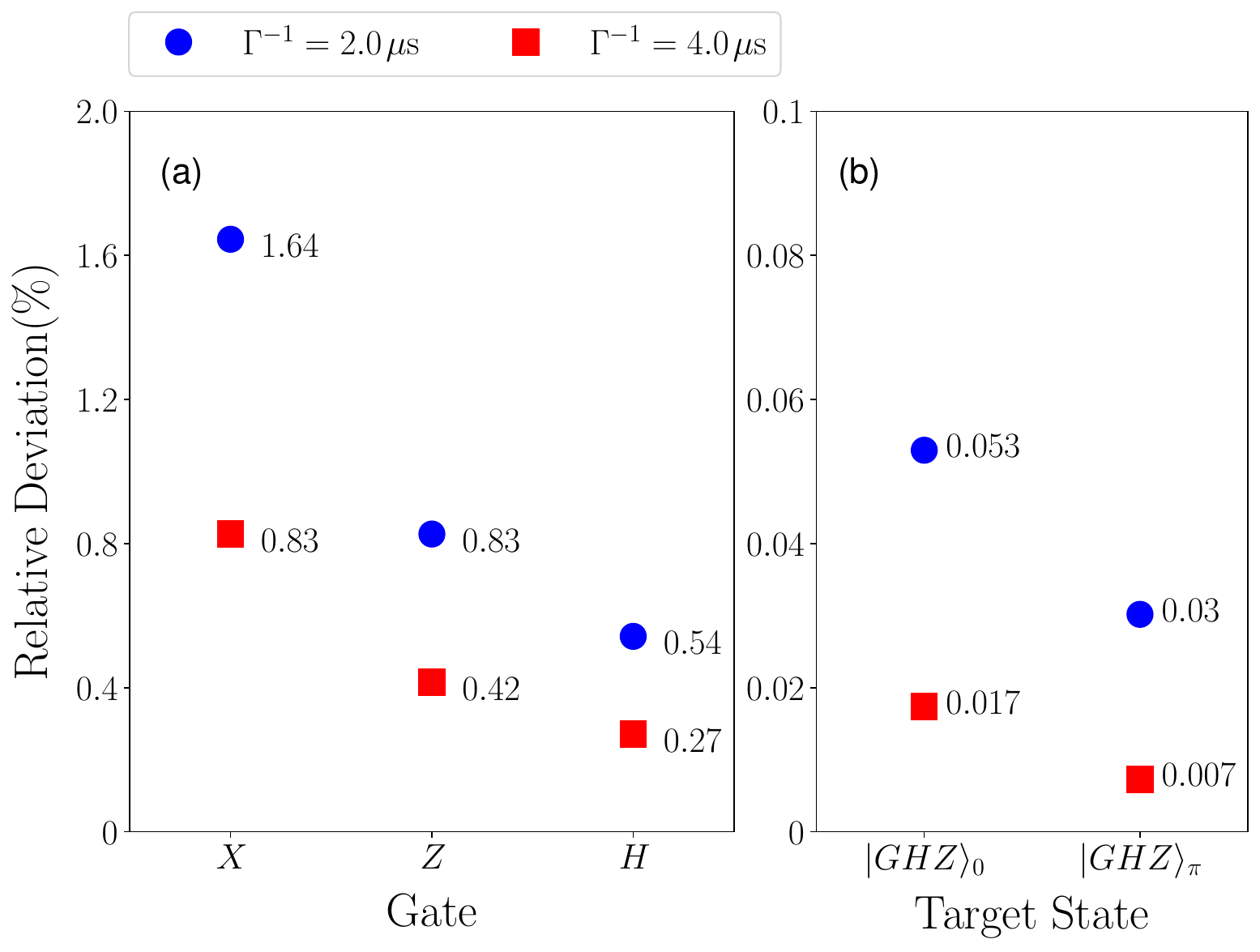}
	\caption{{Relative deviations of the relative fidelity for gate implementations are illustrated in (a), while in (b), relative deviations of the fidelity for GHZ states preparations are shown. We use $B=500$ mT for the calculation for the $Z$ gate, where $N=50$ and $p=5$ are employed in the calculation for the $X$ gate, $H$ gate, and GHZ states. The deviation for implementing the $X$ gate is larger than the other two by the longer evolution time of $\hat{U}_x,$ where the implementation of the $H$ gate enjoys a deviation discount from the initialization in the implementation protocol. From both gate fidelities and state fidelities, it can be seen that the longer coherence time, which can be achieved below room temperature, leads to small deviations in the fidelity profiles.} 
    }\label{fig:fidel_dephasing}
\end{figure}

To determine the effects of the noise, we consider a relative deviation defined by
	\begin{equation}
		\varepsilon = \frac{{F_{\rm ref} - F_{\rm noise}}}{F_{\rm ref}} \label{eq:relative_deviation},
	\end{equation}
where $F_{\rm ref}$ refers to the referent fidelity and $F_{\rm noise}$ is the fidelity with the noisy contribution mentioned above.
The reference fidelities are the same as those computed in the previous section for gates and states, while $F_{\rm noise}$ is when the dephasing effects are included in the dynamics. For illustration, for the $X$ gate, $H$ gate, and GHZ states, we set $N=50$ and $p=5$ where the magnetic field strengths are calculated using Eq.~\eqref{eq:B_op}, while for the $Z$ gate, the magnetic field strength is $B=500$ mT.
For simplicity, we compute the relative deviations at $t=\pi a_z^{-1}$ for the $Z$ gate, and $t={4a_\perp N/\Delta}$ for $X$ gate, Hadamard gate and the GHZ states. We note that, as previously discussed, these time epochs are not optimal times according to the exact evolution $\widetilde{H},$ but they are close to the optimal times with respect to the relevant time scale $a_z^{-1}.$ 
In Fig.~\ref{fig:fidel_dephasing}(a), the relative deviations for the gates $X,$ $Z,$ and Hadamard gate $H$ are given, where one can see the decreases in deviations for the larger coherence time, i.e., lower temperature.
All observed deviations are in the range of a few to several percent, which can be negotiated by further implementation of an optimization scheme, e.g., modifying the filter functions $F_\alpha\cv{t}.$ This suggests that our protocols for the gate implementations are noise-resilient.

{In addition, we observe the implementation of the $X$ suffers the most deviation $1.64 \%$ for $\Gamma^{-1} = 2~\!\rm{\mu s}$ and $0.83 \%$ for $\Gamma^{-1} = 4~\!\rm{\mu s}$. This agrees with the fact that the evolution time required for its implementation is the largest, e.g. $6.116a_z^{-1}= 40.43~\!\rm{ns}$ in our illustrations, compared to $\pi a_z^{-1}= 20.76~\!\rm{ns}$ for the implementation of $Z$ gate.} For the Hadamard gate $H,$ one can see the least deviation from the expected value. This can be explained by the initialization of the electron spin in our protocol, which limits the sample region in the calculation of average gate fidelity. This removes some correlation propagation in the dynamics from the expected value, leading the higher fidelity, as well as lower deviation from the noisy terms in general. However, as in the typical schemes involving projective operators, including initialization or post-selection processes, a higher fidelity can usually be compensated by a lower success possibility, i.e., the successful probability for the electron spin initialization in our case.

A similar deviation can be observed for the GHZ states preparation in Fig.~\ref{fig:fidel_dephasing}(b). The asymmetric nature of the deformation from the dephasing term for the state $\ket{\text{GHZ}}_0$ and $\ket{\text{GHZ}}_\pi$ can also be seen. Although the mechanism for the preparation is similar to the implementation of the $X$ gate, we can see that the fidelity suffers less deviation from the noisy terms. This is because, unlike the gate fidelity, the initial state and the measurement are predetermined. On the one hand, there is no accumulation of the deviation from the sampling process on the state space in the calculation of the fidelity, as in the case of average gate fidelity. On the other hand, similar to the implementation of the Hadamard gate, the projective characteristic of the measurement also removes the coherence in the dynamics from the obtained quantity.

\section{Conclusions}\label{sec:conclusion}
To summarize, we have derived a set of entangling gates on three ${}^{14}$N nitrogen nuclear spins through a VB-center in hBN, utilizing the electron spin of the defect as a control qubit.
Such gates are attained by imposing a strong enough magnetic field along the VB-center axis of symmetry, i.e. perpendicular to the lattice plane, where the $\hat{U}_z$ gate can be obtained by removal of fast oscillation terms and $\hat{U}_x$ can be achieved by the application of control pulses.
We employ the derived unitary rotations to implement the synchronous three-qubit Pauli $X$ gate and $Z$ gate, as well as the Hadamard gate on some relevant three-qubit subspace.
Using a set of realistic system parameters taken from the computational and experimental works, we have demonstrated numerically that such gates can be obtained with high fidelities.
This suggests that our scheme is a promising protocol for gate implementation upon nuclei in hBN, implementable in future experiments.
We further utilize the entangling gates to prepare GHZ states, where a high fidelity above $0.99$ can be observed numerically.
{Finally, we consider the effect of electron dephasing where the relative deviations of the gate fidelities from the noisy terms are less than $2\%,$ and even less than $0.1\%$ for the preparation of GHZ states.}
These results suggest that our protocols for the synchronous three-qubit gates and the preparation of GHZ states are noise-resilient and thus promising for experimental realizations and potentially for quantum computing applications. 
In addition, given the possibility to control several qubits at once with the synchronous gate, these are also interesting protocols for the fault tolerance implications on a realistic platform.

With the implementation of the synchronous gates and the preparation of the GHZ states, our work can be treated as the first step in exploiting the utility of a nuclear spin cluster around a VB-center in hBN as a platform for quantum computation. 
On one end, one can further explore the use of the entangling gates for the implementation of different gates, as well as two-qubit-like gates, which will lead to a complete set of universal gates. On the other end, one can also consider different parameters, e.g., magnetic field and control field, as well as other time scales, to execute different parts of the system dynamics, allowing for the implementation of single-nuclear-spin qubit gates or large-scale information processing. The research in this direction, in our belief, is a promising area that can enable \vb-hBN as a candidate for a quantum computational platform in general.
It is worth mentioning, thanks to the recent advances in preparing highly isotopic purified samples, the investigations performed in this work can even be extended to the case of spin-1/2 clusters of ${}^{15}$N nuclei~\cite{Lee2025}.
Finally, we should add that the rather large magnetic fields (${\sim} 0.5$ T) required for the validity of the rotating wave approximation are currently available in the labs, see, e.g., Ref.~\cite{Murzakhanov2022}. Nevertheless, this requirement partly limits the portability of a device operating based on our proposed protocol, since such magnetic field values can only be prepared in a stationary manner.

Several interesting aspects can improve our proposed protocols. First is the pulse optimization. We have used CPMG pulse sequences for the preparation of $\hat{U}_x.$ This is one of the simplest schemes employed to minimize the unwanted terms in the Hamiltonian to arrive at a workable approximation. This can be replaced by more advanced pulse sequences, where one can prepare in a more optimal fashion. Namely, one can construct a pulse sequence for which the filter functions $F_\alpha\cv{t}$ by minimizing the contribution from $\hat{H}_{\rm RW, odd}$ and $\hat{H}_{\rm CRW},$ e.g., by minimizing $\Vert\hat{H}_{\rm RW, odd}+\hat{H}_{\rm CRW}\Vert_{op}$ where $\Vert{\cdot}\Vert_{op}$ is an operator trace norm. 
Another aspect is the extension of the gates implementation to other gates beyond Pauli gates and rotation gates. In our examples of the implementation of the $X$ gate and $Z$ gate, unlike the preparation of GHZ states, the electron spin does not play a significant role as a control qubit. It is then interesting to examine whether one can utilize the control qubit and prepare an effective two-qubit gate (e.g., electron qubit and any of the nuclear qubits) based on a similar mechanism, and possibly construct a complete universal set of gates on the qubit systems.

Apart from using dynamical decoupling pulses for the gate implementation, one can also use MW-pulses or RF-pulses to control and manipulate nuclear spins \cite{Pla2013}. Furthermore, one can combine dynamical decoupling pulses and MW/RF-pulses to engineer optimal desired gates with high fidelity \cite{Bartling2025,Perlin2019,Dulog2024}. In particular, instead of defining a computational basis from the free Hamiltonian and exploiting the interaction terms as gate actions, one can use the full Hamiltonian, including the interaction, to define a computational basis and address the transitions of the gate actions by MW/RF-pulses. These techniques require more complex control schemes, compared to our protocol, to achieve the desired operations, and they can also lead to long operation times when a considerable number of MW/RF pulses are required. The latter makes those techniques more demanding when the coherence and relaxation times are taken into account. However, they have the advantage of more flexibility for the nuclear spin manipulation. Full investigation of this technique is out of the scope of this work, and it is the subject of a future work.

\begin{acknowledgments}
This work was partially supported by the Yangyang Development Foundation.
We thank Zhenyu Wang for the fruitful discussions and his comments on our paper.
\end{acknowledgments}

\appendix
\section{Rotated Hamiltonian}\label{appen:H_rotate-derivation}
Here we derive the expression of the rotated Hamiltonian Eq.~\eqref{eq:H_rotate} in the main text
Following the conditions we supposed in the main text, the Hamiltonian Eq.~\eqref{eq:H_main} in the rotating frame for the free Hamiltonian
	\begin{equation}
		\hat{H}_0 = - \dfrac{\omega_0\hat{\sigma}_z}{2} - \sum_k\omega_k\hat{I}_k^z + \hat{H}_{\rm C}\cv{t}\label{eq:H_0},
	\end{equation}
can be written in the following form
	\begin{align}
		\widetilde{H}\cv{t} &= F_z\cv{t}\sum_k\cv{\tfrac{A_k^{zz}}{2}}\hat{\sigma}_z\hat{I}_k^z\nn\\
			&\phantom{=.} + \sum_kA_k^\perp\cos\Delta_k t\cv{F_x\cv{t} \hat{\sigma}_x\hat{I}_k^x + F_y\cv{t}\hat{\sigma}_y\hat{I}_k^y}\nn\\
			&\phantom{=.} - \sum_kA_k^\perp\sin\Delta_k t\cv{F_x\cv{t} \hat{\sigma}_x\hat{I}_k^y - F_y\cv{t}\hat{\sigma}_y\hat{I}_k^x}\nn\\
			&\phantom{=.} + \sum_kB_k^\perp\cos\Sigma_k t\cv{F_x\cv{t}\hat{\sigma}_x\hat{I}_k^x -  F_y\cv{t}\hat{\sigma}_y\hat{I}_k^y}\nn\\
			&\phantom{=.} + \sum_kB_k^\perp\sin\Sigma_k t\cv{F_x\cv{t}\hat{\sigma}_x\hat{I}_k^y +  F_y\cv{t}\hat{\sigma}_y\hat{I}_k^x}\nn\\
			&\phantom{=.} - \sum_k\cv{\tfrac{A_k^{xy}}{2\sqrt{2}}}\sin\Sigma_k t \cv{F_x\cv{t}\hat{\sigma}_x\hat{I}_k^x - F_y\cv{t}\hat{\sigma}_y\hat{I}_k^y}\nn\\
			&\phantom{=.} + \sum_k\cv{\tfrac{A_k^{xy}}{2\sqrt{2}}}\cos\Sigma_k t \cv{F_x\cv{t}\hat{\sigma}_x\hat{I}_k^y + F_y\cv{t}\hat{\sigma}_y\hat{I}_k^x},\label{eq:H_int}
	\end{align}
where we write $\Delta_k = \omega_0 - \omega_k,$ $\Sigma_k = \omega_0 + \omega_k$ being rotating wave and counter-rotating wave frequencies for the nuclear spin $k,$ $A_k^\perp = \dfrac{A_k^{xx} + A_k^{yy}}{2\sqrt{2}}$ and  $B_k^\perp = \dfrac{A_k^{xx} - A_k^{yy}}{2\sqrt{2}}.$

Let us use the parameters for $\mathbb{A}_k$ from Ref.~\cite{Gao2022},
	\begin{align}
		\mathbb{A}_1 &= \left(\begin{array}{ccc}
			79.406 & 18.391 & 0\\
			18.391 & 58.170 & 0\\
			0 & 0 & 48.159
		\end{array}\right) \label{eq:A_1}\\
		\mathbb{A}_2 &= \left(\begin{array}{ccc}
			46.944 & 0 & 0\\
			0 & 90.025 & 0\\
			0 & 0 & 48.158
		\end{array}\right) \label{eq:A_2}\\
		\mathbb{A}_3 &= \left(\begin{array}{ccc}
			79.406 & -18.391 & 0\\
			-18.391 & 58.170 & 0\\
			0 & 0 & 48.159
		\end{array}\right) \label{eq:A_3},
	\end{align}
where the elements are expressed in the unit of $2\pi\times{\rm MHz}$. With these parameters, it is observed that several parameters are approximately uniform for all nuclear spins, namely $A_k^{zz}$ and the sum of the in-plane diagonal terms $A_k^{xx} + A_k^{yy}.$ Since $\omega_k = \gamma_nB + A_k^{zz}/2,$ it is then possible to replace the rotating frequencies $\Delta_k$ with their average $\Delta = \sum_k\cv{\omega_0 - \omega_k}/3.$ Likewise, one can use the scaled averages
	\[ a_z = \dfrac{1}{6}\sum_kA_k^{zz},~ a_\perp = \dfrac{1}{6\sqrt{2}}\sum_k\cv{A_k^{xx} + A_k^{yy}},\]
in the places for the coefficients in the rotating terms. Finally to achieve the expression Eq.~\eqref{eq:H_rotate}:
	\begin{align*}
		\widetilde{H}\cv{t} &\approx a_zF_z\cv{t}\hat{\sigma}_z\sum_k\hat{I}_k^z + a_\perp F_x\cv{t} \cos\Delta t\hat{\sigma}_x\sum_k\hat{I}_k^x\nn\\ 
		&\phantom{=.}+ a_\perp F_y\cv{t} \cos\Delta t\hat{\sigma}_y\sum_k\hat{I}_k^y\nn\\ 
		&\phantom{=.}+ \hat{H}_{\rm RW, odd}\cv{t} + \hat{H}_{\rm CRW}\cv{t},
	\end{align*}
we write 
\[\hat{H}_{\rm RW, odd}\cv{t} = -\sum_kA_k^\perp\sin\Delta_k t\cv{F_x\cv{t} \hat{\sigma}_x\hat{I}_k^y - F_y\cv{t}\hat{\sigma}_y\hat{I}_x}\] 
and write $\hat{H}_{\rm CRW}\cv{t}$ to represent all the counter terms in Eq.~\eqref{eq:H_int}, i.e. all the terms associated with frequencies $\Sigma_k.$

\section{Multiple Filter Functions Insertion}\label{appen:filter-insertion}
To simplify the problem, let us consider the cluster Hamiltonian in the form
	\begin{equation}
		\widetilde{H}_{1}\cv{t} = \sum_{\alpha=x,y,z}\hat{\sigma}_\alpha\hat{H}_\alpha\cv{t} +\hat{H}_{\rm C}\cv{t}\label{eq:basic_H},
	\end{equation}
where $\hat{H}_{\rm C}\cv{t}$ is the control field. In our setup, the first term is simply the expression in Eq.~\eqref{eq:H_main} in the rotating frame with respect to the free Hamiltonian Eq.~\eqref{eq:H_0} without the control term, or it can even be the Hamiltonian in the tilted frame in the main line consideration. Here we write the local control field $\hat{H}_{\rm C}\cv{t}$ as
	\begin{align}
		\hat{H}_{\rm C}\cv{t} &= h_z\cv{t}\hat{\sigma}_z + h_x\cv{t}\hat{\sigma}_x  + h_y\cv{t}\hat{\sigma}_y\nn\\
		 &= \mathbf{h}\cv{t}\cdot\boldsymbol{\sigma} = {h}\cv{t}\hat{\mathbf{h}}\cv{t}\cdot\boldsymbol{\sigma} \label{eq:control-field}
	\end{align}
where the first term is the added variation of the magnetic field, and the remaining term corresponds to the microwave field driving the transfer between two levels.

Transform the Hamiltonian Eq.~\eqref{eq:basic_H} into the rotating frame for the control Hamiltonian $\hat{H}_{\rm C}\cv{t}.$ It follows that
	\begin{equation}
		\widetilde{H}\cv{t} = \sum_{\alpha=x,y,z}\hat{\sigma}_\alpha\cv{t}\hat{H}_\alpha\cv{t}	\label{eq:H-rot-control},
	\end{equation}
where $\hat{\sigma}_\alpha\cv{t} = \mathcal{T}e^{i\int_0^t dt {\hat{H}_{\rm C}\cv{t}}}\hat{\sigma}_\alpha \overline{\mathcal{T}}e^{-i\int_0^t dt {\hat{H}_{\rm C}\cv{t}}},$ and, $\mathcal{T}$ and $\overline{\mathcal{T}}$ are a time ordering (the most left is the first event) and time anti ordering operator (the most left is the last event), respectively.

In dynamical decoupling, apart from the oscillating current, the simplest and the most employed type of control functions $h_\alpha\cv{t}$ are the square pulses. In particular, we suppose that 
	\begin{align*}
		\mathbf{h}\cv{t} &=  \sum_{\ell=1}^n \cvb{\Theta\cv{t_\ell}-\Theta\cv{t_{\ell-1}}}\mathbf{h}_\ell\\
		 &= \sum_{\ell=1}^n \cvb{\Theta\cv{t_\ell}-\Theta\cv{t_{\ell-1}}}{h}_\ell\hat{\mathbf{h}}_\ell\cdot\boldsymbol{\sigma},
	\end{align*}
where $\mathbf{h}_\ell$ are constant vectors, $0=t_0\leqslant t_1\leqslant\ldots\leqslant t_n,$ and $\Theta\cv{t} = \left\{\begin{array}{lr} 1, & t\geqslant 0\\ 0, &\text{~otherwise} \end{array} \right.$ is a Heaviside function. Let $\hat{\sigma}_\ell = \hat{\mathbf{h}}_\ell\cdot\boldsymbol{\sigma}$ and $\tau_\ell = t_\ell - t_{\ell-1}.$ Thus,
	\begin{align*}
		\mathcal{T}&e^{i\int_0^t dt {\hat{H}_{\rm C}\cv{t}}}\\ 
		&= \prod_{\ell=1}^n e^{i\tau_\ell h_\ell \hat{\sigma}_\ell} = \prod_{\ell=1}^n \cvb{\iden\cos\cv{h_\ell\tau_\ell} + i\hat{\sigma}_\ell\sin\cv{h_\ell\tau_\ell} }
	\end{align*}
and then
	\begin{equation}
		\hat{\sigma}_\alpha\cv{t} = T_{h_1\tau_1,\sigma_1}\circ\cdots\circ T_{h_n\tau_n,\sigma_n}\cvb{\hat{\sigma}_\alpha}\label{eq:sigma_alpha_t-base},
	\end{equation}
where
	\begin{align}
		T_{h_\ell\tau_\ell,\hat{\sigma}_\ell}\cvb{\hat{\sigma}_\alpha} &=  \hat{\sigma}_\alpha\cos^2\cv{h_\ell\tau_\ell}  + \cv{\hat{\sigma}_\ell\hat{\sigma}_\alpha\hat{\sigma}_\ell}\sin^2\cv{h_\ell\tau_\ell}\nn\\ 
		 &\phantom{=.} +i\cvb{\hat{\sigma}_\ell,\hat{\sigma}_\alpha}\dfrac{\sin\cv{2h_\ell\tau_\ell}}{2}\label{eq:T_ell}.
	\end{align}

The last term in Eq.~\eqref{eq:T_ell} is the crucial term providing a drift from the operator $\hat{\sigma}_\alpha.$ This term can be eliminated by setting $h_\ell\tau_\ell=\pi/2$ for all $\ell,$ and then 
	$ T_{\pi/2,\hat{\sigma}_\ell}\cvb{\hat{\sigma}_\alpha} = \hat{\sigma}_\ell\hat{\sigma}_\alpha\hat{\sigma}_\ell.$ Choosing $\hat{\sigma}_\ell$ for the set $\{\iden,\hat{\sigma}_x,\hat{\sigma}_y,\hat{\sigma}_z\}$ (where $\iden$ stands for $h_\ell=0$ during that time interval,) it follows that
	\begin{align}
		\hat{\sigma}_\alpha\cv{t} &= \cvb{\prod_{\ell=1}^n \hat{\sigma}_\ell}\hat{\sigma}_\alpha\cvb{\prod_{\ell=n}^1 \hat{\sigma}_\ell} = F_\alpha\cv{t}\hat{\sigma}_\alpha\label{eq:sigma_alpha_t}
	\end{align}	 
	where $F_\alpha\cv{t}$ is the filter function defined by
	\begin{equation}
		F_\alpha\cv{t} = \cv{-1}^{\sum_\ell\cv{1-\delta_{\alpha_\ell\alpha}}\cv{1-\delta_{\alpha_\ell 0}}} \label{eq:filter_g_alpha}.
	\end{equation}
and $\alpha_\ell =0,x,y,z$ regarding the choice of the control operator on the interval $\cv{t_{\ell-1},t_\ell}.$ 

\bibliography{main.bbl}

\begin{thebibliography}{75}%
\makeatletter
\providecommand \@ifxundefined [1]{%
 \@ifx{#1\undefined}
}%
\providecommand \@ifnum [1]{%
 \ifnum #1\expandafter \@firstoftwo
 \else \expandafter \@secondoftwo
 \fi
}%
\providecommand \@ifx [1]{%
 \ifx #1\expandafter \@firstoftwo
 \else \expandafter \@secondoftwo
 \fi
}%
\providecommand \natexlab [1]{#1}%
\providecommand \enquote  [1]{``#1''}%
\providecommand \bibnamefont  [1]{#1}%
\providecommand \bibfnamefont [1]{#1}%
\providecommand \citenamefont [1]{#1}%
\providecommand \href@noop [0]{\@secondoftwo}%
\providecommand \href [0]{\begingroup \@sanitize@url \@href}%
\providecommand \@href[1]{\@@startlink{#1}\@@href}%
\providecommand \@@href[1]{\endgroup#1\@@endlink}%
\providecommand \@sanitize@url [0]{\catcode `\\12\catcode `\$12\catcode `\&12\catcode `\#12\catcode `\^12\catcode `\_12\catcode `\%12\relax}%
\providecommand \@@startlink[1]{}%
\providecommand \@@endlink[0]{}%
\providecommand \url  [0]{\begingroup\@sanitize@url \@url }%
\providecommand \@url [1]{\endgroup\@href {#1}{\urlprefix }}%
\providecommand \urlprefix  [0]{URL }%
\providecommand \Eprint [0]{\href }%
\providecommand \doibase [0]{https://doi.org/}%
\providecommand \selectlanguage [0]{\@gobble}%
\providecommand \bibinfo  [0]{\@secondoftwo}%
\providecommand \bibfield  [0]{\@secondoftwo}%
\providecommand \translation [1]{[#1]}%
\providecommand \BibitemOpen [0]{}%
\providecommand \bibitemStop [0]{}%
\providecommand \bibitemNoStop [0]{.\EOS\space}%
\providecommand \EOS [0]{\spacefactor3000\relax}%
\providecommand \BibitemShut  [1]{\csname bibitem#1\endcsname}%
\let\auto@bib@innerbib\@empty
\bibitem [{\citenamefont {Morton}\ \emph {et~al.}(2008)\citenamefont {Morton}, \citenamefont {Tyryshkin}, \citenamefont {Brown}, \citenamefont {Shankar}, \citenamefont {Lovett}, \citenamefont {Ardavan}, \citenamefont {Schenkel}, \citenamefont {Haller}, \citenamefont {Ager},\ and\ \citenamefont {Lyon}}]{Morton2008}%
  \BibitemOpen
  \bibfield  {author} {\bibinfo {author} {\bibfnamefont {J.~J.~L.}\ \bibnamefont {Morton}}, \bibinfo {author} {\bibfnamefont {A.~M.}\ \bibnamefont {Tyryshkin}}, \bibinfo {author} {\bibfnamefont {R.~M.}\ \bibnamefont {Brown}}, \bibinfo {author} {\bibfnamefont {S.}~\bibnamefont {Shankar}}, \bibinfo {author} {\bibfnamefont {B.~W.}\ \bibnamefont {Lovett}}, \bibinfo {author} {\bibfnamefont {A.}~\bibnamefont {Ardavan}}, \bibinfo {author} {\bibfnamefont {T.}~\bibnamefont {Schenkel}}, \bibinfo {author} {\bibfnamefont {E.~E.}\ \bibnamefont {Haller}}, \bibinfo {author} {\bibfnamefont {J.~W.}\ \bibnamefont {Ager}},\ and\ \bibinfo {author} {\bibfnamefont {S.~A.}\ \bibnamefont {Lyon}},\ }\href {https://doi.org/10.1038/nature07295} {\bibfield  {journal} {\bibinfo  {journal} {Nature}\ }\textbf {\bibinfo {volume} {455}},\ \bibinfo {pages} {1085} (\bibinfo {year} {2008})}\BibitemShut {NoStop}%
\bibitem [{\citenamefont {McCamey}\ \emph {et~al.}(2010)\citenamefont {McCamey}, \citenamefont {Van~Tol}, \citenamefont {Morley},\ and\ \citenamefont {Boehme}}]{McCamey2010}%
  \BibitemOpen
  \bibfield  {author} {\bibinfo {author} {\bibfnamefont {D.~R.}\ \bibnamefont {McCamey}}, \bibinfo {author} {\bibfnamefont {J.}~\bibnamefont {Van~Tol}}, \bibinfo {author} {\bibfnamefont {G.~W.}\ \bibnamefont {Morley}},\ and\ \bibinfo {author} {\bibfnamefont {C.}~\bibnamefont {Boehme}},\ }\href {https://doi.org/10.1126/science.1197931} {\bibfield  {journal} {\bibinfo  {journal} {Science}\ }\textbf {\bibinfo {volume} {330}},\ \bibinfo {pages} {1652} (\bibinfo {year} {2010})}\BibitemShut {NoStop}%
\bibitem [{\citenamefont {Muhonen}\ \emph {et~al.}(2014)\citenamefont {Muhonen}, \citenamefont {Dehollain}, \citenamefont {Laucht}, \citenamefont {Hudson}, \citenamefont {Kalra}, \citenamefont {Sekiguchi}, \citenamefont {Itoh}, \citenamefont {Jamieson}, \citenamefont {McCallum}, \citenamefont {Dzurak},\ and\ \citenamefont {Morello}}]{Muhonen2014}%
  \BibitemOpen
  \bibfield  {author} {\bibinfo {author} {\bibfnamefont {J.~T.}\ \bibnamefont {Muhonen}}, \bibinfo {author} {\bibfnamefont {J.~P.}\ \bibnamefont {Dehollain}}, \bibinfo {author} {\bibfnamefont {A.}~\bibnamefont {Laucht}}, \bibinfo {author} {\bibfnamefont {F.~E.}\ \bibnamefont {Hudson}}, \bibinfo {author} {\bibfnamefont {R.}~\bibnamefont {Kalra}}, \bibinfo {author} {\bibfnamefont {T.}~\bibnamefont {Sekiguchi}}, \bibinfo {author} {\bibfnamefont {K.~M.}\ \bibnamefont {Itoh}}, \bibinfo {author} {\bibfnamefont {D.~N.}\ \bibnamefont {Jamieson}}, \bibinfo {author} {\bibfnamefont {J.~C.}\ \bibnamefont {McCallum}}, \bibinfo {author} {\bibfnamefont {A.~S.}\ \bibnamefont {Dzurak}},\ and\ \bibinfo {author} {\bibfnamefont {A.}~\bibnamefont {Morello}},\ }\href {https://doi.org/10.1038/nnano.2014.211} {\bibfield  {journal} {\bibinfo  {journal} {Nat. Nanotechnol.}\ }\textbf {\bibinfo {volume} {9}},\ \bibinfo {pages} {986} (\bibinfo {year} {2014})}\BibitemShut {NoStop}%
\bibitem [{\citenamefont {Witzel}\ and\ \citenamefont {Das~Sarma}(2007)}]{Witzel2007}%
  \BibitemOpen
  \bibfield  {author} {\bibinfo {author} {\bibfnamefont {W.~M.}\ \bibnamefont {Witzel}}\ and\ \bibinfo {author} {\bibfnamefont {S.}~\bibnamefont {Das~Sarma}},\ }\href {https://doi.org/10.1103/physrevb.76.045218} {\bibfield  {journal} {\bibinfo  {journal} {Phys. Rev. B}\ }\textbf {\bibinfo {volume} {76}},\ \bibinfo {pages} {045218} (\bibinfo {year} {2007})}\BibitemShut {NoStop}%
\bibitem [{\citenamefont {Schuetz}\ \emph {et~al.}(2014)\citenamefont {Schuetz}, \citenamefont {Kessler}, \citenamefont {Vandersypen}, \citenamefont {Cirac},\ and\ \citenamefont {Giedke}}]{Schuetz2014}%
  \BibitemOpen
  \bibfield  {author} {\bibinfo {author} {\bibfnamefont {M.~J.~A.}\ \bibnamefont {Schuetz}}, \bibinfo {author} {\bibfnamefont {E.~M.}\ \bibnamefont {Kessler}}, \bibinfo {author} {\bibfnamefont {L.~M.~K.}\ \bibnamefont {Vandersypen}}, \bibinfo {author} {\bibfnamefont {J.~I.}\ \bibnamefont {Cirac}},\ and\ \bibinfo {author} {\bibfnamefont {G.}~\bibnamefont {Giedke}},\ }\href {https://doi.org/10.1103/physrevb.89.195310} {\bibfield  {journal} {\bibinfo  {journal} {Phys. Rev. B}\ }\textbf {\bibinfo {volume} {89}},\ \bibinfo {pages} {195310} (\bibinfo {year} {2014})}\BibitemShut {NoStop}%
\bibitem [{\citenamefont {Rong}\ \emph {et~al.}(2015)\citenamefont {Rong}, \citenamefont {Geng}, \citenamefont {Shi}, \citenamefont {Liu}, \citenamefont {Xu}, \citenamefont {Ma}, \citenamefont {Kong}, \citenamefont {Jiang}, \citenamefont {Wu},\ and\ \citenamefont {Du}}]{Rong2015}%
  \BibitemOpen
  \bibfield  {author} {\bibinfo {author} {\bibfnamefont {X.}~\bibnamefont {Rong}}, \bibinfo {author} {\bibfnamefont {J.}~\bibnamefont {Geng}}, \bibinfo {author} {\bibfnamefont {F.}~\bibnamefont {Shi}}, \bibinfo {author} {\bibfnamefont {Y.}~\bibnamefont {Liu}}, \bibinfo {author} {\bibfnamefont {K.}~\bibnamefont {Xu}}, \bibinfo {author} {\bibfnamefont {W.}~\bibnamefont {Ma}}, \bibinfo {author} {\bibfnamefont {F.}~\bibnamefont {Kong}}, \bibinfo {author} {\bibfnamefont {Z.}~\bibnamefont {Jiang}}, \bibinfo {author} {\bibfnamefont {Y.}~\bibnamefont {Wu}},\ and\ \bibinfo {author} {\bibfnamefont {J.}~\bibnamefont {Du}},\ }\href {https://doi.org/10.1038/ncomms9748} {\bibfield  {journal} {\bibinfo  {journal} {Nat. Commun.}\ }\textbf {\bibinfo {volume} {6}},\ \bibinfo {pages} {8748} (\bibinfo {year} {2015})}\BibitemShut {NoStop}%
\bibitem [{\citenamefont {Abobeih}\ \emph {et~al.}(2022)\citenamefont {Abobeih}, \citenamefont {Wang}, \citenamefont {Randall}, \citenamefont {Loenen}, \citenamefont {Bradley}, \citenamefont {Markham}, \citenamefont {Twitchen}, \citenamefont {Terhal},\ and\ \citenamefont {Taminiau}}]{Abobeih2022}%
  \BibitemOpen
  \bibfield  {author} {\bibinfo {author} {\bibfnamefont {M.~H.}\ \bibnamefont {Abobeih}}, \bibinfo {author} {\bibfnamefont {Y.}~\bibnamefont {Wang}}, \bibinfo {author} {\bibfnamefont {J.}~\bibnamefont {Randall}}, \bibinfo {author} {\bibfnamefont {S.~J.~H.}\ \bibnamefont {Loenen}}, \bibinfo {author} {\bibfnamefont {C.~E.}\ \bibnamefont {Bradley}}, \bibinfo {author} {\bibfnamefont {M.}~\bibnamefont {Markham}}, \bibinfo {author} {\bibfnamefont {D.~J.}\ \bibnamefont {Twitchen}}, \bibinfo {author} {\bibfnamefont {B.~M.}\ \bibnamefont {Terhal}},\ and\ \bibinfo {author} {\bibfnamefont {T.~H.}\ \bibnamefont {Taminiau}},\ }\href {https://doi.org/10.1038/s41586-022-04819-6} {\bibfield  {journal} {\bibinfo  {journal} {Nature}\ }\textbf {\bibinfo {volume} {606}},\ \bibinfo {pages} {884} (\bibinfo {year} {2022})}\BibitemShut {NoStop}%
\bibitem [{\citenamefont {Denning}\ \emph {et~al.}(2019)\citenamefont {Denning}, \citenamefont {Gangloff}, \citenamefont {Atature}, \citenamefont {Mork},\ and\ \citenamefont {Le~Gall}}]{Denning2019}%
  \BibitemOpen
  \bibfield  {author} {\bibinfo {author} {\bibfnamefont {E.~V.}\ \bibnamefont {Denning}}, \bibinfo {author} {\bibfnamefont {D.~A.}\ \bibnamefont {Gangloff}}, \bibinfo {author} {\bibfnamefont {M.}~\bibnamefont {Atature}}, \bibinfo {author} {\bibfnamefont {J.}~\bibnamefont {Mork}},\ and\ \bibinfo {author} {\bibfnamefont {C.}~\bibnamefont {Le~Gall}},\ }\href {https://doi.org/10.1103/physrevlett.123.140502} {\bibfield  {journal} {\bibinfo  {journal} {Phys. Rev. Lett.}\ }\textbf {\bibinfo {volume} {123}},\ \bibinfo {pages} {140502} (\bibinfo {year} {2019})}\BibitemShut {NoStop}%
\bibitem [{\citenamefont {Parthasarathy}\ \emph {et~al.}(2023)\citenamefont {Parthasarathy}, \citenamefont {Kallinger}, \citenamefont {Kaiser}, \citenamefont {Berwian}, \citenamefont {Dasari}, \citenamefont {Friedrich},\ and\ \citenamefont {Nagy}}]{Parthasarathy2023}%
  \BibitemOpen
  \bibfield  {author} {\bibinfo {author} {\bibfnamefont {S.~K.}\ \bibnamefont {Parthasarathy}}, \bibinfo {author} {\bibfnamefont {B.}~\bibnamefont {Kallinger}}, \bibinfo {author} {\bibfnamefont {F.}~\bibnamefont {Kaiser}}, \bibinfo {author} {\bibfnamefont {P.}~\bibnamefont {Berwian}}, \bibinfo {author} {\bibfnamefont {D.~B.}\ \bibnamefont {Dasari}}, \bibinfo {author} {\bibfnamefont {J.}~\bibnamefont {Friedrich}},\ and\ \bibinfo {author} {\bibfnamefont {R.}~\bibnamefont {Nagy}},\ }\href {https://doi.org/10.1103/physrevapplied.19.034026} {\bibfield  {journal} {\bibinfo  {journal} {Phys. Rev. Applied}\ }\textbf {\bibinfo {volume} {19}},\ \bibinfo {pages} {034026} (\bibinfo {year} {2023})}\BibitemShut {NoStop}%
\bibitem [{\citenamefont {Zhao}\ \emph {et~al.}(2011)\citenamefont {Zhao}, \citenamefont {Hu}, \citenamefont {Ho}, \citenamefont {Wan},\ and\ \citenamefont {Liu}}]{Zhao2011}%
  \BibitemOpen
  \bibfield  {author} {\bibinfo {author} {\bibfnamefont {N.}~\bibnamefont {Zhao}}, \bibinfo {author} {\bibfnamefont {J.-L.}\ \bibnamefont {Hu}}, \bibinfo {author} {\bibfnamefont {S.-W.}\ \bibnamefont {Ho}}, \bibinfo {author} {\bibfnamefont {J.~T.~K.}\ \bibnamefont {Wan}},\ and\ \bibinfo {author} {\bibfnamefont {R.~B.}\ \bibnamefont {Liu}},\ }\href {https://doi.org/10.1038/nnano.2011.22} {\bibfield  {journal} {\bibinfo  {journal} {Nat. Nanotechnol.}\ }\textbf {\bibinfo {volume} {6}},\ \bibinfo {pages} {242} (\bibinfo {year} {2011})}\BibitemShut {NoStop}%
\bibitem [{\citenamefont {Kolkowitz}\ \emph {et~al.}(2012)\citenamefont {Kolkowitz}, \citenamefont {Unterreithmeier}, \citenamefont {Bennett},\ and\ \citenamefont {Lukin}}]{Kolkowitz2012}%
  \BibitemOpen
  \bibfield  {author} {\bibinfo {author} {\bibfnamefont {S.}~\bibnamefont {Kolkowitz}}, \bibinfo {author} {\bibfnamefont {Q.~P.}\ \bibnamefont {Unterreithmeier}}, \bibinfo {author} {\bibfnamefont {S.~D.}\ \bibnamefont {Bennett}},\ and\ \bibinfo {author} {\bibfnamefont {M.~D.}\ \bibnamefont {Lukin}},\ }\href {https://doi.org/10.1103/PhysRevLett.109.137601} {\bibfield  {journal} {\bibinfo  {journal} {Phys. Rev. Lett.}\ }\textbf {\bibinfo {volume} {109}},\ \bibinfo {pages} {137601} (\bibinfo {year} {2012})}\BibitemShut {NoStop}%
\bibitem [{\citenamefont {Taminiau}\ \emph {et~al.}(2012)\citenamefont {Taminiau}, \citenamefont {Wagenaar}, \citenamefont {van~der Sar}, \citenamefont {Jelezko}, \citenamefont {Dobrovitski},\ and\ \citenamefont {Hanson}}]{Taminiau2012}%
  \BibitemOpen
  \bibfield  {author} {\bibinfo {author} {\bibfnamefont {T.~H.}\ \bibnamefont {Taminiau}}, \bibinfo {author} {\bibfnamefont {J.~J.~T.}\ \bibnamefont {Wagenaar}}, \bibinfo {author} {\bibfnamefont {T.}~\bibnamefont {van~der Sar}}, \bibinfo {author} {\bibfnamefont {F.}~\bibnamefont {Jelezko}}, \bibinfo {author} {\bibfnamefont {V.~V.}\ \bibnamefont {Dobrovitski}},\ and\ \bibinfo {author} {\bibfnamefont {R.}~\bibnamefont {Hanson}},\ }\href {https://doi.org/10.1103/PhysRevLett.109.137602} {\bibfield  {journal} {\bibinfo  {journal} {Phys. Rev. Lett.}\ }\textbf {\bibinfo {volume} {109}},\ \bibinfo {pages} {137602} (\bibinfo {year} {2012})}\BibitemShut {NoStop}%
\bibitem [{\citenamefont {Smeltzer}\ \emph {et~al.}(2009)\citenamefont {Smeltzer}, \citenamefont {McIntyre},\ and\ \citenamefont {Childress}}]{Smeltzer2009}%
  \BibitemOpen
  \bibfield  {author} {\bibinfo {author} {\bibfnamefont {B.}~\bibnamefont {Smeltzer}}, \bibinfo {author} {\bibfnamefont {J.}~\bibnamefont {McIntyre}},\ and\ \bibinfo {author} {\bibfnamefont {L.}~\bibnamefont {Childress}},\ }\href {https://doi.org/10.1103/physreva.80.050302} {\bibfield  {journal} {\bibinfo  {journal} {Phys. Rev. A}\ }\textbf {\bibinfo {volume} {80}},\ \bibinfo {pages} {050302} (\bibinfo {year} {2009})}\BibitemShut {NoStop}%
\bibitem [{\citenamefont {Busaite}\ \emph {et~al.}(2020)\citenamefont {Busaite}, \citenamefont {Lazda}, \citenamefont {Berzins}, \citenamefont {Auzinsh}, \citenamefont {Ferber},\ and\ \citenamefont {Gahbauer}}]{Busaite2020}%
  \BibitemOpen
  \bibfield  {author} {\bibinfo {author} {\bibfnamefont {L.}~\bibnamefont {Busaite}}, \bibinfo {author} {\bibfnamefont {R.}~\bibnamefont {Lazda}}, \bibinfo {author} {\bibfnamefont {A.}~\bibnamefont {Berzins}}, \bibinfo {author} {\bibfnamefont {M.}~\bibnamefont {Auzinsh}}, \bibinfo {author} {\bibfnamefont {R.}~\bibnamefont {Ferber}},\ and\ \bibinfo {author} {\bibfnamefont {F.}~\bibnamefont {Gahbauer}},\ }\href {https://doi.org/10.1103/physrevb.102.224101} {\bibfield  {journal} {\bibinfo  {journal} {Phys. Rev. B}\ }\textbf {\bibinfo {volume} {102}},\ \bibinfo {pages} {224101} (\bibinfo {year} {2020})}\BibitemShut {NoStop}%
\bibitem [{\citenamefont {Chen}\ \emph {et~al.}(2017)\citenamefont {Chen}, \citenamefont {Schwarz},\ and\ \citenamefont {Plenio}}]{ChenQ2017}%
  \BibitemOpen
  \bibfield  {author} {\bibinfo {author} {\bibfnamefont {Q.}~\bibnamefont {Chen}}, \bibinfo {author} {\bibfnamefont {I.}~\bibnamefont {Schwarz}},\ and\ \bibinfo {author} {\bibfnamefont {M.~B.}\ \bibnamefont {Plenio}},\ }\href {https://doi.org/10.1103/PhysRevLett.119.010801} {\bibfield  {journal} {\bibinfo  {journal} {Phys. Rev. Lett.}\ }\textbf {\bibinfo {volume} {119}},\ \bibinfo {pages} {010801} (\bibinfo {year} {2017})}\BibitemShut {NoStop}%
\bibitem [{\citenamefont {Haase}\ \emph {et~al.}(2018)\citenamefont {Haase}, \citenamefont {Wang}, \citenamefont {Casanova},\ and\ \citenamefont {Plenio}}]{Haase2018}%
  \BibitemOpen
  \bibfield  {author} {\bibinfo {author} {\bibfnamefont {J.~F.}\ \bibnamefont {Haase}}, \bibinfo {author} {\bibfnamefont {Z.-Y.}\ \bibnamefont {Wang}}, \bibinfo {author} {\bibfnamefont {J.}~\bibnamefont {Casanova}},\ and\ \bibinfo {author} {\bibfnamefont {M.~B.}\ \bibnamefont {Plenio}},\ }\href {https://doi.org/10.1103/PhysRevLett.121.050402} {\bibfield  {journal} {\bibinfo  {journal} {Phys. Rev. Lett.}\ }\textbf {\bibinfo {volume} {121}},\ \bibinfo {pages} {050402} (\bibinfo {year} {2018})}\BibitemShut {NoStop}%
\bibitem [{\citenamefont {Perlin}\ \emph {et~al.}(2018)\citenamefont {Perlin}, \citenamefont {Wang}, \citenamefont {Casanova},\ and\ \citenamefont {Plenio}}]{Perlin2019}%
  \BibitemOpen
  \bibfield  {author} {\bibinfo {author} {\bibfnamefont {M.~A.}\ \bibnamefont {Perlin}}, \bibinfo {author} {\bibfnamefont {Z.-Y.}\ \bibnamefont {Wang}}, \bibinfo {author} {\bibfnamefont {J.}~\bibnamefont {Casanova}},\ and\ \bibinfo {author} {\bibfnamefont {M.~B.}\ \bibnamefont {Plenio}},\ }\href {https://doi.org/10.1088/2058-9565/aade5c} {\bibfield  {journal} {\bibinfo  {journal} {Quantum Sci. Technol.}\ }\textbf {\bibinfo {volume} {4}},\ \bibinfo {pages} {015007} (\bibinfo {year} {2018})}\BibitemShut {NoStop}%
\bibitem [{\citenamefont {Abobeih}\ \emph {et~al.}(2019)\citenamefont {Abobeih}, \citenamefont {Randall}, \citenamefont {Bradley}, \citenamefont {Bartling}, \citenamefont {Bakker}, \citenamefont {Degen}, \citenamefont {Markham}, \citenamefont {Twitchen},\ and\ \citenamefont {Taminiau}}]{Abobeih2019}%
  \BibitemOpen
  \bibfield  {author} {\bibinfo {author} {\bibfnamefont {M.~H.}\ \bibnamefont {Abobeih}}, \bibinfo {author} {\bibfnamefont {J.}~\bibnamefont {Randall}}, \bibinfo {author} {\bibfnamefont {C.~E.}\ \bibnamefont {Bradley}}, \bibinfo {author} {\bibfnamefont {H.~P.}\ \bibnamefont {Bartling}}, \bibinfo {author} {\bibfnamefont {M.~A.}\ \bibnamefont {Bakker}}, \bibinfo {author} {\bibfnamefont {M.~J.}\ \bibnamefont {Degen}}, \bibinfo {author} {\bibfnamefont {M.}~\bibnamefont {Markham}}, \bibinfo {author} {\bibfnamefont {D.~J.}\ \bibnamefont {Twitchen}},\ and\ \bibinfo {author} {\bibfnamefont {T.~H.}\ \bibnamefont {Taminiau}},\ }\href {https://doi.org/10.1038/s41586-019-1834-7} {\bibfield  {journal} {\bibinfo  {journal} {Nature}\ }\textbf {\bibinfo {volume} {576}},\ \bibinfo {pages} {411} (\bibinfo {year} {2019})}\BibitemShut {NoStop}%
\bibitem [{\citenamefont {Goldman}\ \emph {et~al.}(2020)\citenamefont {Goldman}, \citenamefont {Patti}, \citenamefont {Levonian}, \citenamefont {Yelin},\ and\ \citenamefont {Lukin}}]{Goldman2020}%
  \BibitemOpen
  \bibfield  {author} {\bibinfo {author} {\bibfnamefont {M.~L.}\ \bibnamefont {Goldman}}, \bibinfo {author} {\bibfnamefont {T.~L.}\ \bibnamefont {Patti}}, \bibinfo {author} {\bibfnamefont {D.}~\bibnamefont {Levonian}}, \bibinfo {author} {\bibfnamefont {S.~F.}\ \bibnamefont {Yelin}},\ and\ \bibinfo {author} {\bibfnamefont {M.~D.}\ \bibnamefont {Lukin}},\ }\href {https://doi.org/10.1103/physrevlett.124.153203} {\bibfield  {journal} {\bibinfo  {journal} {Phys. Rev. Lett.}\ }\textbf {\bibinfo {volume} {124}},\ \bibinfo {pages} {153203} (\bibinfo {year} {2020})}\BibitemShut {NoStop}%
\bibitem [{\citenamefont {Tratzmiller}\ \emph {et~al.}(2021)\citenamefont {Tratzmiller}, \citenamefont {Haase}, \citenamefont {Wang},\ and\ \citenamefont {Plenio}}]{Tratzmiller2021}%
  \BibitemOpen
  \bibfield  {author} {\bibinfo {author} {\bibfnamefont {B.}~\bibnamefont {Tratzmiller}}, \bibinfo {author} {\bibfnamefont {J.~F.}\ \bibnamefont {Haase}}, \bibinfo {author} {\bibfnamefont {Z.}~\bibnamefont {Wang}},\ and\ \bibinfo {author} {\bibfnamefont {M.~B.}\ \bibnamefont {Plenio}},\ }\href {https://doi.org/10.1103/PhysRevA.103.012607} {\bibfield  {journal} {\bibinfo  {journal} {Phys. Rev. A}\ }\textbf {\bibinfo {volume} {103}},\ \bibinfo {pages} {012607} (\bibinfo {year} {2021})}\BibitemShut {NoStop}%
\bibitem [{\citenamefont {Soshenko}\ \emph {et~al.}(2021)\citenamefont {Soshenko}, \citenamefont {Bolshedvorskii}, \citenamefont {Rubinas}, \citenamefont {Sorokin}, \citenamefont {Smolyaninov}, \citenamefont {Vorobyov},\ and\ \citenamefont {Akimov}}]{Soshenko2021}%
  \BibitemOpen
  \bibfield  {author} {\bibinfo {author} {\bibfnamefont {V.~V.}\ \bibnamefont {Soshenko}}, \bibinfo {author} {\bibfnamefont {S.~V.}\ \bibnamefont {Bolshedvorskii}}, \bibinfo {author} {\bibfnamefont {O.}~\bibnamefont {Rubinas}}, \bibinfo {author} {\bibfnamefont {V.~N.}\ \bibnamefont {Sorokin}}, \bibinfo {author} {\bibfnamefont {A.~N.}\ \bibnamefont {Smolyaninov}}, \bibinfo {author} {\bibfnamefont {V.~V.}\ \bibnamefont {Vorobyov}},\ and\ \bibinfo {author} {\bibfnamefont {A.~V.}\ \bibnamefont {Akimov}},\ }\href {https://doi.org/10.1103/physrevlett.126.197702} {\bibfield  {journal} {\bibinfo  {journal} {Phys. Rev. Lett.}\ }\textbf {\bibinfo {volume} {126}},\ \bibinfo {pages} {197702} (\bibinfo {year} {2021})}\BibitemShut {NoStop}%
\bibitem [{\citenamefont {Jiang}\ and\ \citenamefont {Chen}(2022)}]{JiangJ2022}%
  \BibitemOpen
  \bibfield  {author} {\bibinfo {author} {\bibfnamefont {J.}~\bibnamefont {Jiang}}\ and\ \bibinfo {author} {\bibfnamefont {Q.}~\bibnamefont {Chen}},\ }\href {https://doi.org/10.1103/PhysRevA.105.042426} {\bibfield  {journal} {\bibinfo  {journal} {Phys. Rev. A}\ }\textbf {\bibinfo {volume} {105}},\ \bibinfo {pages} {042426} (\bibinfo {year} {2022})}\BibitemShut {NoStop}%
\bibitem [{\citenamefont {Munuera-Javaloy}\ \emph {et~al.}(2023)\citenamefont {Munuera-Javaloy}, \citenamefont {Tobalina},\ and\ \citenamefont {Casanova}}]{MunueraJavaloy2023}%
  \BibitemOpen
  \bibfield  {author} {\bibinfo {author} {\bibfnamefont {C.}~\bibnamefont {Munuera-Javaloy}}, \bibinfo {author} {\bibfnamefont {A.}~\bibnamefont {Tobalina}},\ and\ \bibinfo {author} {\bibfnamefont {J.}~\bibnamefont {Casanova}},\ }\href {https://doi.org/10.1103/PhysRevLett.130.133603} {\bibfield  {journal} {\bibinfo  {journal} {Phys. Rev. Lett.}\ }\textbf {\bibinfo {volume} {130}},\ \bibinfo {pages} {133603} (\bibinfo {year} {2023})}\BibitemShut {NoStop}%
\bibitem [{\citenamefont {Zeng}\ \emph {et~al.}(2024)\citenamefont {Zeng}, \citenamefont {Yu}, \citenamefont {Plenio},\ and\ \citenamefont {Wang}}]{Zeng2024}%
  \BibitemOpen
  \bibfield  {author} {\bibinfo {author} {\bibfnamefont {K.}~\bibnamefont {Zeng}}, \bibinfo {author} {\bibfnamefont {X.}~\bibnamefont {Yu}}, \bibinfo {author} {\bibfnamefont {M.~B.}\ \bibnamefont {Plenio}},\ and\ \bibinfo {author} {\bibfnamefont {Z.-Y.}\ \bibnamefont {Wang}},\ }\href {https://doi.org/10.1103/PhysRevLett.132.250801} {\bibfield  {journal} {\bibinfo  {journal} {Phys. Rev. Lett.}\ }\textbf {\bibinfo {volume} {132}},\ \bibinfo {pages} {250801} (\bibinfo {year} {2024})}\BibitemShut {NoStop}%
\bibitem [{\citenamefont {Xu}\ \emph {et~al.}(2024)\citenamefont {Xu}, \citenamefont {Xie},\ and\ \citenamefont {Wang}}]{Xu2024}%
  \BibitemOpen
  \bibfield  {author} {\bibinfo {author} {\bibfnamefont {S.}~\bibnamefont {Xu}}, \bibinfo {author} {\bibfnamefont {C.}~\bibnamefont {Xie}},\ and\ \bibinfo {author} {\bibfnamefont {Z.-Y.}\ \bibnamefont {Wang}},\ }\href {https://doi.org/10.1103/PhysRevA.109.L020601} {\bibfield  {journal} {\bibinfo  {journal} {Phys. Rev. A}\ }\textbf {\bibinfo {volume} {109}},\ \bibinfo {pages} {L020601} (\bibinfo {year} {2024})}\BibitemShut {NoStop}%
\bibitem [{\citenamefont {Mizuno}\ \emph {et~al.}(2024)\citenamefont {Mizuno}, \citenamefont {Fujisaki}, \citenamefont {Tomioka}, \citenamefont {Ishiwata}, \citenamefont {Onoda}, \citenamefont {Iwasaki}, \citenamefont {Arai},\ and\ \citenamefont {Hatano}}]{Mizuno2024}%
  \BibitemOpen
  \bibfield  {author} {\bibinfo {author} {\bibfnamefont {K.}~\bibnamefont {Mizuno}}, \bibinfo {author} {\bibfnamefont {I.}~\bibnamefont {Fujisaki}}, \bibinfo {author} {\bibfnamefont {H.}~\bibnamefont {Tomioka}}, \bibinfo {author} {\bibfnamefont {H.}~\bibnamefont {Ishiwata}}, \bibinfo {author} {\bibfnamefont {S.}~\bibnamefont {Onoda}}, \bibinfo {author} {\bibfnamefont {T.}~\bibnamefont {Iwasaki}}, \bibinfo {author} {\bibfnamefont {K.}~\bibnamefont {Arai}},\ and\ \bibinfo {author} {\bibfnamefont {M.}~\bibnamefont {Hatano}},\ }\href {https://doi.org/10.1088/2399-6528/ad2b8b} {\bibfield  {journal} {\bibinfo  {journal} {J. Phys. Commun.}\ }\textbf {\bibinfo {volume} {8}},\ \bibinfo {pages} {035002} (\bibinfo {year} {2024})}\BibitemShut {NoStop}%
\bibitem [{\citenamefont {Bartling}\ \emph {et~al.}(2025)\citenamefont {Bartling}, \citenamefont {Yun}, \citenamefont {Schymik}, \citenamefont {van Riggelen}, \citenamefont {Enthoven}, \citenamefont {van Ommen}, \citenamefont {Babaie}, \citenamefont {Sebastiano}, \citenamefont {Markham}, \citenamefont {Twitchen},\ and\ \citenamefont {Taminiau}}]{Bartling2025}%
  \BibitemOpen
  \bibfield  {author} {\bibinfo {author} {\bibfnamefont {H.}~\bibnamefont {Bartling}}, \bibinfo {author} {\bibfnamefont {J.}~\bibnamefont {Yun}}, \bibinfo {author} {\bibfnamefont {K.}~\bibnamefont {Schymik}}, \bibinfo {author} {\bibfnamefont {M.}~\bibnamefont {van Riggelen}}, \bibinfo {author} {\bibfnamefont {L.}~\bibnamefont {Enthoven}}, \bibinfo {author} {\bibfnamefont {H.}~\bibnamefont {van Ommen}}, \bibinfo {author} {\bibfnamefont {M.}~\bibnamefont {Babaie}}, \bibinfo {author} {\bibfnamefont {F.}~\bibnamefont {Sebastiano}}, \bibinfo {author} {\bibfnamefont {M.}~\bibnamefont {Markham}}, \bibinfo {author} {\bibfnamefont {D.}~\bibnamefont {Twitchen}},\ and\ \bibinfo {author} {\bibfnamefont {T.}~\bibnamefont {Taminiau}},\ }\href {https://doi.org/10.1103/physrevapplied.23.034052} {\bibfield  {journal} {\bibinfo  {journal} {Phys. Rev. Appl.}\ }\textbf {\bibinfo {volume} {23}},\ \bibinfo {pages} {034052} (\bibinfo {year} {2025})}\BibitemShut {NoStop}%
\bibitem [{\citenamefont {Jaeger}\ \emph {et~al.}(2024)\citenamefont {Jaeger}, \citenamefont {Kwon}, \citenamefont {Keller}, \citenamefont {Maier}, \citenamefont {Bronn}, \citenamefont {Finsterhoelzl}, \citenamefont {Burkard}, \citenamefont {Buettner}, \citenamefont {Eberle}, \citenamefont {Haehnel}, \citenamefont {Vorobyov},\ and\ \citenamefont {Wrachtrup}}]{Jaeger2024}%
  \BibitemOpen
  \bibfield  {author} {\bibinfo {author} {\bibfnamefont {T.}~\bibnamefont {Jaeger}}, \bibinfo {author} {\bibfnamefont {M.}~\bibnamefont {Kwon}}, \bibinfo {author} {\bibfnamefont {M.}~\bibnamefont {Keller}}, \bibinfo {author} {\bibfnamefont {R.}~\bibnamefont {Maier}}, \bibinfo {author} {\bibfnamefont {N.}~\bibnamefont {Bronn}}, \bibinfo {author} {\bibfnamefont {R.}~\bibnamefont {Finsterhoelzl}}, \bibinfo {author} {\bibfnamefont {G.}~\bibnamefont {Burkard}}, \bibinfo {author} {\bibfnamefont {L.}~\bibnamefont {Buettner}}, \bibinfo {author} {\bibfnamefont {R.}~\bibnamefont {Eberle}}, \bibinfo {author} {\bibfnamefont {D.}~\bibnamefont {Haehnel}}, \bibinfo {author} {\bibfnamefont {V.}~\bibnamefont {Vorobyov}},\ and\ \bibinfo {author} {\bibfnamefont {J.}~\bibnamefont {Wrachtrup}},\ }\Eprint {https://arxiv.org/abs/2412.12959} {arXiv:2412.12959 [quant-ph]}  (\bibinfo {year} {2024})\BibitemShut {NoStop}%
\bibitem [{\citenamefont {Dulog}\ and\ \citenamefont {Plenio}(2024)}]{Dulog2024}%
  \BibitemOpen
  \bibfield  {author} {\bibinfo {author} {\bibfnamefont {D.}~\bibnamefont {Dulog}}\ and\ \bibinfo {author} {\bibfnamefont {M.~B.}\ \bibnamefont {Plenio}},\ }\Eprint {https://arxiv.org/abs/2411.18450} {arXiv:2411.18450 [quant-ph]}  (\bibinfo {year} {2024})\BibitemShut {NoStop}%
\bibitem [{\citenamefont {Lee}\ \emph {et~al.}(2025)\citenamefont {Lee}, \citenamefont {Liu}, \citenamefont {Zhang}, \citenamefont {Kim}, \citenamefont {Gong}, \citenamefont {Du}, \citenamefont {Pham}, \citenamefont {Poirier}, \citenamefont {Hao}, \citenamefont {Edgar}, \citenamefont {Kim}, \citenamefont {Zu}, \citenamefont {Davis},\ and\ \citenamefont {Yao}}]{Lee2025}%
  \BibitemOpen
  \bibfield  {author} {\bibinfo {author} {\bibfnamefont {W.}~\bibnamefont {Lee}}, \bibinfo {author} {\bibfnamefont {V.}~\bibnamefont {Liu}}, \bibinfo {author} {\bibfnamefont {Z.}~\bibnamefont {Zhang}}, \bibinfo {author} {\bibfnamefont {S.}~\bibnamefont {Kim}}, \bibinfo {author} {\bibfnamefont {R.}~\bibnamefont {Gong}}, \bibinfo {author} {\bibfnamefont {X.}~\bibnamefont {Du}}, \bibinfo {author} {\bibfnamefont {K.}~\bibnamefont {Pham}}, \bibinfo {author} {\bibfnamefont {T.}~\bibnamefont {Poirier}}, \bibinfo {author} {\bibfnamefont {Z.}~\bibnamefont {Hao}}, \bibinfo {author} {\bibfnamefont {J.}~\bibnamefont {Edgar}}, \bibinfo {author} {\bibfnamefont {P.}~\bibnamefont {Kim}}, \bibinfo {author} {\bibfnamefont {C.}~\bibnamefont {Zu}}, \bibinfo {author} {\bibfnamefont {E.}~\bibnamefont {Davis}},\ and\ \bibinfo {author} {\bibfnamefont {N.}~\bibnamefont {Yao}},\ }\href {https://doi.org/10.1103/physrevlett.134.096202} {\bibfield  {journal} {\bibinfo  {journal} {Physical Review Letters}\ }\textbf {\bibinfo {volume}
  {134}},\ \bibinfo {pages} {096202} (\bibinfo {year} {2025})}\BibitemShut {NoStop}%
\bibitem [{\citenamefont {Tran}\ \emph {et~al.}(2015)\citenamefont {Tran}, \citenamefont {Bray}, \citenamefont {Ford}, \citenamefont {Toth},\ and\ \citenamefont {Aharonovich}}]{Tran2015}%
  \BibitemOpen
  \bibfield  {author} {\bibinfo {author} {\bibfnamefont {T.~T.}\ \bibnamefont {Tran}}, \bibinfo {author} {\bibfnamefont {K.}~\bibnamefont {Bray}}, \bibinfo {author} {\bibfnamefont {M.~J.}\ \bibnamefont {Ford}}, \bibinfo {author} {\bibfnamefont {M.}~\bibnamefont {Toth}},\ and\ \bibinfo {author} {\bibfnamefont {I.}~\bibnamefont {Aharonovich}},\ }\href {https://doi.org/10.1038/nnano.2015.242} {\bibfield  {journal} {\bibinfo  {journal} {Nat. Nanotechnol.}\ }\textbf {\bibinfo {volume} {11}},\ \bibinfo {pages} {37} (\bibinfo {year} {2015})}\BibitemShut {NoStop}%
\bibitem [{\citenamefont {Gottscholl}\ \emph {et~al.}(2021)\citenamefont {Gottscholl}, \citenamefont {Diez}, \citenamefont {Soltamov}, \citenamefont {Kasper}, \citenamefont {Sperlich}, \citenamefont {Kianinia}, \citenamefont {Bradac}, \citenamefont {Aharonovich},\ and\ \citenamefont {Dyakonov}}]{Gottscholl2021}%
  \BibitemOpen
  \bibfield  {author} {\bibinfo {author} {\bibfnamefont {A.}~\bibnamefont {Gottscholl}}, \bibinfo {author} {\bibfnamefont {M.}~\bibnamefont {Diez}}, \bibinfo {author} {\bibfnamefont {V.}~\bibnamefont {Soltamov}}, \bibinfo {author} {\bibfnamefont {C.}~\bibnamefont {Kasper}}, \bibinfo {author} {\bibfnamefont {A.}~\bibnamefont {Sperlich}}, \bibinfo {author} {\bibfnamefont {M.}~\bibnamefont {Kianinia}}, \bibinfo {author} {\bibfnamefont {C.}~\bibnamefont {Bradac}}, \bibinfo {author} {\bibfnamefont {I.}~\bibnamefont {Aharonovich}},\ and\ \bibinfo {author} {\bibfnamefont {V.}~\bibnamefont {Dyakonov}},\ }\href {https://doi.org/10.1126/sciadv.abf3630} {\bibfield  {journal} {\bibinfo  {journal} {Sci. Adv.}\ }\textbf {\bibinfo {volume} {7}},\ \bibinfo {pages} {eabf3630} (\bibinfo {year} {2021})}\BibitemShut {NoStop}%
\bibitem [{\citenamefont {Stern}\ \emph {et~al.}(2022)\citenamefont {Stern}, \citenamefont {Gu}, \citenamefont {Jarman}, \citenamefont {Eizagirre~Barker}, \citenamefont {Mendelson}, \citenamefont {Chugh}, \citenamefont {Schott}, \citenamefont {Tan}, \citenamefont {Sirringhaus}, \citenamefont {Aharonovich},\ and\ \citenamefont {Atat\"{u}re}}]{Stern2022}%
  \BibitemOpen
  \bibfield  {author} {\bibinfo {author} {\bibfnamefont {H.~L.}\ \bibnamefont {Stern}}, \bibinfo {author} {\bibfnamefont {Q.}~\bibnamefont {Gu}}, \bibinfo {author} {\bibfnamefont {J.}~\bibnamefont {Jarman}}, \bibinfo {author} {\bibfnamefont {S.}~\bibnamefont {Eizagirre~Barker}}, \bibinfo {author} {\bibfnamefont {N.}~\bibnamefont {Mendelson}}, \bibinfo {author} {\bibfnamefont {D.}~\bibnamefont {Chugh}}, \bibinfo {author} {\bibfnamefont {S.}~\bibnamefont {Schott}}, \bibinfo {author} {\bibfnamefont {H.~H.}\ \bibnamefont {Tan}}, \bibinfo {author} {\bibfnamefont {H.}~\bibnamefont {Sirringhaus}}, \bibinfo {author} {\bibfnamefont {I.}~\bibnamefont {Aharonovich}},\ and\ \bibinfo {author} {\bibfnamefont {M.}~\bibnamefont {Atat\"{u}re}},\ }\href {https://doi.org/10.1038/s41467-022-28169-z} {\bibfield  {journal} {\bibinfo  {journal} {Nat. Commun.}\ }\textbf {\bibinfo {volume} {13}},\ \bibinfo {pages} {618} (\bibinfo {year} {2022})}\BibitemShut {NoStop}%
\bibitem [{\citenamefont {Mu}\ \emph {et~al.}(2022)\citenamefont {Mu}, \citenamefont {Cai}, \citenamefont {Chen}, \citenamefont {Kenny}, \citenamefont {Jiang}, \citenamefont {Ru}, \citenamefont {Lyu}, \citenamefont {Koh}, \citenamefont {Liu}, \citenamefont {Aharonovich},\ and\ \citenamefont {Gao}}]{Mu2022}%
  \BibitemOpen
  \bibfield  {author} {\bibinfo {author} {\bibfnamefont {Z.}~\bibnamefont {Mu}}, \bibinfo {author} {\bibfnamefont {H.}~\bibnamefont {Cai}}, \bibinfo {author} {\bibfnamefont {D.}~\bibnamefont {Chen}}, \bibinfo {author} {\bibfnamefont {J.}~\bibnamefont {Kenny}}, \bibinfo {author} {\bibfnamefont {Z.}~\bibnamefont {Jiang}}, \bibinfo {author} {\bibfnamefont {S.}~\bibnamefont {Ru}}, \bibinfo {author} {\bibfnamefont {X.}~\bibnamefont {Lyu}}, \bibinfo {author} {\bibfnamefont {T.~S.}\ \bibnamefont {Koh}}, \bibinfo {author} {\bibfnamefont {X.}~\bibnamefont {Liu}}, \bibinfo {author} {\bibfnamefont {I.}~\bibnamefont {Aharonovich}},\ and\ \bibinfo {author} {\bibfnamefont {W.}~\bibnamefont {Gao}},\ }\href {https://doi.org/10.1103/PhysRevLett.128.216402} {\bibfield  {journal} {\bibinfo  {journal} {Phys. Rev. Lett.}\ }\textbf {\bibinfo {volume} {128}},\ \bibinfo {pages} {216402} (\bibinfo {year} {2022})}\BibitemShut {NoStop}%
\bibitem [{\citenamefont {Abdi}\ \emph {et~al.}(2018)\citenamefont {Abdi}, \citenamefont {Chou}, \citenamefont {Gali},\ and\ \citenamefont {Plenio}}]{Abdi2018}%
  \BibitemOpen
  \bibfield  {author} {\bibinfo {author} {\bibfnamefont {M.}~\bibnamefont {Abdi}}, \bibinfo {author} {\bibfnamefont {J.-P.}\ \bibnamefont {Chou}}, \bibinfo {author} {\bibfnamefont {A.}~\bibnamefont {Gali}},\ and\ \bibinfo {author} {\bibfnamefont {M.~B.}\ \bibnamefont {Plenio}},\ }\href {https://doi.org/10.1021/acsphotonics.7b01442} {\bibfield  {journal} {\bibinfo  {journal} {ACS Photonics}\ }\textbf {\bibinfo {volume} {5}},\ \bibinfo {pages} {1967} (\bibinfo {year} {2018})}\BibitemShut {NoStop}%
\bibitem [{\citenamefont {Sajid}\ \emph {et~al.}(2020)\citenamefont {Sajid}, \citenamefont {Ford},\ and\ \citenamefont {Reimers}}]{Sajid2020}%
  \BibitemOpen
  \bibfield  {author} {\bibinfo {author} {\bibfnamefont {A.}~\bibnamefont {Sajid}}, \bibinfo {author} {\bibfnamefont {M.~J.}\ \bibnamefont {Ford}},\ and\ \bibinfo {author} {\bibfnamefont {J.~R.}\ \bibnamefont {Reimers}},\ }\href {https://doi.org/10.1088/1361-6633/ab6310} {\bibfield  {journal} {\bibinfo  {journal} {Rep. Prog. Phys.}\ }\textbf {\bibinfo {volume} {83}},\ \bibinfo {pages} {044501} (\bibinfo {year} {2020})}\BibitemShut {NoStop}%
\bibitem [{\citenamefont {Gottscholl}\ \emph {et~al.}(2020)\citenamefont {Gottscholl}, \citenamefont {Kianinia}, \citenamefont {Soltamov}, \citenamefont {Orlinskii}, \citenamefont {Mamin}, \citenamefont {Bradac}, \citenamefont {Kasper}, \citenamefont {Krambrock}, \citenamefont {Sperlich}, \citenamefont {Toth}, \citenamefont {Aharonovich},\ and\ \citenamefont {Dyakonov}}]{Gottscholl2020}%
  \BibitemOpen
  \bibfield  {author} {\bibinfo {author} {\bibfnamefont {A.}~\bibnamefont {Gottscholl}}, \bibinfo {author} {\bibfnamefont {M.}~\bibnamefont {Kianinia}}, \bibinfo {author} {\bibfnamefont {V.}~\bibnamefont {Soltamov}}, \bibinfo {author} {\bibfnamefont {S.}~\bibnamefont {Orlinskii}}, \bibinfo {author} {\bibfnamefont {G.}~\bibnamefont {Mamin}}, \bibinfo {author} {\bibfnamefont {C.}~\bibnamefont {Bradac}}, \bibinfo {author} {\bibfnamefont {C.}~\bibnamefont {Kasper}}, \bibinfo {author} {\bibfnamefont {K.}~\bibnamefont {Krambrock}}, \bibinfo {author} {\bibfnamefont {A.}~\bibnamefont {Sperlich}}, \bibinfo {author} {\bibfnamefont {M.}~\bibnamefont {Toth}}, \bibinfo {author} {\bibfnamefont {I.}~\bibnamefont {Aharonovich}},\ and\ \bibinfo {author} {\bibfnamefont {V.}~\bibnamefont {Dyakonov}},\ }\href {https://doi.org/10.1038/s41563-020-0619-6} {\bibfield  {journal} {\bibinfo  {journal} {Nat. Mater.}\ }\textbf {\bibinfo {volume} {19}},\ \bibinfo {pages} {540} (\bibinfo {year} {2020})}\BibitemShut {NoStop}%
\bibitem [{\citenamefont {{\c C}akan}\ \emph {et~al.}(2025)\citenamefont {{\c C}akan}, \citenamefont {Cholsuk}, \citenamefont {Gale}, \citenamefont {Kianinia}, \citenamefont {Pa{\c c}al}, \citenamefont {Ate{\c s}}, \citenamefont {Aharonovich}, \citenamefont {Toth},\ and\ \citenamefont {Vogl}}]{Cakan2025}%
  \BibitemOpen
  \bibfield  {author} {\bibinfo {author} {\bibfnamefont {A.}~\bibnamefont {{\c C}akan}}, \bibinfo {author} {\bibfnamefont {C.}~\bibnamefont {Cholsuk}}, \bibinfo {author} {\bibfnamefont {A.}~\bibnamefont {Gale}}, \bibinfo {author} {\bibfnamefont {M.}~\bibnamefont {Kianinia}}, \bibinfo {author} {\bibfnamefont {S.}~\bibnamefont {Pa{\c c}al}}, \bibinfo {author} {\bibfnamefont {S.}~\bibnamefont {Ate{\c s}}}, \bibinfo {author} {\bibfnamefont {I.}~\bibnamefont {Aharonovich}}, \bibinfo {author} {\bibfnamefont {M.}~\bibnamefont {Toth}},\ and\ \bibinfo {author} {\bibfnamefont {T.}~\bibnamefont {Vogl}},\ }\href {https://doi.org/https://doi.org/10.1002/adom.202402508} {\bibfield  {journal} {\bibinfo  {journal} {Advanced Optical Materials}\ }\textbf {\bibinfo {volume} {13}},\ \bibinfo {pages} {2402508} (\bibinfo {year} {2025})}\BibitemShut {NoStop}%
\bibitem [{\citenamefont {Wolfowicz}\ \emph {et~al.}(2021)\citenamefont {Wolfowicz}, \citenamefont {Heremans}, \citenamefont {Anderson}, \citenamefont {Kanai}, \citenamefont {Seo}, \citenamefont {Gali}, \citenamefont {Galli},\ and\ \citenamefont {Awschalom}}]{Wolfowicz2021}%
  \BibitemOpen
  \bibfield  {author} {\bibinfo {author} {\bibfnamefont {G.}~\bibnamefont {Wolfowicz}}, \bibinfo {author} {\bibfnamefont {F.~J.}\ \bibnamefont {Heremans}}, \bibinfo {author} {\bibfnamefont {C.~P.}\ \bibnamefont {Anderson}}, \bibinfo {author} {\bibfnamefont {S.}~\bibnamefont {Kanai}}, \bibinfo {author} {\bibfnamefont {H.}~\bibnamefont {Seo}}, \bibinfo {author} {\bibfnamefont {A.}~\bibnamefont {Gali}}, \bibinfo {author} {\bibfnamefont {G.}~\bibnamefont {Galli}},\ and\ \bibinfo {author} {\bibfnamefont {D.~D.}\ \bibnamefont {Awschalom}},\ }\href {https://doi.org/10.1038/s41578-021-00306-y} {\bibfield  {journal} {\bibinfo  {journal} {Nat. Rev. Mater.}\ }\textbf {\bibinfo {volume} {6}},\ \bibinfo {pages} {906} (\bibinfo {year} {2021})}\BibitemShut {NoStop}%
\bibitem [{\citenamefont {Chejanovsky}\ \emph {et~al.}(2016)\citenamefont {Chejanovsky}, \citenamefont {Rezai}, \citenamefont {Paolucci}, \citenamefont {Kim}, \citenamefont {Rendler}, \citenamefont {Rouabeh}, \citenamefont {F{\'a}varo~de Oliveira}, \citenamefont {Herlinger}, \citenamefont {Denisenko}, \citenamefont {Yang}, \citenamefont {Gerhardt}, \citenamefont {Finkler}, \citenamefont {Smet},\ and\ \citenamefont {Wrachtrup}}]{Chejanovsky2016}%
  \BibitemOpen
  \bibfield  {author} {\bibinfo {author} {\bibfnamefont {N.}~\bibnamefont {Chejanovsky}}, \bibinfo {author} {\bibfnamefont {M.}~\bibnamefont {Rezai}}, \bibinfo {author} {\bibfnamefont {F.}~\bibnamefont {Paolucci}}, \bibinfo {author} {\bibfnamefont {Y.}~\bibnamefont {Kim}}, \bibinfo {author} {\bibfnamefont {T.}~\bibnamefont {Rendler}}, \bibinfo {author} {\bibfnamefont {W.}~\bibnamefont {Rouabeh}}, \bibinfo {author} {\bibfnamefont {F.}~\bibnamefont {F{\'a}varo~de Oliveira}}, \bibinfo {author} {\bibfnamefont {P.}~\bibnamefont {Herlinger}}, \bibinfo {author} {\bibfnamefont {A.}~\bibnamefont {Denisenko}}, \bibinfo {author} {\bibfnamefont {S.}~\bibnamefont {Yang}}, \bibinfo {author} {\bibfnamefont {I.}~\bibnamefont {Gerhardt}}, \bibinfo {author} {\bibfnamefont {A.}~\bibnamefont {Finkler}}, \bibinfo {author} {\bibfnamefont {J.~H.}\ \bibnamefont {Smet}},\ and\ \bibinfo {author} {\bibfnamefont {J.}~\bibnamefont {Wrachtrup}},\ }\href {https://doi.org/10.1021/acs.nanolett.6b03268} {\bibfield  {journal} {\bibinfo
  {journal} {Nano Lett.}\ }\textbf {\bibinfo {volume} {16}},\ \bibinfo {pages} {7037} (\bibinfo {year} {2016})}\BibitemShut {NoStop}%
\bibitem [{\citenamefont {Proscia}\ \emph {et~al.}(2018)\citenamefont {Proscia}, \citenamefont {Shotan}, \citenamefont {Jayakumar}, \citenamefont {Reddy}, \citenamefont {Cohen}, \citenamefont {Dollar}, \citenamefont {Alkauskas}, \citenamefont {Doherty}, \citenamefont {Meriles},\ and\ \citenamefont {Menon}}]{Proscia2018}%
  \BibitemOpen
  \bibfield  {author} {\bibinfo {author} {\bibfnamefont {N.~V.}\ \bibnamefont {Proscia}}, \bibinfo {author} {\bibfnamefont {Z.}~\bibnamefont {Shotan}}, \bibinfo {author} {\bibfnamefont {H.}~\bibnamefont {Jayakumar}}, \bibinfo {author} {\bibfnamefont {P.}~\bibnamefont {Reddy}}, \bibinfo {author} {\bibfnamefont {C.}~\bibnamefont {Cohen}}, \bibinfo {author} {\bibfnamefont {M.}~\bibnamefont {Dollar}}, \bibinfo {author} {\bibfnamefont {A.}~\bibnamefont {Alkauskas}}, \bibinfo {author} {\bibfnamefont {M.}~\bibnamefont {Doherty}}, \bibinfo {author} {\bibfnamefont {C.~A.}\ \bibnamefont {Meriles}},\ and\ \bibinfo {author} {\bibfnamefont {V.~M.}\ \bibnamefont {Menon}},\ }\href {https://doi.org/10.1364/OPTICA.5.001128} {\bibfield  {journal} {\bibinfo  {journal} {Optica}\ }\textbf {\bibinfo {volume} {5}},\ \bibinfo {pages} {1128} (\bibinfo {year} {2018})}\BibitemShut {NoStop}%
\bibitem [{\citenamefont {Exarhos}\ \emph {et~al.}(2019)\citenamefont {Exarhos}, \citenamefont {Hopper}, \citenamefont {Patel}, \citenamefont {Doherty},\ and\ \citenamefont {Bassett}}]{Exarhos2019}%
  \BibitemOpen
  \bibfield  {author} {\bibinfo {author} {\bibfnamefont {A.~L.}\ \bibnamefont {Exarhos}}, \bibinfo {author} {\bibfnamefont {D.~A.}\ \bibnamefont {Hopper}}, \bibinfo {author} {\bibfnamefont {R.~N.}\ \bibnamefont {Patel}}, \bibinfo {author} {\bibfnamefont {M.~W.}\ \bibnamefont {Doherty}},\ and\ \bibinfo {author} {\bibfnamefont {L.~C.}\ \bibnamefont {Bassett}},\ }\href {https://doi.org/10.1038/s41467-018-08185-8} {\bibfield  {journal} {\bibinfo  {journal} {Nat. Commun.}\ }\textbf {\bibinfo {volume} {10}},\ \bibinfo {pages} {222} (\bibinfo {year} {2019})}\BibitemShut {NoStop}%
\bibitem [{\citenamefont {Tabesh}\ \emph {et~al.}(2021)\citenamefont {Tabesh}, \citenamefont {Hassanzada}, \citenamefont {Hadian}, \citenamefont {Hashemi}, \citenamefont {Abdolhosseini~Sarsari},\ and\ \citenamefont {Abdi}}]{Tabesh2021}%
  \BibitemOpen
  \bibfield  {author} {\bibinfo {author} {\bibfnamefont {F.~T.}\ \bibnamefont {Tabesh}}, \bibinfo {author} {\bibfnamefont {Q.}~\bibnamefont {Hassanzada}}, \bibinfo {author} {\bibfnamefont {M.}~\bibnamefont {Hadian}}, \bibinfo {author} {\bibfnamefont {A.}~\bibnamefont {Hashemi}}, \bibinfo {author} {\bibfnamefont {I.}~\bibnamefont {Abdolhosseini~Sarsari}},\ and\ \bibinfo {author} {\bibfnamefont {M.}~\bibnamefont {Abdi}},\ }\href {https://doi.org/10.1088/2058-9565/ac2f4d} {\bibfield  {journal} {\bibinfo  {journal} {Quantum Sci. Technol.}\ }\textbf {\bibinfo {volume} {7}},\ \bibinfo {pages} {015002} (\bibinfo {year} {2021})}\BibitemShut {NoStop}%
\bibitem [{\citenamefont {Liu}\ and\ \citenamefont {Hersam}(2019)}]{Liu2019}%
  \BibitemOpen
  \bibfield  {author} {\bibinfo {author} {\bibfnamefont {X.}~\bibnamefont {Liu}}\ and\ \bibinfo {author} {\bibfnamefont {M.~C.}\ \bibnamefont {Hersam}},\ }\href {https://doi.org/10.1038/s41578-019-0136-x} {\bibfield  {journal} {\bibinfo  {journal} {Nat. Rev. Mater.}\ }\textbf {\bibinfo {volume} {4}},\ \bibinfo {pages} {669} (\bibinfo {year} {2019})}\BibitemShut {NoStop}%
\bibitem [{\citenamefont {Abdi}(2021)}]{Abdi2021}%
  \BibitemOpen
  \bibfield  {author} {\bibinfo {author} {\bibfnamefont {M.}~\bibnamefont {Abdi}},\ }\href {https://doi.org/10.1103/physreva.103.043520} {\bibfield  {journal} {\bibinfo  {journal} {Phys. Rev. A}\ }\textbf {\bibinfo {volume} {103}},\ \bibinfo {pages} {043520} (\bibinfo {year} {2021})}\BibitemShut {NoStop}%
\bibitem [{\citenamefont {Kianinia}\ \emph {et~al.}(2020)\citenamefont {Kianinia}, \citenamefont {White}, \citenamefont {Fr\"{o}ch}, \citenamefont {Bradac},\ and\ \citenamefont {Aharonovich}}]{Kianinia2020}%
  \BibitemOpen
  \bibfield  {author} {\bibinfo {author} {\bibfnamefont {M.}~\bibnamefont {Kianinia}}, \bibinfo {author} {\bibfnamefont {S.}~\bibnamefont {White}}, \bibinfo {author} {\bibfnamefont {J.~E.}\ \bibnamefont {Fr\"{o}ch}}, \bibinfo {author} {\bibfnamefont {C.}~\bibnamefont {Bradac}},\ and\ \bibinfo {author} {\bibfnamefont {I.}~\bibnamefont {Aharonovich}},\ }\href {https://doi.org/10.1021/acsphotonics.0c00614} {\bibfield  {journal} {\bibinfo  {journal} {ACS Photonics}\ }\textbf {\bibinfo {volume} {7}},\ \bibinfo {pages} {2147} (\bibinfo {year} {2020})}\BibitemShut {NoStop}%
\bibitem [{\citenamefont {Shaik}\ and\ \citenamefont {Palla}(2021)}]{Shaik2021}%
  \BibitemOpen
  \bibfield  {author} {\bibinfo {author} {\bibfnamefont {A.~B.}\ \bibnamefont {Shaik}}\ and\ \bibinfo {author} {\bibfnamefont {P.}~\bibnamefont {Palla}},\ }\href {https://doi.org/10.1038/s41598-021-90804-4} {\bibfield  {journal} {\bibinfo  {journal} {Sci. Rep.}\ }\textbf {\bibinfo {volume} {11}},\ \bibinfo {pages} {12285} (\bibinfo {year} {2021})}\BibitemShut {NoStop}%
\bibitem [{\citenamefont {Gao}\ \emph {et~al.}(2021)\citenamefont {Gao}, \citenamefont {Jiang}, \citenamefont {Llacsahuanga~Allcca}, \citenamefont {Shen}, \citenamefont {Sadi}, \citenamefont {Solanki}, \citenamefont {Ju}, \citenamefont {Xu}, \citenamefont {Upadhyaya}, \citenamefont {Chen}, \citenamefont {Bhave},\ and\ \citenamefont {Li}}]{Gao2021}%
  \BibitemOpen
  \bibfield  {author} {\bibinfo {author} {\bibfnamefont {X.}~\bibnamefont {Gao}}, \bibinfo {author} {\bibfnamefont {B.}~\bibnamefont {Jiang}}, \bibinfo {author} {\bibfnamefont {A.~E.}\ \bibnamefont {Llacsahuanga~Allcca}}, \bibinfo {author} {\bibfnamefont {K.}~\bibnamefont {Shen}}, \bibinfo {author} {\bibfnamefont {M.~A.}\ \bibnamefont {Sadi}}, \bibinfo {author} {\bibfnamefont {A.~B.}\ \bibnamefont {Solanki}}, \bibinfo {author} {\bibfnamefont {P.}~\bibnamefont {Ju}}, \bibinfo {author} {\bibfnamefont {Z.}~\bibnamefont {Xu}}, \bibinfo {author} {\bibfnamefont {P.}~\bibnamefont {Upadhyaya}}, \bibinfo {author} {\bibfnamefont {Y.~P.}\ \bibnamefont {Chen}}, \bibinfo {author} {\bibfnamefont {S.~A.}\ \bibnamefont {Bhave}},\ and\ \bibinfo {author} {\bibfnamefont {T.}~\bibnamefont {Li}},\ }\href {https://doi.org/10.1021/acs.nanolett.1c02495} {\bibfield  {journal} {\bibinfo  {journal} {Nano Lett.}\ }\textbf {\bibinfo {volume} {21}},\ \bibinfo {pages} {7708} (\bibinfo {year} {2021})}\BibitemShut {NoStop}%
\bibitem [{\citenamefont {Mendelson}\ \emph {et~al.}(2021)\citenamefont {Mendelson}, \citenamefont {Ritika}, \citenamefont {Kianinia}, \citenamefont {Scott}, \citenamefont {Kim}, \citenamefont {Fr\"{o}ch}, \citenamefont {Gazzana}, \citenamefont {Westerhausen}, \citenamefont {Xiao}, \citenamefont {Mohajerani}, \citenamefont {Strauf}, \citenamefont {Toth}, \citenamefont {Aharonovich},\ and\ \citenamefont {Xu}}]{Mendelson2021}%
  \BibitemOpen
  \bibfield  {author} {\bibinfo {author} {\bibfnamefont {N.}~\bibnamefont {Mendelson}}, \bibinfo {author} {\bibfnamefont {R.}~\bibnamefont {Ritika}}, \bibinfo {author} {\bibfnamefont {M.}~\bibnamefont {Kianinia}}, \bibinfo {author} {\bibfnamefont {J.}~\bibnamefont {Scott}}, \bibinfo {author} {\bibfnamefont {S.}~\bibnamefont {Kim}}, \bibinfo {author} {\bibfnamefont {J.~E.}\ \bibnamefont {Fr\"{o}ch}}, \bibinfo {author} {\bibfnamefont {C.}~\bibnamefont {Gazzana}}, \bibinfo {author} {\bibfnamefont {M.}~\bibnamefont {Westerhausen}}, \bibinfo {author} {\bibfnamefont {L.}~\bibnamefont {Xiao}}, \bibinfo {author} {\bibfnamefont {S.~S.}\ \bibnamefont {Mohajerani}}, \bibinfo {author} {\bibfnamefont {S.}~\bibnamefont {Strauf}}, \bibinfo {author} {\bibfnamefont {M.}~\bibnamefont {Toth}}, \bibinfo {author} {\bibfnamefont {I.}~\bibnamefont {Aharonovich}},\ and\ \bibinfo {author} {\bibfnamefont {Z.}~\bibnamefont {Xu}},\ }\href {https://doi.org/10.1002/adma.202106046} {\bibfield  {journal} {\bibinfo  {journal} {Adv. Mater.}\
  }\textbf {\bibinfo {volume} {34}},\ \bibinfo {pages} {2106046} (\bibinfo {year} {2021})}\BibitemShut {NoStop}%
\bibitem [{\citenamefont {Guo}\ \emph {et~al.}(2022)\citenamefont {Guo}, \citenamefont {Liu}, \citenamefont {Li}, \citenamefont {Yang}, \citenamefont {Yu}, \citenamefont {Meng}, \citenamefont {Wang}, \citenamefont {Zeng}, \citenamefont {Yan}, \citenamefont {Li}, \citenamefont {Wang}, \citenamefont {Xu}, \citenamefont {Wang}, \citenamefont {Tang}, \citenamefont {Li},\ and\ \citenamefont {Guo}}]{Guo2022}%
  \BibitemOpen
  \bibfield  {author} {\bibinfo {author} {\bibfnamefont {N.-J.}\ \bibnamefont {Guo}}, \bibinfo {author} {\bibfnamefont {W.}~\bibnamefont {Liu}}, \bibinfo {author} {\bibfnamefont {Z.-P.}\ \bibnamefont {Li}}, \bibinfo {author} {\bibfnamefont {Y.-Z.}\ \bibnamefont {Yang}}, \bibinfo {author} {\bibfnamefont {S.}~\bibnamefont {Yu}}, \bibinfo {author} {\bibfnamefont {Y.}~\bibnamefont {Meng}}, \bibinfo {author} {\bibfnamefont {Z.-A.}\ \bibnamefont {Wang}}, \bibinfo {author} {\bibfnamefont {X.-D.}\ \bibnamefont {Zeng}}, \bibinfo {author} {\bibfnamefont {F.-F.}\ \bibnamefont {Yan}}, \bibinfo {author} {\bibfnamefont {Q.}~\bibnamefont {Li}}, \bibinfo {author} {\bibfnamefont {J.-F.}\ \bibnamefont {Wang}}, \bibinfo {author} {\bibfnamefont {J.-S.}\ \bibnamefont {Xu}}, \bibinfo {author} {\bibfnamefont {Y.-T.}\ \bibnamefont {Wang}}, \bibinfo {author} {\bibfnamefont {J.-S.}\ \bibnamefont {Tang}}, \bibinfo {author} {\bibfnamefont {C.-F.}\ \bibnamefont {Li}},\ and\ \bibinfo {author} {\bibfnamefont {G.-C.}\ \bibnamefont {Guo}},\
  }\href {https://doi.org/10.1021/acsomega.1c04564} {\bibfield  {journal} {\bibinfo  {journal} {ACS Omega}\ }\textbf {\bibinfo {volume} {7}},\ \bibinfo {pages} {1733} (\bibinfo {year} {2022})}\BibitemShut {NoStop}%
\bibitem [{\citenamefont {Vaidya}\ \emph {et~al.}(2023)\citenamefont {Vaidya}, \citenamefont {Gao}, \citenamefont {Dikshit}, \citenamefont {Aharonovich},\ and\ \citenamefont {Li}}]{Vaidya2023}%
  \BibitemOpen
  \bibfield  {author} {\bibinfo {author} {\bibfnamefont {S.}~\bibnamefont {Vaidya}}, \bibinfo {author} {\bibfnamefont {X.}~\bibnamefont {Gao}}, \bibinfo {author} {\bibfnamefont {S.}~\bibnamefont {Dikshit}}, \bibinfo {author} {\bibfnamefont {I.}~\bibnamefont {Aharonovich}},\ and\ \bibinfo {author} {\bibfnamefont {T.}~\bibnamefont {Li}},\ }\href {https://doi.org/10.1080/23746149.2023.2206049} {\bibfield  {journal} {\bibinfo  {journal} {Adv. Phys.: X}\ }\textbf {\bibinfo {volume} {8}},\ \bibinfo {pages} {2206049} (\bibinfo {year} {2023})}\BibitemShut {NoStop}%
\bibitem [{\citenamefont {Das}\ \emph {et~al.}(2024)\citenamefont {Das}, \citenamefont {Melendez}, \citenamefont {Kao}, \citenamefont {Garc\'{\i}a-Monge}, \citenamefont {Russell}, \citenamefont {Li}, \citenamefont {Watanabe}, \citenamefont {Taniguchi}, \citenamefont {Edgar}, \citenamefont {Katoch}, \citenamefont {Yang}, \citenamefont {Hammel},\ and\ \citenamefont {Singh}}]{Das2024}%
  \BibitemOpen
  \bibfield  {author} {\bibinfo {author} {\bibfnamefont {S.}~\bibnamefont {Das}}, \bibinfo {author} {\bibfnamefont {A.~L.}\ \bibnamefont {Melendez}}, \bibinfo {author} {\bibfnamefont {I.-H.}\ \bibnamefont {Kao}}, \bibinfo {author} {\bibfnamefont {J.~A.}\ \bibnamefont {Garc\'{\i}a-Monge}}, \bibinfo {author} {\bibfnamefont {D.}~\bibnamefont {Russell}}, \bibinfo {author} {\bibfnamefont {J.}~\bibnamefont {Li}}, \bibinfo {author} {\bibfnamefont {K.}~\bibnamefont {Watanabe}}, \bibinfo {author} {\bibfnamefont {T.}~\bibnamefont {Taniguchi}}, \bibinfo {author} {\bibfnamefont {J.~H.}\ \bibnamefont {Edgar}}, \bibinfo {author} {\bibfnamefont {J.}~\bibnamefont {Katoch}}, \bibinfo {author} {\bibfnamefont {F.}~\bibnamefont {Yang}}, \bibinfo {author} {\bibfnamefont {P.~C.}\ \bibnamefont {Hammel}},\ and\ \bibinfo {author} {\bibfnamefont {S.}~\bibnamefont {Singh}},\ }\href {https://doi.org/10.1103/PhysRevLett.133.166704} {\bibfield  {journal} {\bibinfo  {journal} {Phys. Rev. Lett.}\ }\textbf {\bibinfo {volume} {133}},\
  \bibinfo {pages} {166704} (\bibinfo {year} {2024})}\BibitemShut {NoStop}%
\bibitem [{\citenamefont {Wong}\ \emph {et~al.}(2015)\citenamefont {Wong}, \citenamefont {Velasco}, \citenamefont {Ju}, \citenamefont {Lee}, \citenamefont {Kahn}, \citenamefont {Tsai}, \citenamefont {Germany}, \citenamefont {Taniguchi}, \citenamefont {Watanabe}, \citenamefont {Zettl}, \citenamefont {Wang},\ and\ \citenamefont {Crommie}}]{Wong2015}%
  \BibitemOpen
  \bibfield  {author} {\bibinfo {author} {\bibfnamefont {D.}~\bibnamefont {Wong}}, \bibinfo {author} {\bibfnamefont {J.}~\bibnamefont {Velasco}}, \bibinfo {author} {\bibfnamefont {L.}~\bibnamefont {Ju}}, \bibinfo {author} {\bibfnamefont {J.}~\bibnamefont {Lee}}, \bibinfo {author} {\bibfnamefont {S.}~\bibnamefont {Kahn}}, \bibinfo {author} {\bibfnamefont {H.-Z.}\ \bibnamefont {Tsai}}, \bibinfo {author} {\bibfnamefont {C.}~\bibnamefont {Germany}}, \bibinfo {author} {\bibfnamefont {T.}~\bibnamefont {Taniguchi}}, \bibinfo {author} {\bibfnamefont {K.}~\bibnamefont {Watanabe}}, \bibinfo {author} {\bibfnamefont {A.}~\bibnamefont {Zettl}}, \bibinfo {author} {\bibfnamefont {F.}~\bibnamefont {Wang}},\ and\ \bibinfo {author} {\bibfnamefont {M.~F.}\ \bibnamefont {Crommie}},\ }\href {https://doi.org/10.1038/nnano.2015.188} {\bibfield  {journal} {\bibinfo  {journal} {Nat. Nanotechnol.}\ }\textbf {\bibinfo {volume} {10}},\ \bibinfo {pages} {949} (\bibinfo {year} {2015})}\BibitemShut {NoStop}%
\bibitem [{\citenamefont {Iv\'{a}dy}\ \emph {et~al.}(2020)\citenamefont {Iv\'{a}dy}, \citenamefont {Barcza}, \citenamefont {Thiering}, \citenamefont {Li}, \citenamefont {Hamdi}, \citenamefont {Chou}, \citenamefont {Legeza},\ and\ \citenamefont {Gali}}]{Ivady2020}%
  \BibitemOpen
  \bibfield  {author} {\bibinfo {author} {\bibfnamefont {V.}~\bibnamefont {Iv\'{a}dy}}, \bibinfo {author} {\bibfnamefont {G.}~\bibnamefont {Barcza}}, \bibinfo {author} {\bibfnamefont {G.}~\bibnamefont {Thiering}}, \bibinfo {author} {\bibfnamefont {S.}~\bibnamefont {Li}}, \bibinfo {author} {\bibfnamefont {H.}~\bibnamefont {Hamdi}}, \bibinfo {author} {\bibfnamefont {J.-P.}\ \bibnamefont {Chou}}, \bibinfo {author} {\bibfnamefont {O.}~\bibnamefont {Legeza}},\ and\ \bibinfo {author} {\bibfnamefont {A.}~\bibnamefont {Gali}},\ }\href {https://doi.org/10.1038/s41524-020-0305-x} {\bibfield  {journal} {\bibinfo  {journal} {npj Comput. Mater.}\ }\textbf {\bibinfo {volume} {6}},\ \bibinfo {pages} {41} (\bibinfo {year} {2020})}\BibitemShut {NoStop}%
\bibitem [{\citenamefont {Gao}\ \emph {et~al.}(2022)\citenamefont {Gao}, \citenamefont {Vaidya}, \citenamefont {Li}, \citenamefont {Ju}, \citenamefont {Jiang}, \citenamefont {Xu}, \citenamefont {Allcca}, \citenamefont {Shen}, \citenamefont {Taniguchi}, \citenamefont {Watanabe}, \citenamefont {Bhave}, \citenamefont {Chen}, \citenamefont {Ping},\ and\ \citenamefont {Li}}]{Gao2022}%
  \BibitemOpen
  \bibfield  {author} {\bibinfo {author} {\bibfnamefont {X.}~\bibnamefont {Gao}}, \bibinfo {author} {\bibfnamefont {S.}~\bibnamefont {Vaidya}}, \bibinfo {author} {\bibfnamefont {K.}~\bibnamefont {Li}}, \bibinfo {author} {\bibfnamefont {P.}~\bibnamefont {Ju}}, \bibinfo {author} {\bibfnamefont {B.}~\bibnamefont {Jiang}}, \bibinfo {author} {\bibfnamefont {Z.}~\bibnamefont {Xu}}, \bibinfo {author} {\bibfnamefont {A.~E.~L.}\ \bibnamefont {Allcca}}, \bibinfo {author} {\bibfnamefont {K.}~\bibnamefont {Shen}}, \bibinfo {author} {\bibfnamefont {T.}~\bibnamefont {Taniguchi}}, \bibinfo {author} {\bibfnamefont {K.}~\bibnamefont {Watanabe}}, \bibinfo {author} {\bibfnamefont {S.~A.}\ \bibnamefont {Bhave}}, \bibinfo {author} {\bibfnamefont {Y.~P.}\ \bibnamefont {Chen}}, \bibinfo {author} {\bibfnamefont {Y.}~\bibnamefont {Ping}},\ and\ \bibinfo {author} {\bibfnamefont {T.}~\bibnamefont {Li}},\ }\href {https://doi.org/10.1038/s41563-022-01329-8} {\bibfield  {journal} {\bibinfo  {journal} {Nat. Mater.}\ }\textbf {\bibinfo
  {volume} {21}},\ \bibinfo {pages} {1024} (\bibinfo {year} {2022})}\BibitemShut {NoStop}%
\bibitem [{\citenamefont {Tabesh}\ \emph {et~al.}(2023)\citenamefont {Tabesh}, \citenamefont {Fani}, \citenamefont {Pedernales}, \citenamefont {Plenio},\ and\ \citenamefont {Abdi}}]{Tabesh2023}%
  \BibitemOpen
  \bibfield  {author} {\bibinfo {author} {\bibfnamefont {F.~T.}\ \bibnamefont {Tabesh}}, \bibinfo {author} {\bibfnamefont {M.}~\bibnamefont {Fani}}, \bibinfo {author} {\bibfnamefont {J.~S.}\ \bibnamefont {Pedernales}}, \bibinfo {author} {\bibfnamefont {M.~B.}\ \bibnamefont {Plenio}},\ and\ \bibinfo {author} {\bibfnamefont {M.}~\bibnamefont {Abdi}},\ }\href {https://doi.org/10.1103/PhysRevB.107.214307} {\bibfield  {journal} {\bibinfo  {journal} {Phys. Rev. B}\ }\textbf {\bibinfo {volume} {107}},\ \bibinfo {pages} {214307} (\bibinfo {year} {2023})}\BibitemShut {NoStop}%
\bibitem [{\citenamefont {Murzakhanov}\ \emph {et~al.}(2022)\citenamefont {Murzakhanov}, \citenamefont {Mamin}, \citenamefont {Orlinskii}, \citenamefont {Gerstmann}, \citenamefont {Schmidt}, \citenamefont {Biktagirov}, \citenamefont {Aharonovich}, \citenamefont {Gottscholl}, \citenamefont {Sperlich}, \citenamefont {Dyakonov},\ and\ \citenamefont {Soltamov}}]{Murzakhanov2022}%
  \BibitemOpen
  \bibfield  {author} {\bibinfo {author} {\bibfnamefont {F.~F.}\ \bibnamefont {Murzakhanov}}, \bibinfo {author} {\bibfnamefont {G.~V.}\ \bibnamefont {Mamin}}, \bibinfo {author} {\bibfnamefont {S.~B.}\ \bibnamefont {Orlinskii}}, \bibinfo {author} {\bibfnamefont {U.}~\bibnamefont {Gerstmann}}, \bibinfo {author} {\bibfnamefont {W.~G.}\ \bibnamefont {Schmidt}}, \bibinfo {author} {\bibfnamefont {T.}~\bibnamefont {Biktagirov}}, \bibinfo {author} {\bibfnamefont {I.}~\bibnamefont {Aharonovich}}, \bibinfo {author} {\bibfnamefont {A.}~\bibnamefont {Gottscholl}}, \bibinfo {author} {\bibfnamefont {A.}~\bibnamefont {Sperlich}}, \bibinfo {author} {\bibfnamefont {V.}~\bibnamefont {Dyakonov}},\ and\ \bibinfo {author} {\bibfnamefont {V.~A.}\ \bibnamefont {Soltamov}},\ }\href {https://doi.org/10.1021/acs.nanolett.1c04610} {\bibfield  {journal} {\bibinfo  {journal} {Nano Lett.}\ }\textbf {\bibinfo {volume} {22}},\ \bibinfo {pages} {2718} (\bibinfo {year} {2022})}\BibitemShut {NoStop}%
\bibitem [{\citenamefont {Doherty}\ \emph {et~al.}(2013)\citenamefont {Doherty}, \citenamefont {Manson}, \citenamefont {Delaney}, \citenamefont {Jelezko}, \citenamefont {Wrachtrup},\ and\ \citenamefont {Hollenberg}}]{Doherty2013}%
  \BibitemOpen
  \bibfield  {author} {\bibinfo {author} {\bibfnamefont {M.~W.}\ \bibnamefont {Doherty}}, \bibinfo {author} {\bibfnamefont {N.~B.}\ \bibnamefont {Manson}}, \bibinfo {author} {\bibfnamefont {P.}~\bibnamefont {Delaney}}, \bibinfo {author} {\bibfnamefont {F.}~\bibnamefont {Jelezko}}, \bibinfo {author} {\bibfnamefont {J.}~\bibnamefont {Wrachtrup}},\ and\ \bibinfo {author} {\bibfnamefont {L.~C.}\ \bibnamefont {Hollenberg}},\ }\href {https://doi.org/https://doi.org/10.1016/j.physrep.2013.02.001} {\bibfield  {journal} {\bibinfo  {journal} {Phys. Rep.}\ }\textbf {\bibinfo {volume} {528}},\ \bibinfo {pages} {1} (\bibinfo {year} {2013})},\ \bibinfo {note} {the nitrogen-vacancy colour centre in diamond}\BibitemShut {NoStop}%
\bibitem [{\citenamefont {Liu}\ \emph {et~al.}(2022)\citenamefont {Liu}, \citenamefont {Iv{\'a}dy}, \citenamefont {Li}, \citenamefont {Yang}, \citenamefont {Yu}, \citenamefont {Meng}, \citenamefont {Wang}, \citenamefont {Guo}, \citenamefont {Yan}, \citenamefont {Li}, \citenamefont {Wang}, \citenamefont {Xu}, \citenamefont {Liu}, \citenamefont {Zhou}, \citenamefont {Dong}, \citenamefont {Chen}, \citenamefont {Sun}, \citenamefont {Wang}, \citenamefont {Tang}, \citenamefont {Gali}, \citenamefont {Li},\ and\ \citenamefont {Guo}}]{Liu2022}%
  \BibitemOpen
  \bibfield  {author} {\bibinfo {author} {\bibfnamefont {W.}~\bibnamefont {Liu}}, \bibinfo {author} {\bibfnamefont {V.}~\bibnamefont {Iv{\'a}dy}}, \bibinfo {author} {\bibfnamefont {Z.-P.}\ \bibnamefont {Li}}, \bibinfo {author} {\bibfnamefont {Y.-Z.}\ \bibnamefont {Yang}}, \bibinfo {author} {\bibfnamefont {S.}~\bibnamefont {Yu}}, \bibinfo {author} {\bibfnamefont {Y.}~\bibnamefont {Meng}}, \bibinfo {author} {\bibfnamefont {Z.-A.}\ \bibnamefont {Wang}}, \bibinfo {author} {\bibfnamefont {N.-J.}\ \bibnamefont {Guo}}, \bibinfo {author} {\bibfnamefont {F.-F.}\ \bibnamefont {Yan}}, \bibinfo {author} {\bibfnamefont {Q.}~\bibnamefont {Li}}, \bibinfo {author} {\bibfnamefont {J.-F.}\ \bibnamefont {Wang}}, \bibinfo {author} {\bibfnamefont {J.-S.}\ \bibnamefont {Xu}}, \bibinfo {author} {\bibfnamefont {X.}~\bibnamefont {Liu}}, \bibinfo {author} {\bibfnamefont {Z.-Q.}\ \bibnamefont {Zhou}}, \bibinfo {author} {\bibfnamefont {Y.}~\bibnamefont {Dong}}, \bibinfo {author} {\bibfnamefont {X.-D.}\ \bibnamefont {Chen}}, \bibinfo
  {author} {\bibfnamefont {F.-W.}\ \bibnamefont {Sun}}, \bibinfo {author} {\bibfnamefont {Y.-T.}\ \bibnamefont {Wang}}, \bibinfo {author} {\bibfnamefont {J.-S.}\ \bibnamefont {Tang}}, \bibinfo {author} {\bibfnamefont {A.}~\bibnamefont {Gali}}, \bibinfo {author} {\bibfnamefont {C.-F.}\ \bibnamefont {Li}},\ and\ \bibinfo {author} {\bibfnamefont {G.-C.}\ \bibnamefont {Guo}},\ }\href {https://doi.org/10.1038/s41467-022-33399-2} {\bibfield  {journal} {\bibinfo  {journal} {Nat. Commun.}\ }\textbf {\bibinfo {volume} {13}},\ \bibinfo {pages} {5713} (\bibinfo {year} {2022})}\BibitemShut {NoStop}%
\bibitem [{\citenamefont {Nielsen}\ and\ \citenamefont {Chuang}(2010)}]{Nielsen_Chuang_2010}%
  \BibitemOpen
  \bibfield  {author} {\bibinfo {author} {\bibfnamefont {M.~A.}\ \bibnamefont {Nielsen}}\ and\ \bibinfo {author} {\bibfnamefont {I.~L.}\ \bibnamefont {Chuang}},\ }\href@noop {} {\emph {\bibinfo {title} {Quantum Computation and Quantum Information: 10th Anniversary Edition}}}\ (\bibinfo  {publisher} {Cambridge University Press},\ \bibinfo {year} {2010})\BibitemShut {NoStop}%
\bibitem [{\citenamefont {Gilchrist}\ \emph {et~al.}(2005)\citenamefont {Gilchrist}, \citenamefont {Langford},\ and\ \citenamefont {Nielsen}}]{Gilchrist2005}%
  \BibitemOpen
  \bibfield  {author} {\bibinfo {author} {\bibfnamefont {A.}~\bibnamefont {Gilchrist}}, \bibinfo {author} {\bibfnamefont {N.~K.}\ \bibnamefont {Langford}},\ and\ \bibinfo {author} {\bibfnamefont {M.~A.}\ \bibnamefont {Nielsen}},\ }\href {https://doi.org/10.1103/PhysRevA.71.062310} {\bibfield  {journal} {\bibinfo  {journal} {Phys. Rev. A}\ }\textbf {\bibinfo {volume} {71}},\ \bibinfo {pages} {062310} (\bibinfo {year} {2005})}\BibitemShut {NoStop}%
\bibitem [{\citenamefont {Johansson}\ \emph {et~al.}(2012)\citenamefont {Johansson}, \citenamefont {Nation},\ and\ \citenamefont {Nori}}]{QuTip1}%
  \BibitemOpen
  \bibfield  {author} {\bibinfo {author} {\bibfnamefont {J.}~\bibnamefont {Johansson}}, \bibinfo {author} {\bibfnamefont {P.}~\bibnamefont {Nation}},\ and\ \bibinfo {author} {\bibfnamefont {F.}~\bibnamefont {Nori}},\ }\href {https://doi.org/https://doi.org/10.1016/j.cpc.2012.02.021} {\bibfield  {journal} {\bibinfo  {journal} {Comput. Phys. Commun.}\ }\textbf {\bibinfo {volume} {183}},\ \bibinfo {pages} {1760} (\bibinfo {year} {2012})}\BibitemShut {NoStop}%
\bibitem [{\citenamefont {Johansson}\ \emph {et~al.}(2013)\citenamefont {Johansson}, \citenamefont {Nation},\ and\ \citenamefont {Nori}}]{QuTip2}%
  \BibitemOpen
  \bibfield  {author} {\bibinfo {author} {\bibfnamefont {J.}~\bibnamefont {Johansson}}, \bibinfo {author} {\bibfnamefont {P.}~\bibnamefont {Nation}},\ and\ \bibinfo {author} {\bibfnamefont {F.}~\bibnamefont {Nori}},\ }\href {https://doi.org/https://doi.org/10.1016/j.cpc.2012.11.019} {\bibfield  {journal} {\bibinfo  {journal} {Comput. Phys. Commun.}\ }\textbf {\bibinfo {volume} {184}},\ \bibinfo {pages} {1234} (\bibinfo {year} {2013})}\BibitemShut {NoStop}%
\bibitem [{\citenamefont {Greenberger}\ \emph {et~al.}(1990)\citenamefont {Greenberger}, \citenamefont {Horne}, \citenamefont {Shimony},\ and\ \citenamefont {Zeilinger}}]{Greenberger1990}%
  \BibitemOpen
  \bibfield  {author} {\bibinfo {author} {\bibfnamefont {D.~M.}\ \bibnamefont {Greenberger}}, \bibinfo {author} {\bibfnamefont {M.~A.}\ \bibnamefont {Horne}}, \bibinfo {author} {\bibfnamefont {A.}~\bibnamefont {Shimony}},\ and\ \bibinfo {author} {\bibfnamefont {A.}~\bibnamefont {Zeilinger}},\ }\href {https://doi.org/10.1119/1.16243} {\bibfield  {journal} {\bibinfo  {journal} {Am. J. Phys.}\ }\textbf {\bibinfo {volume} {58}},\ \bibinfo {pages} {1131} (\bibinfo {year} {1990})}\BibitemShut {NoStop}%
\bibitem [{\citenamefont {Jozsa}(1994)}]{Jozsa1994}%
  \BibitemOpen
  \bibfield  {author} {\bibinfo {author} {\bibfnamefont {R.}~\bibnamefont {Jozsa}},\ }\href {https://doi.org/10.1080/09500349414552171} {\bibfield  {journal} {\bibinfo  {journal} {J. Mod. Opt.}\ }\textbf {\bibinfo {volume} {41}},\ \bibinfo {pages} {2315} (\bibinfo {year} {1994})}\BibitemShut {NoStop}%
\bibitem [{\citenamefont {Tabesh}\ \emph {et~al.}(2025)\citenamefont {Tabesh}, \citenamefont {Rahimi-Keshari},\ and\ \citenamefont {Abdi}}]{Tabesh2025}%
  \BibitemOpen
  \bibfield  {author} {\bibinfo {author} {\bibfnamefont {F.~T.}\ \bibnamefont {Tabesh}}, \bibinfo {author} {\bibfnamefont {S.}~\bibnamefont {Rahimi-Keshari}},\ and\ \bibinfo {author} {\bibfnamefont {M.}~\bibnamefont {Abdi}},\ }\Eprint {https://arxiv.org/abs/2501.08055} {arXiv:2501.08055 [quant-ph]}  (\bibinfo {year} {2025})\BibitemShut {NoStop}%
\bibitem [{\citenamefont {Ye}\ \emph {et~al.}(2019)\citenamefont {Ye}, \citenamefont {Seo},\ and\ \citenamefont {Galli}}]{Ye2019}%
  \BibitemOpen
  \bibfield  {author} {\bibinfo {author} {\bibfnamefont {M.}~\bibnamefont {Ye}}, \bibinfo {author} {\bibfnamefont {H.}~\bibnamefont {Seo}},\ and\ \bibinfo {author} {\bibfnamefont {G.}~\bibnamefont {Galli}},\ }\bibfield  {journal} {\bibinfo  {journal} {npj Comput. Mater.}\ }\textbf {\bibinfo {volume} {5}},\ \href {https://doi.org/10.1038/s41524-019-0182-3} {10.1038/s41524-019-0182-3} (\bibinfo {year} {2019})\BibitemShut {NoStop}%
\bibitem [{\citenamefont {Lee}\ \emph {et~al.}(2022)\citenamefont {Lee}, \citenamefont {Park},\ and\ \citenamefont {Seo}}]{Lee2022}%
  \BibitemOpen
  \bibfield  {author} {\bibinfo {author} {\bibfnamefont {J.}~\bibnamefont {Lee}}, \bibinfo {author} {\bibfnamefont {H.}~\bibnamefont {Park}},\ and\ \bibinfo {author} {\bibfnamefont {H.}~\bibnamefont {Seo}},\ }\bibfield  {journal} {\bibinfo  {journal} {npj 2D Materials and Applications}\ }\textbf {\bibinfo {volume} {6}},\ \href {https://doi.org/10.1038/s41699-022-00336-2} {10.1038/s41699-022-00336-2} (\bibinfo {year} {2022})\BibitemShut {NoStop}%
\bibitem [{\citenamefont {Zhao}\ \emph {et~al.}(2012)\citenamefont {Zhao}, \citenamefont {Ho},\ and\ \citenamefont {Liu}}]{Zhao2012}%
  \BibitemOpen
  \bibfield  {author} {\bibinfo {author} {\bibfnamefont {N.}~\bibnamefont {Zhao}}, \bibinfo {author} {\bibfnamefont {S.-W.}\ \bibnamefont {Ho}},\ and\ \bibinfo {author} {\bibfnamefont {R.-B.}\ \bibnamefont {Liu}},\ }\bibfield  {journal} {\bibinfo  {journal} {Physical Review B}\ }\textbf {\bibinfo {volume} {85}},\ \href {https://doi.org/10.1103/physrevb.85.115303} {10.1103/physrevb.85.115303} (\bibinfo {year} {2012})\BibitemShut {NoStop}%
\bibitem [{\citenamefont {Park}\ \emph {et~al.}(2022)\citenamefont {Park}, \citenamefont {Lee}, \citenamefont {Han}, \citenamefont {Oh},\ and\ \citenamefont {Seo}}]{Park2022}%
  \BibitemOpen
  \bibfield  {author} {\bibinfo {author} {\bibfnamefont {H.}~\bibnamefont {Park}}, \bibinfo {author} {\bibfnamefont {J.}~\bibnamefont {Lee}}, \bibinfo {author} {\bibfnamefont {S.}~\bibnamefont {Han}}, \bibinfo {author} {\bibfnamefont {S.}~\bibnamefont {Oh}},\ and\ \bibinfo {author} {\bibfnamefont {H.}~\bibnamefont {Seo}},\ }\bibfield  {journal} {\bibinfo  {journal} {npj Quantum Information}\ }\textbf {\bibinfo {volume} {8}},\ \href {https://doi.org/10.1038/s41534-022-00665-6} {10.1038/s41534-022-00665-6} (\bibinfo {year} {2022})\BibitemShut {NoStop}%
\bibitem [{\citenamefont {Radishev}\ \emph {et~al.}(2021)\citenamefont {Radishev}, \citenamefont {Lobaev}, \citenamefont {Bogdanov}, \citenamefont {Gorbachev}, \citenamefont {Vikharev},\ and\ \citenamefont {Drozdov}}]{Radishev2021}%
  \BibitemOpen
  \bibfield  {author} {\bibinfo {author} {\bibfnamefont {D.}~\bibnamefont {Radishev}}, \bibinfo {author} {\bibfnamefont {M.}~\bibnamefont {Lobaev}}, \bibinfo {author} {\bibfnamefont {S.}~\bibnamefont {Bogdanov}}, \bibinfo {author} {\bibfnamefont {A.}~\bibnamefont {Gorbachev}}, \bibinfo {author} {\bibfnamefont {A.}~\bibnamefont {Vikharev}},\ and\ \bibinfo {author} {\bibfnamefont {M.}~\bibnamefont {Drozdov}},\ }\href {https://doi.org/10.1016/j.jlumin.2021.118404} {\bibfield  {journal} {\bibinfo  {journal} {Journal of Luminescence}\ }\textbf {\bibinfo {volume} {239}},\ \bibinfo {pages} {118404} (\bibinfo {year} {2021})}\BibitemShut {NoStop}%
\bibitem [{\citenamefont {Chrostoski}\ \emph {et~al.}(2022)\citenamefont {Chrostoski}, \citenamefont {Kehayias},\ and\ \citenamefont {Santamore}}]{Chrostoski2022}%
  \BibitemOpen
  \bibfield  {author} {\bibinfo {author} {\bibfnamefont {P.}~\bibnamefont {Chrostoski}}, \bibinfo {author} {\bibfnamefont {P.}~\bibnamefont {Kehayias}},\ and\ \bibinfo {author} {\bibfnamefont {D.~H.}\ \bibnamefont {Santamore}},\ }\href {https://doi.org/10.1103/PhysRevB.106.235311} {\bibfield  {journal} {\bibinfo  {journal} {Phys. Rev. B}\ }\textbf {\bibinfo {volume} {106}},\ \bibinfo {pages} {235311} (\bibinfo {year} {2022})}\BibitemShut {NoStop}%
\bibitem [{\citenamefont {Anderson}\ \emph {et~al.}(2022)\citenamefont {Anderson}, \citenamefont {Glen}, \citenamefont {Zeledon}, \citenamefont {Bourassa}, \citenamefont {Jin}, \citenamefont {Zhu}, \citenamefont {Vorwerk}, \citenamefont {Crook}, \citenamefont {Abe}, \citenamefont {Ul-Hassan}, \citenamefont {Ohshima}, \citenamefont {Son}, \citenamefont {Galli},\ and\ \citenamefont {Awschalom}}]{Anderson2022}%
  \BibitemOpen
  \bibfield  {author} {\bibinfo {author} {\bibfnamefont {C.~P.}\ \bibnamefont {Anderson}}, \bibinfo {author} {\bibfnamefont {E.~O.}\ \bibnamefont {Glen}}, \bibinfo {author} {\bibfnamefont {C.}~\bibnamefont {Zeledon}}, \bibinfo {author} {\bibfnamefont {A.}~\bibnamefont {Bourassa}}, \bibinfo {author} {\bibfnamefont {Y.}~\bibnamefont {Jin}}, \bibinfo {author} {\bibfnamefont {Y.}~\bibnamefont {Zhu}}, \bibinfo {author} {\bibfnamefont {C.}~\bibnamefont {Vorwerk}}, \bibinfo {author} {\bibfnamefont {A.~L.}\ \bibnamefont {Crook}}, \bibinfo {author} {\bibfnamefont {H.}~\bibnamefont {Abe}}, \bibinfo {author} {\bibfnamefont {J.}~\bibnamefont {Ul-Hassan}}, \bibinfo {author} {\bibfnamefont {T.}~\bibnamefont {Ohshima}}, \bibinfo {author} {\bibfnamefont {N.~T.}\ \bibnamefont {Son}}, \bibinfo {author} {\bibfnamefont {G.}~\bibnamefont {Galli}},\ and\ \bibinfo {author} {\bibfnamefont {D.~D.}\ \bibnamefont {Awschalom}},\ }\href {https://doi.org/10.1126/sciadv.abm5912} {\bibfield  {journal} {\bibinfo  {journal} {Science
  Advances}\ }\textbf {\bibinfo {volume} {8}},\ \bibinfo {pages} {eabm5912} (\bibinfo {year} {2022})}\BibitemShut {NoStop}%
\bibitem [{\citenamefont {Ramsay}\ \emph {et~al.}(2023)\citenamefont {Ramsay}, \citenamefont {Hekmati}, \citenamefont {Patrickson}, \citenamefont {Baber}, \citenamefont {Arvidsson-Shukur}, \citenamefont {Bennett},\ and\ \citenamefont {Luxmoore}}]{Ramsay2023}%
  \BibitemOpen
  \bibfield  {author} {\bibinfo {author} {\bibfnamefont {A.~J.}\ \bibnamefont {Ramsay}}, \bibinfo {author} {\bibfnamefont {R.}~\bibnamefont {Hekmati}}, \bibinfo {author} {\bibfnamefont {C.~J.}\ \bibnamefont {Patrickson}}, \bibinfo {author} {\bibfnamefont {S.}~\bibnamefont {Baber}}, \bibinfo {author} {\bibfnamefont {D.~R.~M.}\ \bibnamefont {Arvidsson-Shukur}}, \bibinfo {author} {\bibfnamefont {A.~J.}\ \bibnamefont {Bennett}},\ and\ \bibinfo {author} {\bibfnamefont {I.~J.}\ \bibnamefont {Luxmoore}},\ }\href {https://doi.org/10.1038/s41467-023-36196-7} {\bibfield  {journal} {\bibinfo  {journal} {Nat. Commun.}\ }\textbf {\bibinfo {volume} {14}},\ \bibinfo {pages} {461} (\bibinfo {year} {2023})}\BibitemShut {NoStop}%
\bibitem [{\citenamefont {Pla}\ \emph {et~al.}(2013)\citenamefont {Pla}, \citenamefont {Tan}, \citenamefont {Dehollain}, \citenamefont {Lim}, \citenamefont {Morton}, \citenamefont {Zwanenburg}, \citenamefont {Jamieson}, \citenamefont {Dzurak},\ and\ \citenamefont {Morello}}]{Pla2013}%
  \BibitemOpen
  \bibfield  {author} {\bibinfo {author} {\bibfnamefont {J.~J.}\ \bibnamefont {Pla}}, \bibinfo {author} {\bibfnamefont {K.~Y.}\ \bibnamefont {Tan}}, \bibinfo {author} {\bibfnamefont {J.~P.}\ \bibnamefont {Dehollain}}, \bibinfo {author} {\bibfnamefont {W.~H.}\ \bibnamefont {Lim}}, \bibinfo {author} {\bibfnamefont {J.~J.~L.}\ \bibnamefont {Morton}}, \bibinfo {author} {\bibfnamefont {F.~A.}\ \bibnamefont {Zwanenburg}}, \bibinfo {author} {\bibfnamefont {D.~N.}\ \bibnamefont {Jamieson}}, \bibinfo {author} {\bibfnamefont {A.~S.}\ \bibnamefont {Dzurak}},\ and\ \bibinfo {author} {\bibfnamefont {A.}~\bibnamefont {Morello}},\ }\href {https://doi.org/10.1038/nature12011} {\bibfield  {journal} {\bibinfo  {journal} {Nature}\ }\textbf {\bibinfo {volume} {496}},\ \bibinfo {pages} {334} (\bibinfo {year} {2013})}\BibitemShut {NoStop}%
\end{thebibliography}%

\end{document}